\documentclass[twoside,reqno]{bjp}
\usepackage{epsfig,cite}
\usepackage{graphics}
\usepackage{amssymb,amsmath}
\usepackage{times}
\begin{document}

\sloppy \raggedbottom

 \setcounter{page}{1}

\title{Proxy-SU(3): A symmetry for heavy nuclei}

\runningheads{Proxy-SU(3): A symmetry for heavy nuclei}{D. Bonatsos, I.E. Assimakis, N. Minkov, A. Martinou, S. K. Peroulis, et al.}

\begin{start}

\author{D. Bonatsos}{1},
\coauthor{I.E. Assimakis}{1},
\coauthor{N. Minkov}{2},
\coauthor{A. Martinou}{1},
\coauthor{S. K. Peroulis}{1},
\coauthor{S. Sarantopoulou}{1},
\coauthor{R.B. Cakirli}{3},
\coauthor{R.F. Casten}{4,5},
\coauthor{K. Blaum}{6}

\address{Institute of Nuclear and Particle Physics, National Centre for Scientific Research ``Demokritos'', GR-15310 Aghia Paraskevi, Attiki, Greece}{1}

\address{Institute of Nuclear Research and Nuclear Energy, Bulgarian Academy of Sciences, 72 Tzarigrad Road, 1784 Sofia, Bulgaria}{2}

\address{Department of Physics, University of Istanbul, 34134 Istanbul, Turkey}{3} 

\address{Wright Laboratory, Yale University, New Haven, Connecticut 06520, USA}{4}

\address{Facility for Rare Isotope Beams, 640 South Shaw Lane, Michigan State University, East Lansing, MI 48824 USA}{5}

\address{Max-Planck-Institut f\"{u}r Kernphysik, Saupfercheckweg 1, D-69117 Heidelberg, Germany}{6}

\received{31 October 2017}

\begin{Abstract} 

The SU(3) symmetry realized by J. P. Elliott in the sd nuclear shell is destroyed in heavier shells by the strong spin-orbit interaction. On the other hand, the SU(3) symmetry has been used for the description of heavy nuclei in terms of bosons in the framework of the Interacting Boson Approximation, as well as in terms of fermions using the pseudo-SU(3) approximation. A new fermionic approximation, called the proxy-SU(3), has been recently introduced and applied to the even rare earths. We show that the applicability of proxy-SU(3) can be extended to even nuclei in the 28-50 proton shell, to even superheavy elements, as well as to odd-odd and odd rare earths. Parameter free predictions for the $\beta$ and $\gamma$ deformation parameters are presented and compared to alternative theoretical predictions and to existing data. 

\end{Abstract}

\PACS {21.60.Fw, 21.60.Ev, 21.60.Cs}
\end{start}

\section{Introduction}

A new algebraic approach to heavy deformed nuclei, based on fermionic symmetries and called the proxy-SU(3) scheme, has been introduced recently \cite{proxy1,proxy2}. Its basic assumptions and microscopic justification have been discussed in Ref. \cite{proxy1} and are further considered in the present Workshop in Ref. \cite{Assimak}. A first success of the proxy-SU(3) scheme is the explanation of the prolate over oblate dominance in deformed nuclei,
which has been considered in Refs. \cite{proxy2,EPJA} and is further discussed in the present Workshop in Ref. \cite{Saranto}. The border of the prolate to oblate transition is also determined \cite{proxy2,Saranto}. In addition, parameter-free predictions for the deformation parameters $\beta$ and $\gamma$ for even rare earths have been predicted in Ref. \cite{proxy2} and successfully compared to Relativistic Mean Field predictions \cite{Lalazissis} and to existing data \cite{Raman}. 

In the present work we obtain parameter-free predictions for the deformation parameters $\beta$ and $\gamma$ for nuclei in the 28-50 proton shell, as well as for even superheavy elements and we compare them to alternative theoretical predictions and to existing data. Furthermore, we apply the proxy-SU(3) scheme in odd-odd rare earths and odd rare earths 
and compare the results to existing theoretical predictions.  

\section{Numerical results}

\begin{figure}[b]
\centering{\epsfig{file=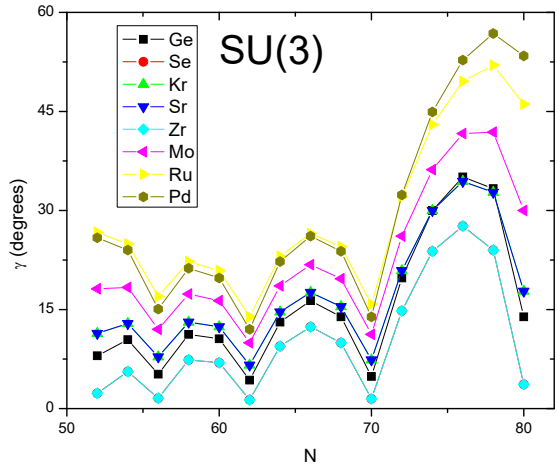,width=95mm}}
\caption{Proxy SU(3) predictions for $Z=32$-46 for $\gamma$. See Section \ref{Z2850} for further discussion.}
\end{figure}

\subsection{Even nuclei in the 28-50 proton shell}\label{Z2850}

This shell is of particular current interest, because of the presence of the $Z=40$ subshell closure and the appearance of shape coexistence \cite{Wood,Heyde} around it. Parameter-free proxy-SU(3) predictions for the $\gamma$ deformation parameter are shown in Fig. 1 and are compared to D1S Gogny calculations \cite{Gogny} in Fig. 2, while in Fig. 3 the parameter free proxy-SU(3) predictions for the $\beta$ deformation parameter are shown and compared to theoretical predictions by Relativistic Mean Field with the NL3 parametrization\cite{Lalazissis}, D1S Gogny interaction \cite{Gogny}, and the FRDM (2012) mass tables \cite{Moller}, as well as to available data \cite {Raman}. 

In general, good agreement is observed for the $\beta$ deformation parameter between the proxy-SU(3) predictions and the predictions of alternative theories and the data.
Deviations are stronger in the Sr, Zr, and Mo isotopes ($Z=38-42$), which lie on or close 
to the $Z=40$ subshell closure, which is not taken into account in the proxy-SU(3) scheme
in any specific way.  

In the case of the $\gamma$ deformation parameter, proxy-SU(3) shows a tendency to values near 30 degrees, indicating triaxial shapes, in the region around $N=74$, while it climbs at even higher values, approaching 60 degrees (indicating oblate shapes) in the Ru and Pd isotopes 
in this region. Further study of these results is needed, taking into account that nuclei 
near the end of the neutron shell are not very well deformed, as indicated, for example, 
by their $R_{4/2}=E(4_1^+)/E(2_1^+)$ ratios \cite{ENSDF}.

\subsection{Odd-odd and even-odd rare earths}\label{oddodd}

Proxy-SU(3) results have been obtained for odd-odd and even-odd rare earths 
 with $Z$ in the $sdg$ proxy-SU(3) shell (which is an approximation of the 50-82 shell) and neutrons in the $pfh$ proxy-SU(3) shell (which is an approximation of the 82-126 shell). 

Parameter-free proxy-SU(3) predictions for the $\beta$ and $\gamma$ deformation parameters 
for odd-odd nuclei are shown in Figs. 4 and 5, compared to predictions for $\beta$ reported in the mass table FRDM(2012) \cite{Moller}, where they have been calculated within the finite-range droplet macroscopic model and the folded-Yukawa single-particle microscopic model. 
Good agreement is observed in general, with the largest deviations appearing for Au ($Z=79$), 
i.e., near the end of the 50-82 proton shell. 

A similar set of figures for odd-$N$  rare earths appears in Figs. 6 and 7. Again, the higher 
deviations appear near the end of the 50-82 proton shell, at the Pt ($Z=78$) isotopes. 

\subsection{Even superheavy elements}\label{eSHE}

Parameter independent proxy-SU(3) results have been obtained for superheavy elements (SHE) with $Z$ in the $pfh$ proxy-SU(3) shell (which is an approximation of the 82-126 shell) and neutrons in the sdgi proxy-SU(3) shell (which is an approximation of the 126-184 shell), 
as well as in the pfhj proxy-SU(3) shell (which is an approximation of the 184-258 shell).  
For the illustrative and pedagogical purposes of this work, we take the relevant shells for the actinides and super heavy nuclei as $Z = 82$-126,  $N = 126$-184, and $N=184$-258, although the upper bounds are by no means certain and microscopic calculations give many varying scenarios. Results are shown for $100\leq Z \leq 114$.
In order to have results from alternative calculations to compare our results with, 
we confine ourselves to $128\leq N\leq 220$.

We compare our results to predictions contained in the following sources.

Extended results for $10\leq Z\leq 110$ and $N\leq 200$ with the D1S Gogny interaction 
are given in \cite{Gogny} for the mean ground state $\beta$ deformation, as well as 
for the mean ground state $\gamma$ deformation.  

Extended results for the proton deformation $\beta_p$ and the neutron deformation 
$\beta_n$ with covariant density functional theory for $96\leq Z \leq 130$
and $N$ from the proton drip line up to $N=196$ are given in Ref. \cite{Ring} for the functionals PC-PK1 and DD-PC1. 

Extended results for the deformation $\beta$ within a microscopic-macroscopic method (MMM) for $98\leq Z \leq 126$ and $134\leq N \leq 192$ are given in Ref. \cite{Skalski}. 

Extended results for the deformation $\beta$ up to $A=339$ are reported in the mass table FRDM(2012) \cite{Moller}, calculated within the finite-range droplet macroscopic model and the folded-Yukawa single-particle microscopic model. 

The results are summarized in Figs. 8 and 9. Overall good agreement is observed between the 
parameter-free proxy-SU(3) predictions and the alternative calculations. 

\section{Conclusion} 

In the present work, the applicability of the proxy-SU(3) scheme is tested, through parameter independent predictions for the deformation parameters $\beta$ and $\gamma$ for even nuclei 
in the 28-50 proton shell and even superheavy elements, as well as in odd-odd nuclei 
and odd-$N$ rare earths. In general, good agreement is obtained with results of alternative
calculations, as well as with existing data. Some deviations seen, as for example around the $Z=40$ subshell closure, which is playing a central role in shape coexistence in the relevant region, call for further investigations. 

\section*{Acknowledgements}

Work partly supported by the Bulgarian National Science Fund (BNSF) under Contract No. DFNI-E02/6, by the US DOE under Grant No. DE-FG02- 91ER-40609, and by the MSU-FRIB laboratory, by the Max Planck Partner group, TUBA-GEBIP, and by the Istanbul University Scientific Research Project No. 54135.



\begin{figure}[b]
\epsfig{file=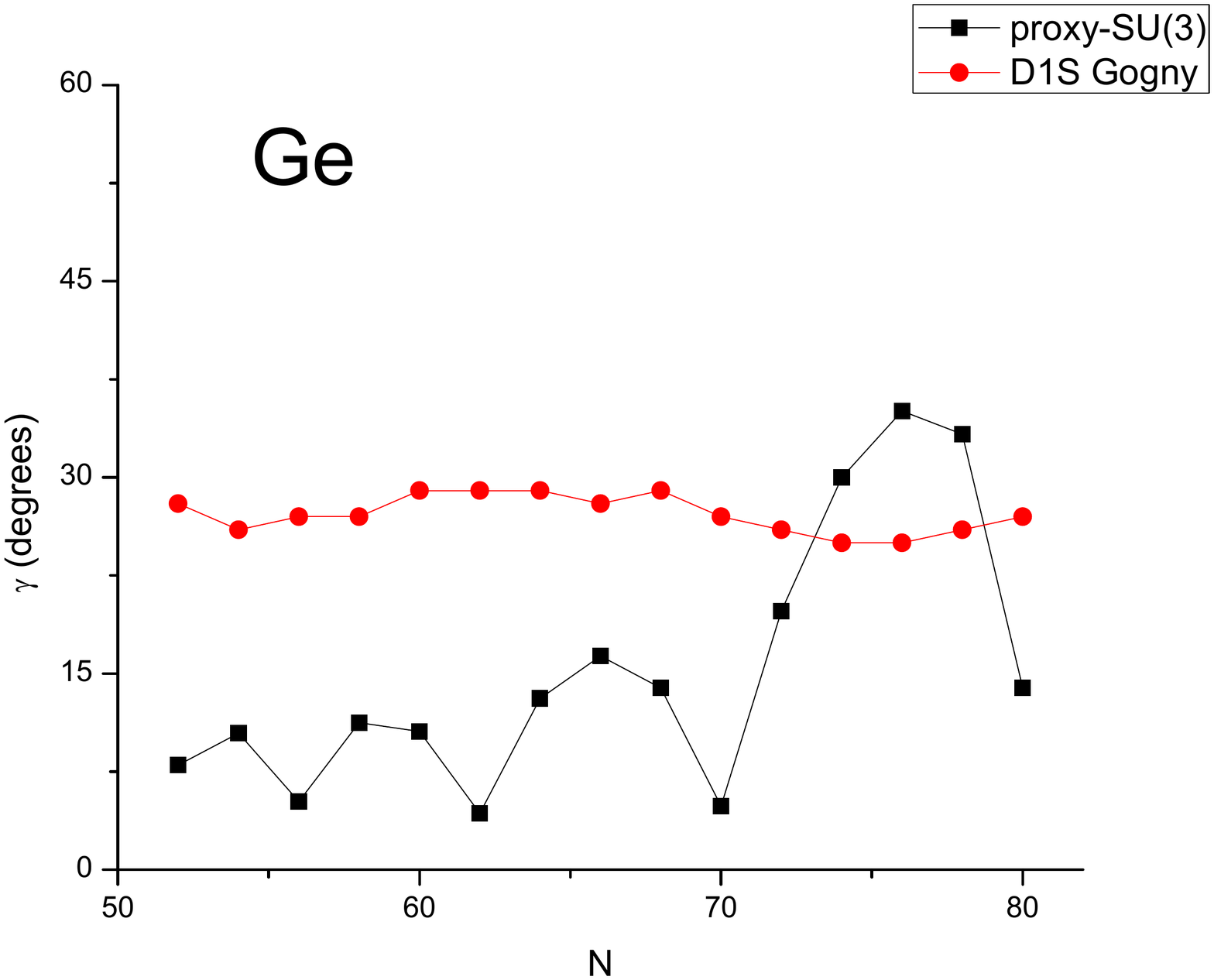,width=50mm}
\epsfig{file=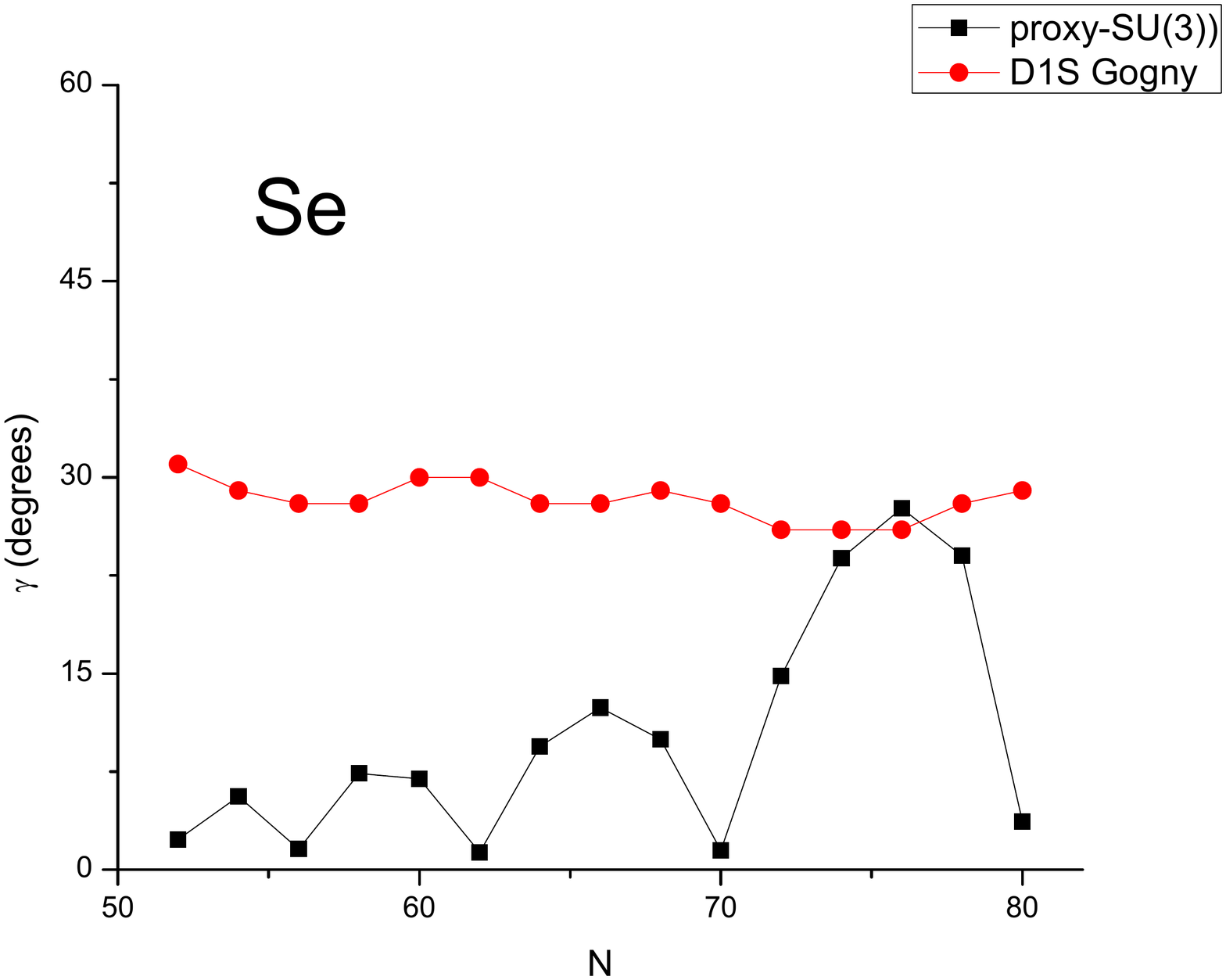,width=50mm}

\epsfig{file=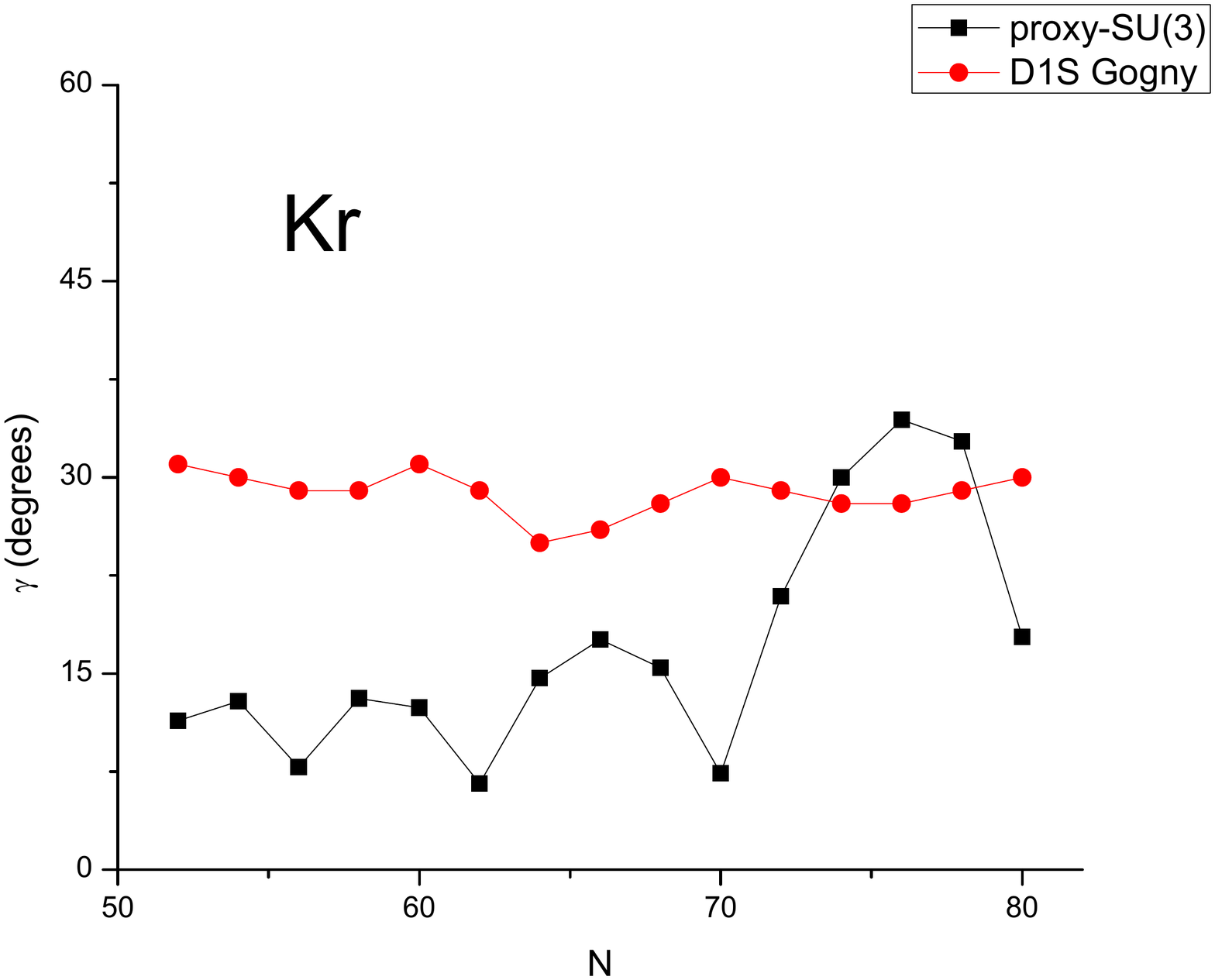,width=50mm}
\epsfig{file=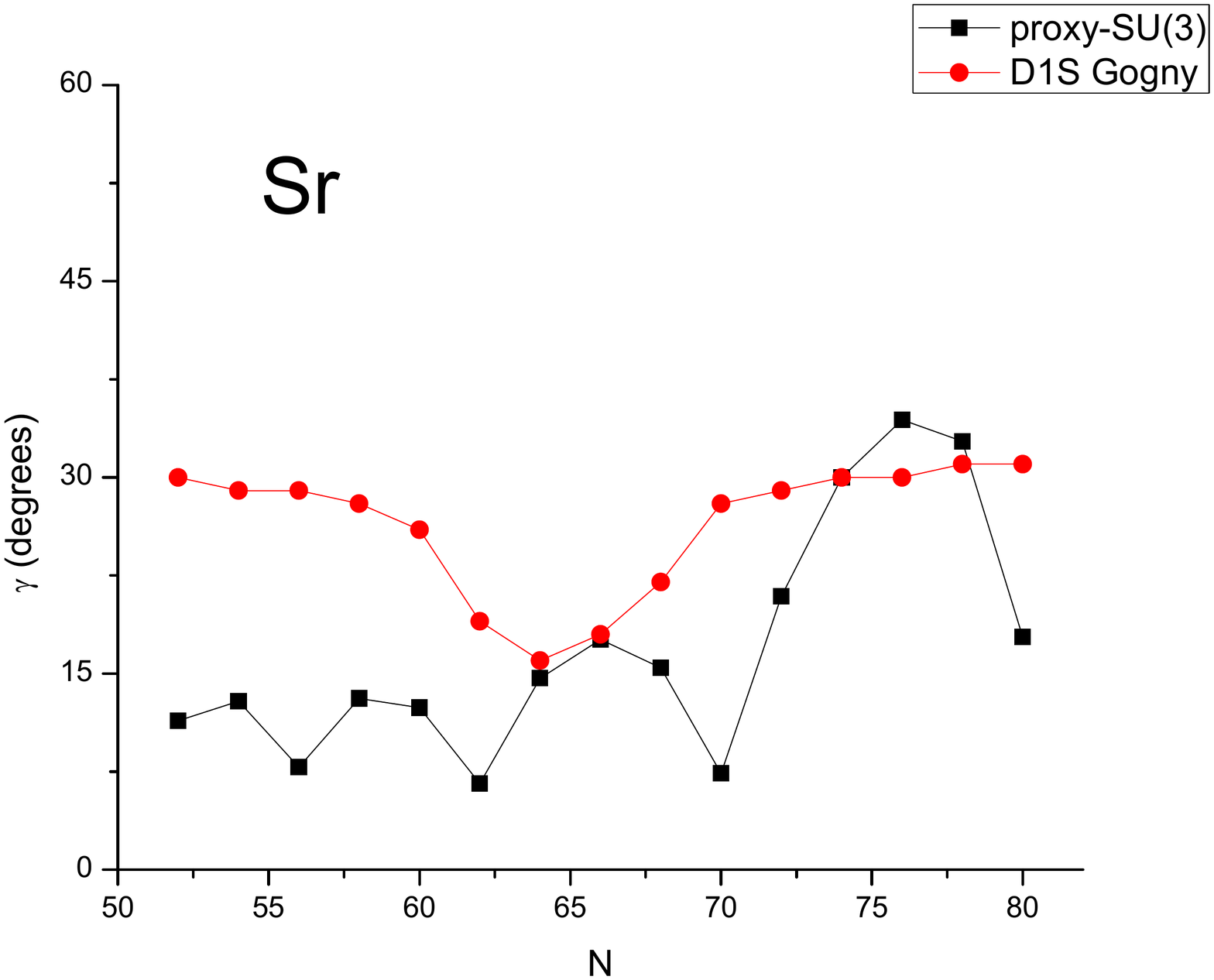,width=50mm}

\epsfig{file=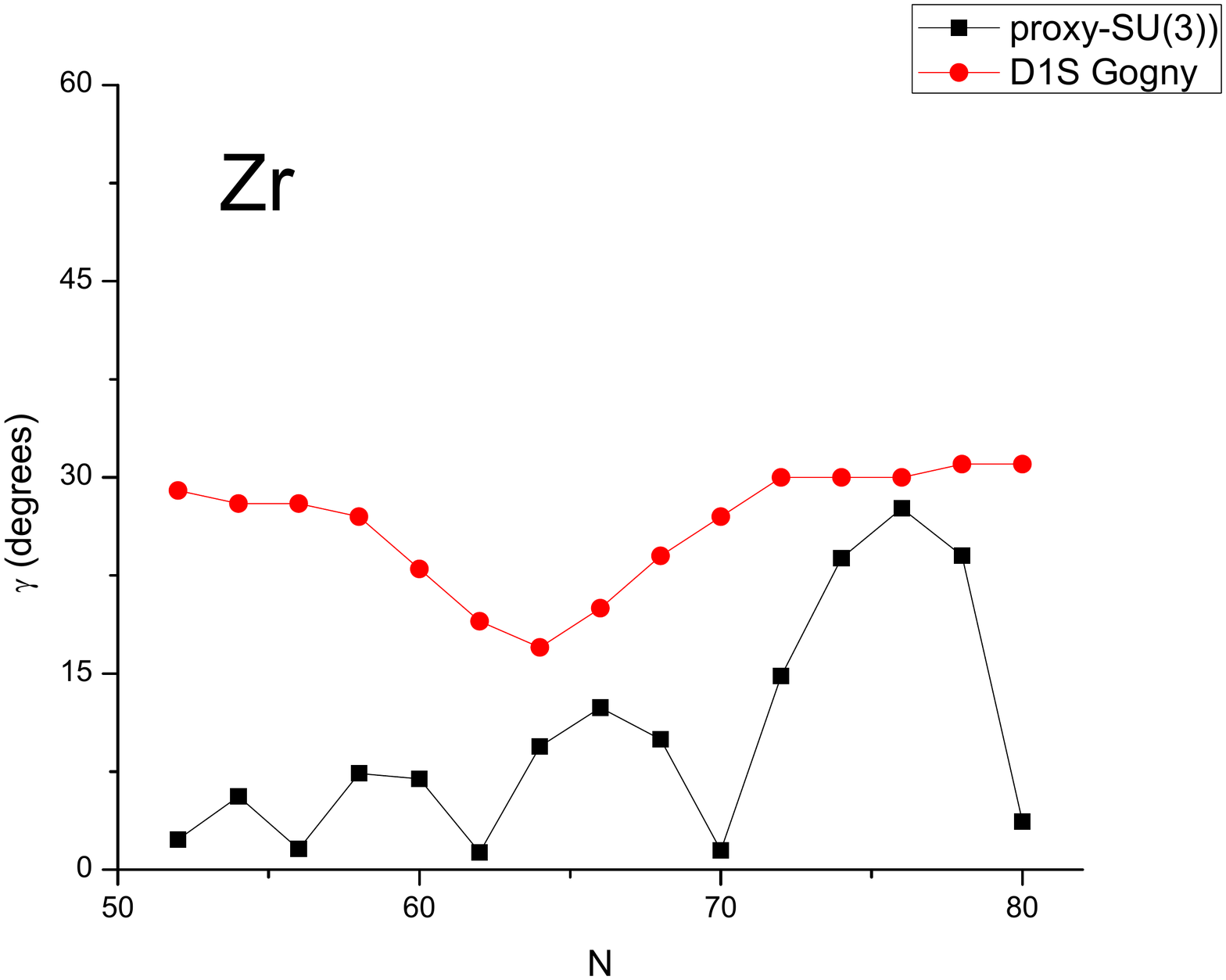,width=50mm}
\epsfig{file=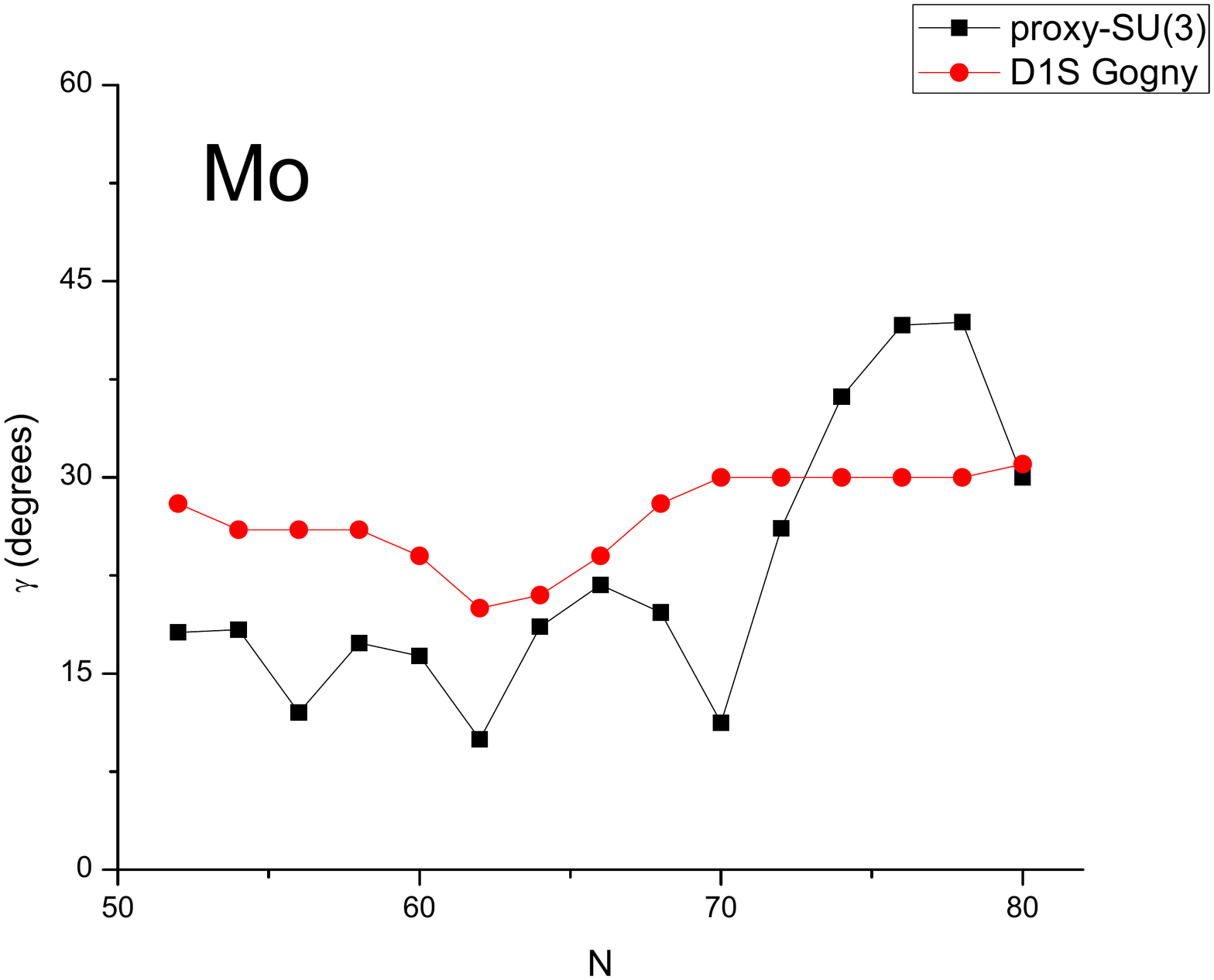,width=50mm}

\epsfig{file=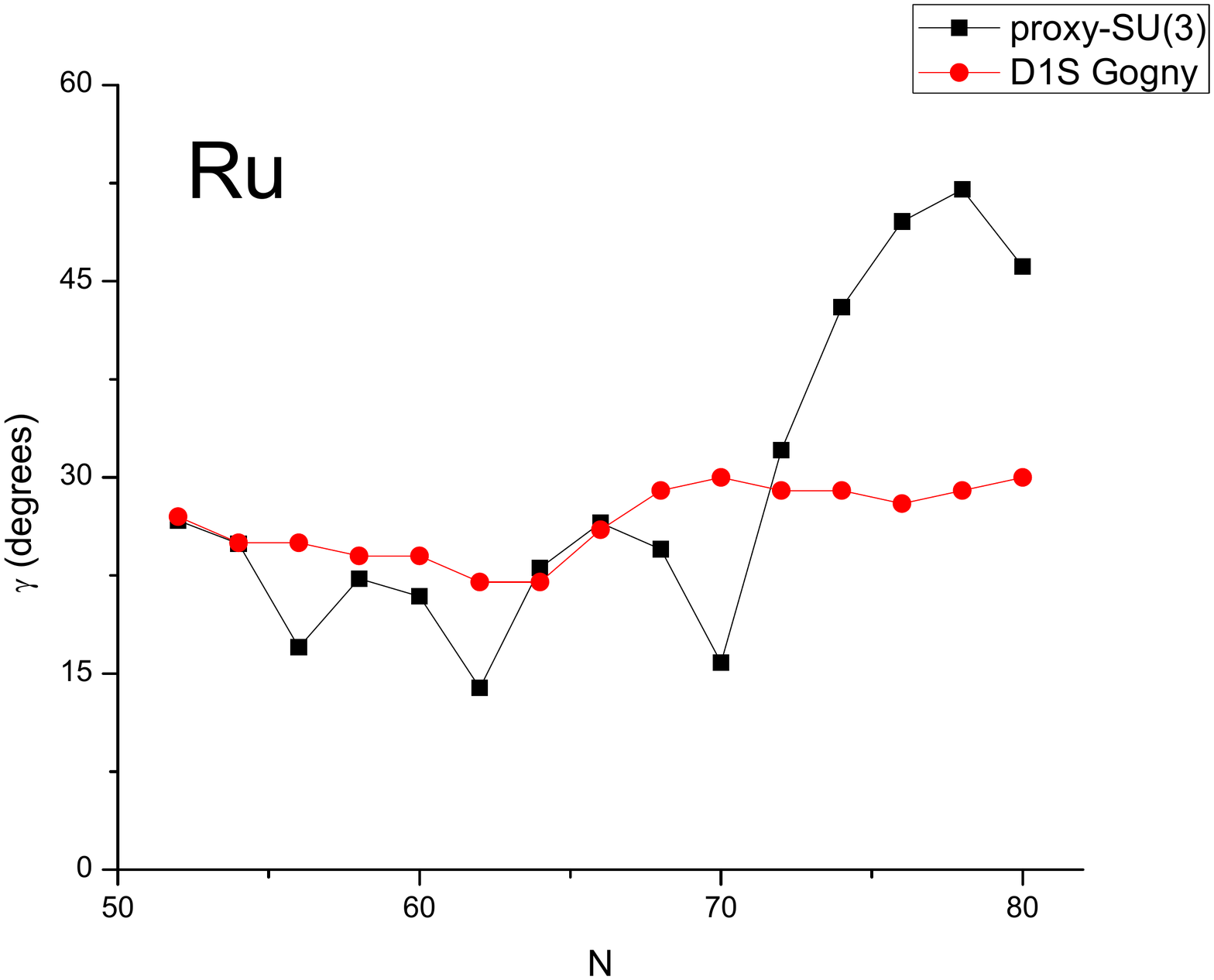,width=50mm}
\epsfig{file=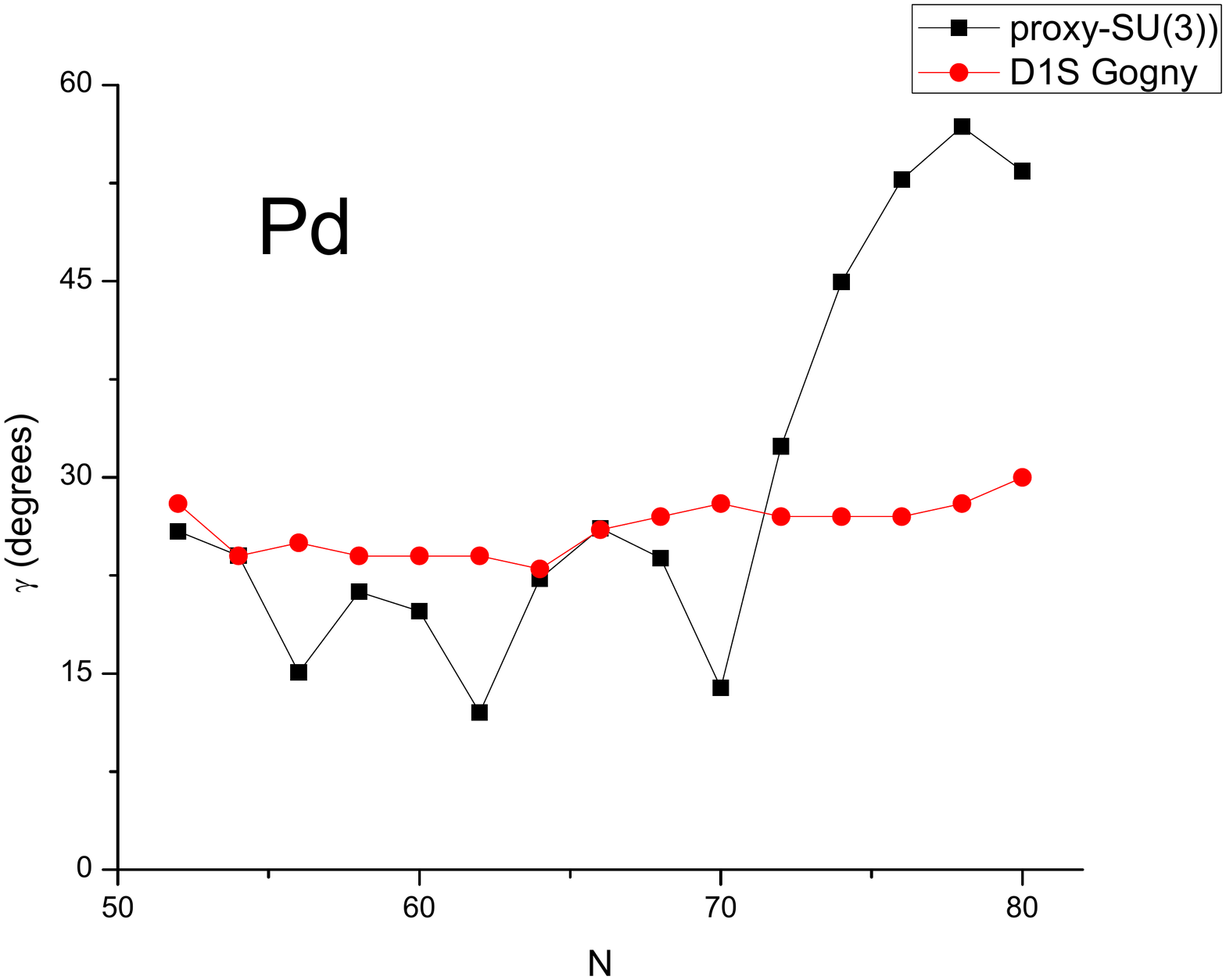,width=50mm}

\caption{Proxy SU(3) predictions for $Z=32$-46 for $\gamma$, compared with
D1S-Gogny calculations (D1S-Gogny) \cite{Gogny}. See Section \ref{Z2850} for further discussion.}

\end{figure}


\begin{figure}[b]
\epsfig{file=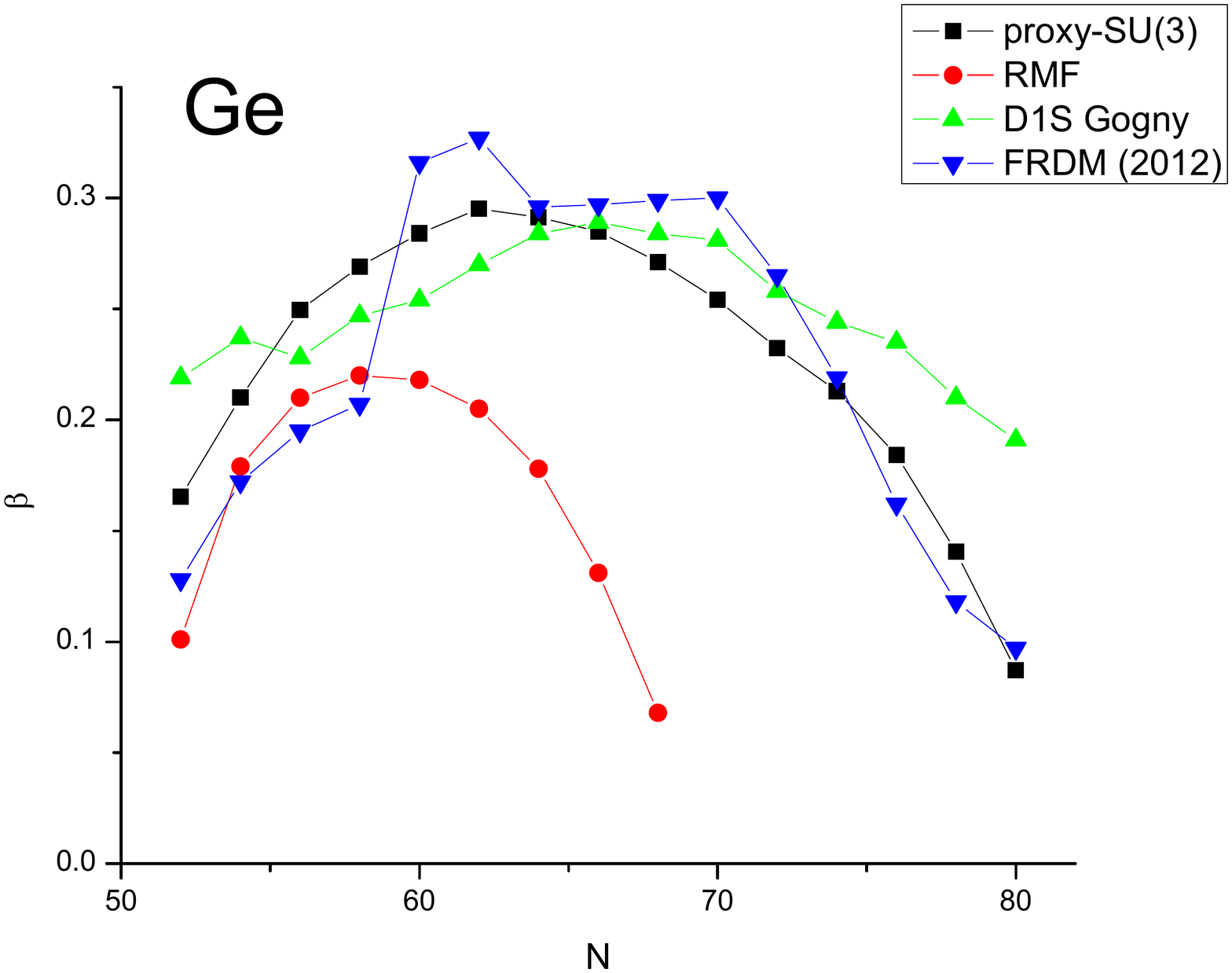,width=50mm}
\epsfig{file=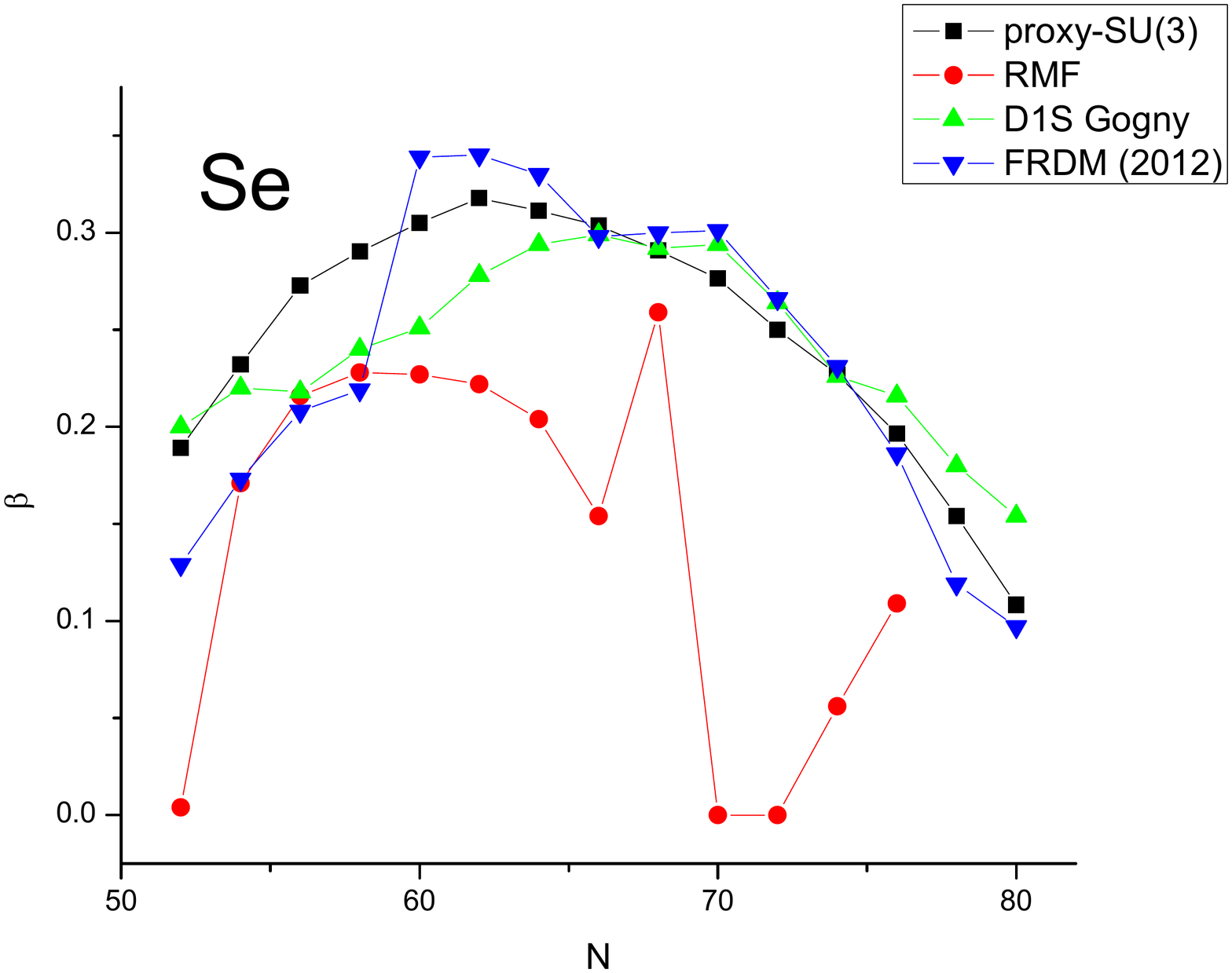,width=50mm}

\epsfig{file=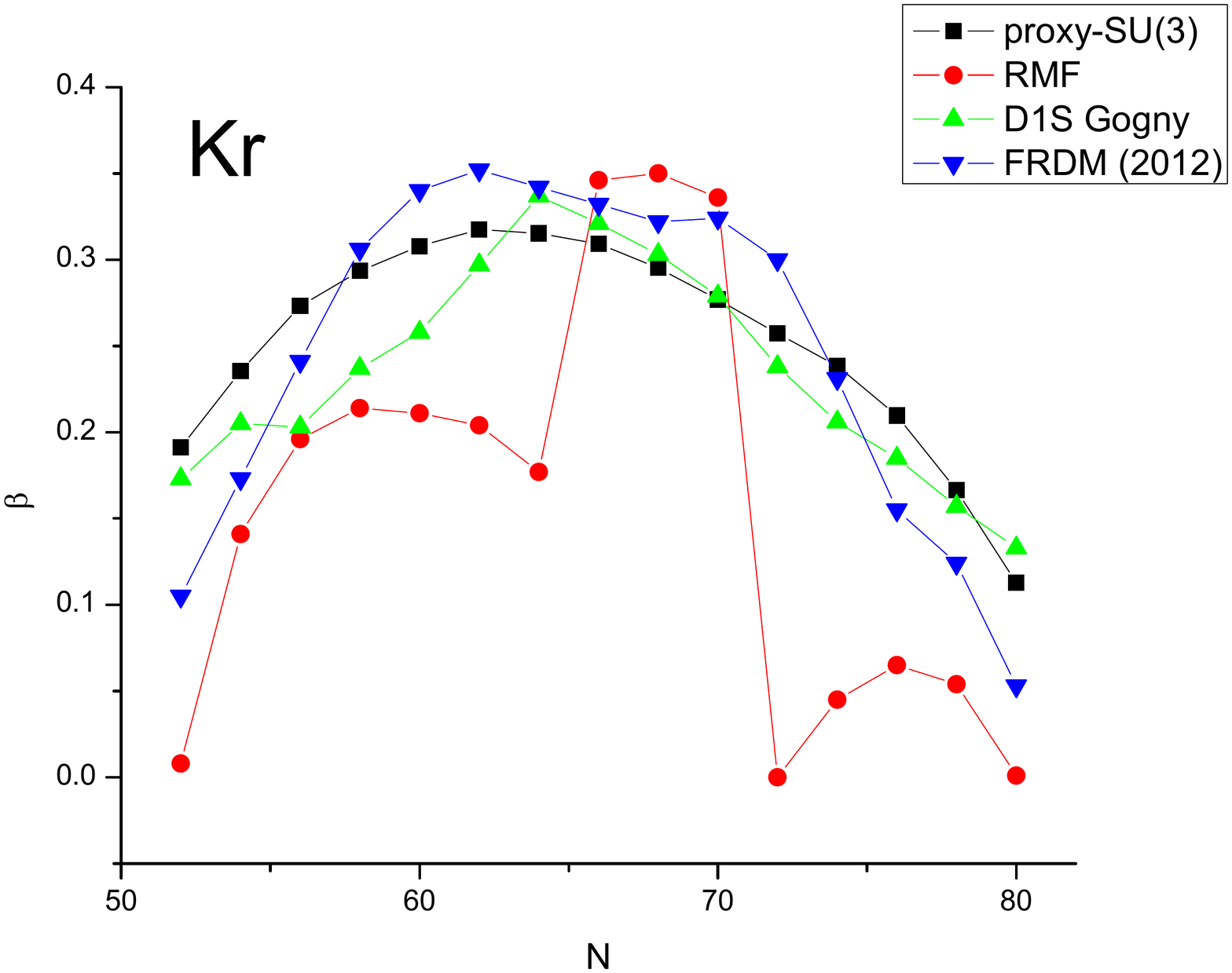,width=50mm}
\epsfig{file=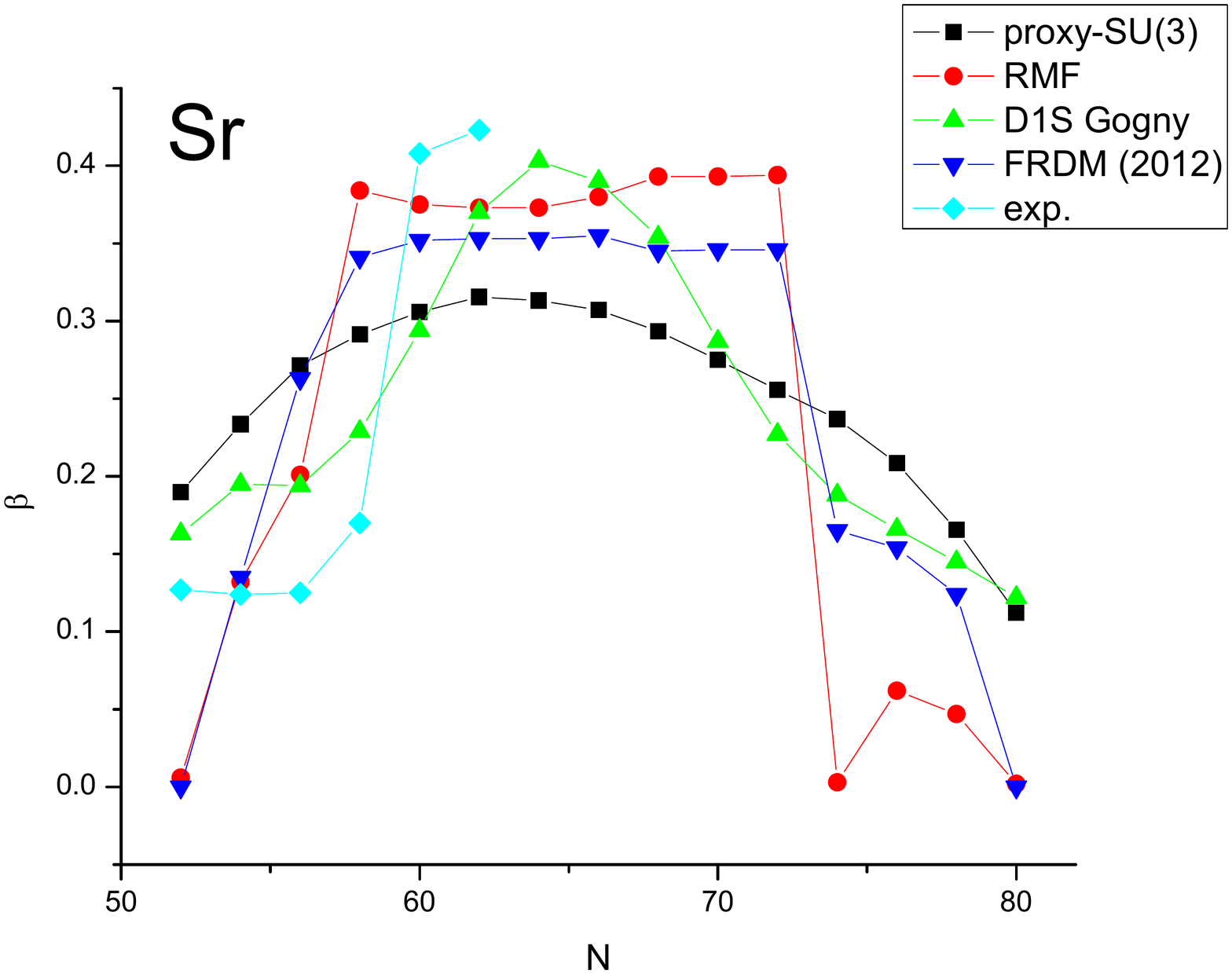,width=50mm}

\epsfig{file=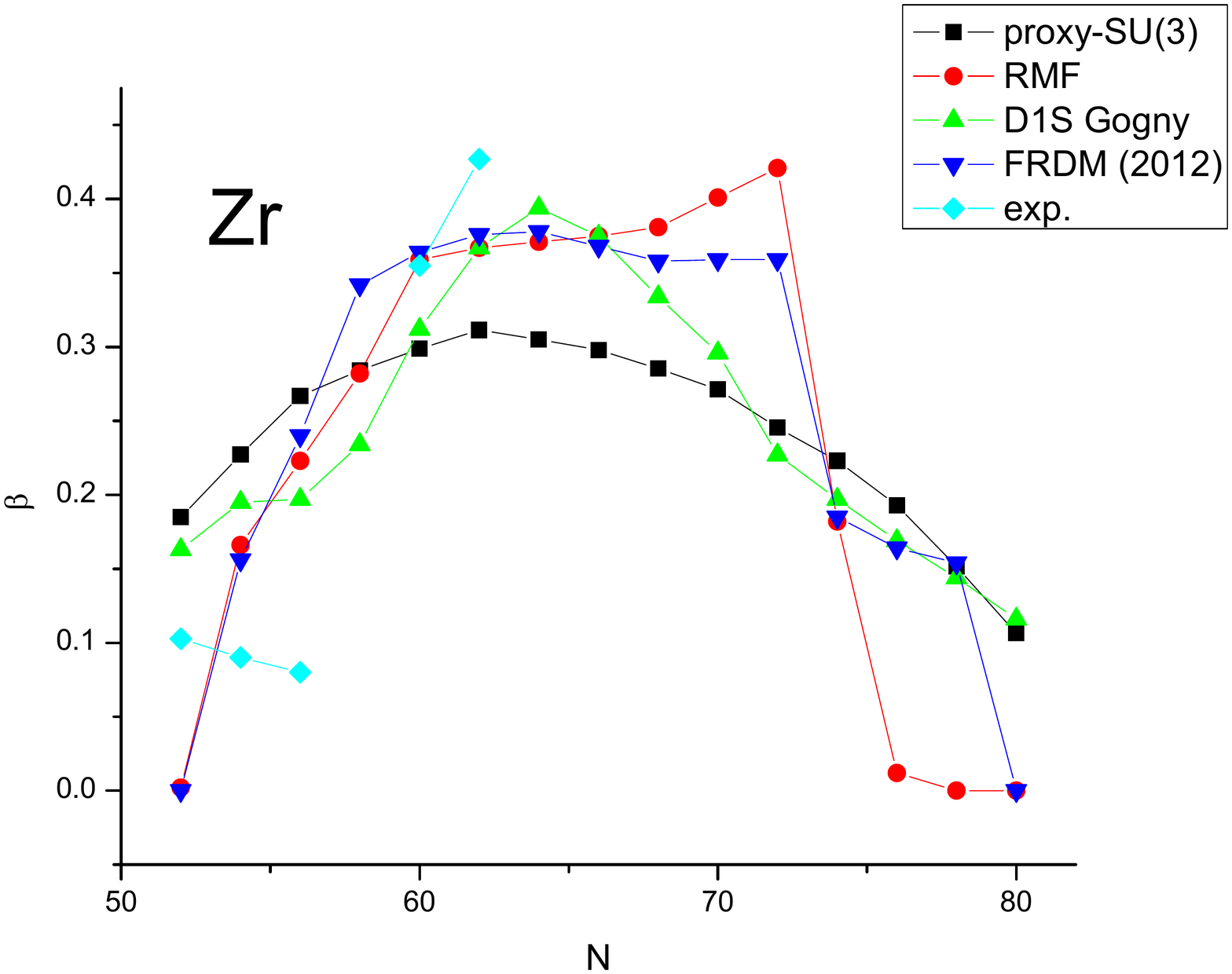,width=50mm}
\epsfig{file=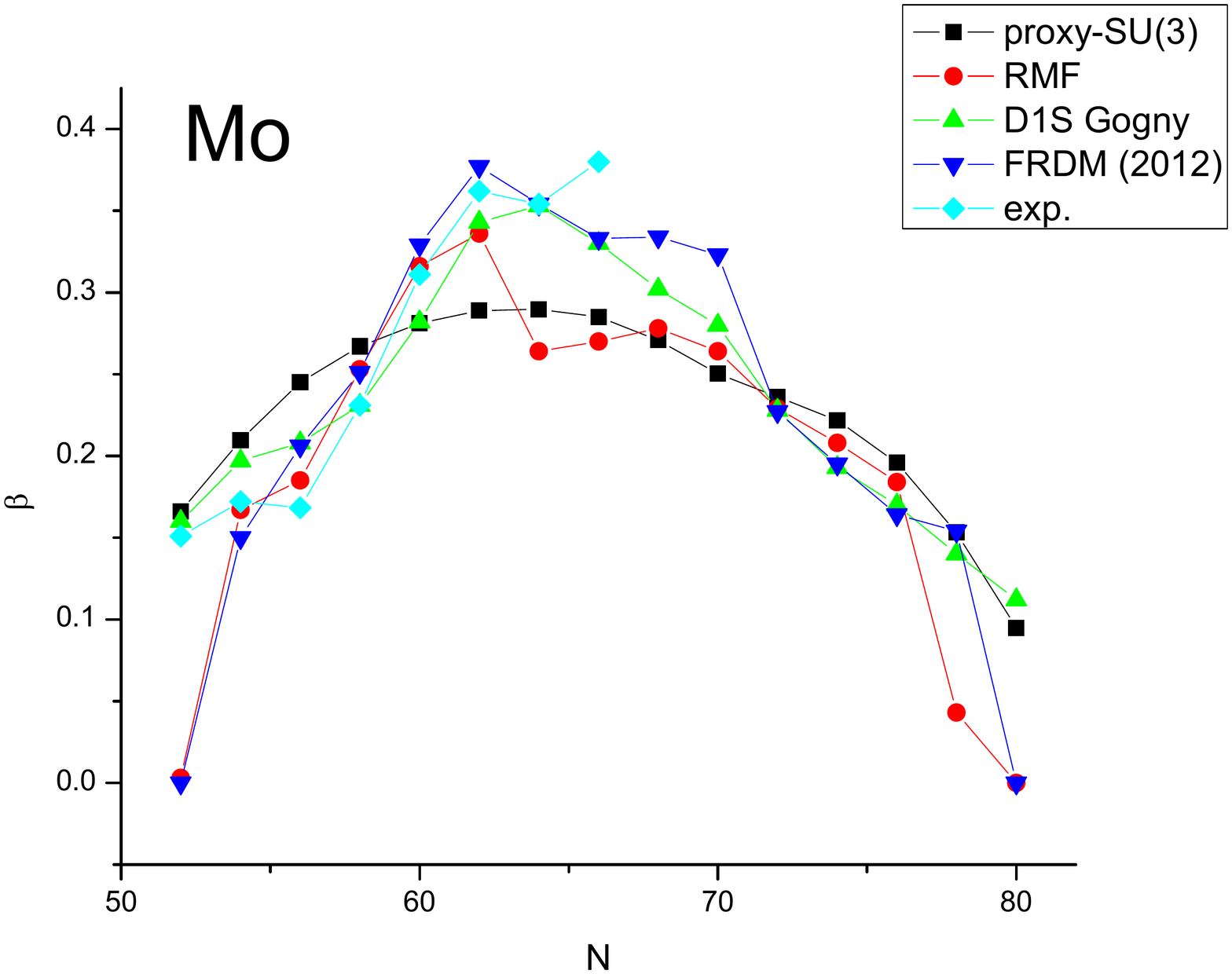,width=50mm}

\epsfig{file=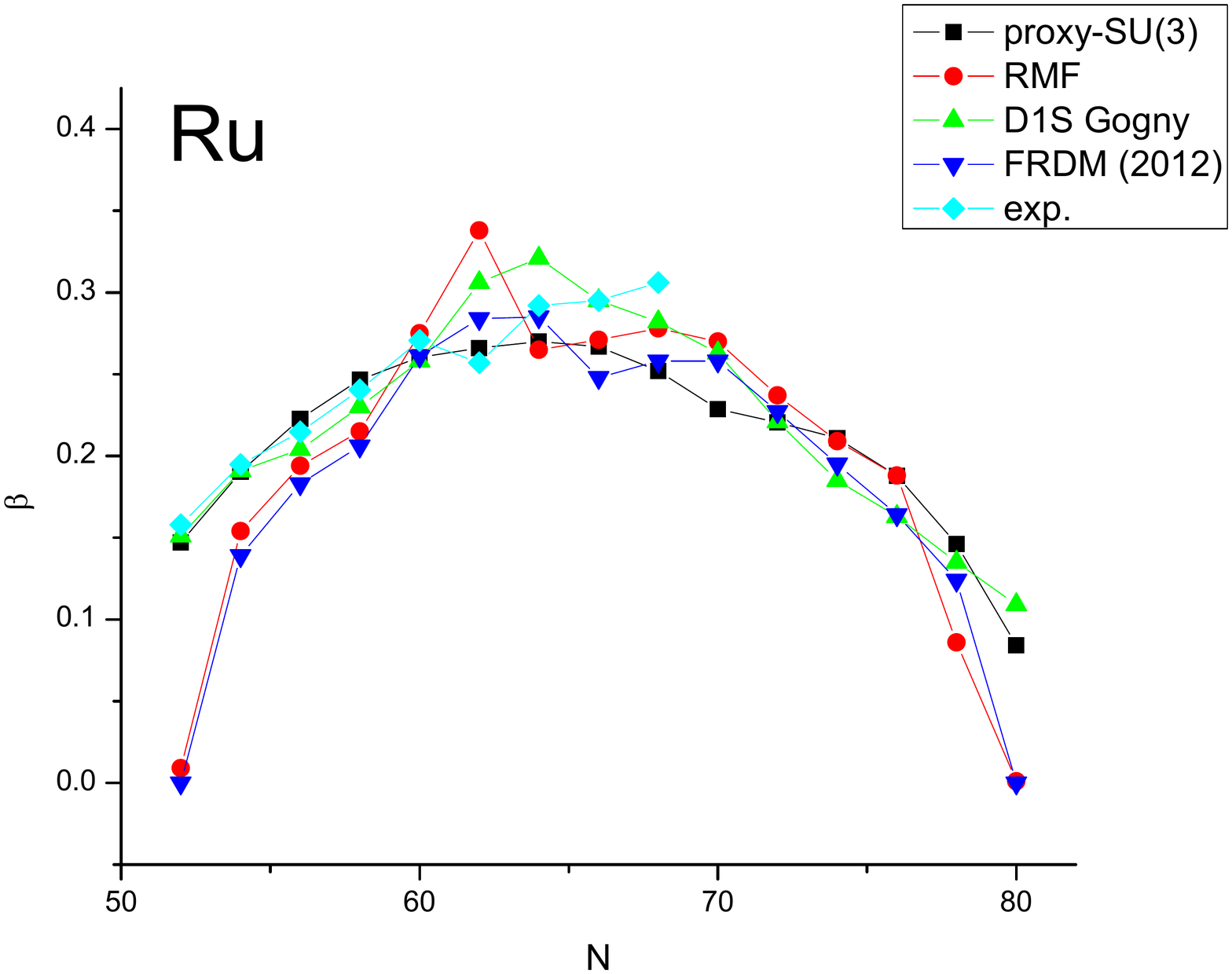,width=50mm}
\epsfig{file=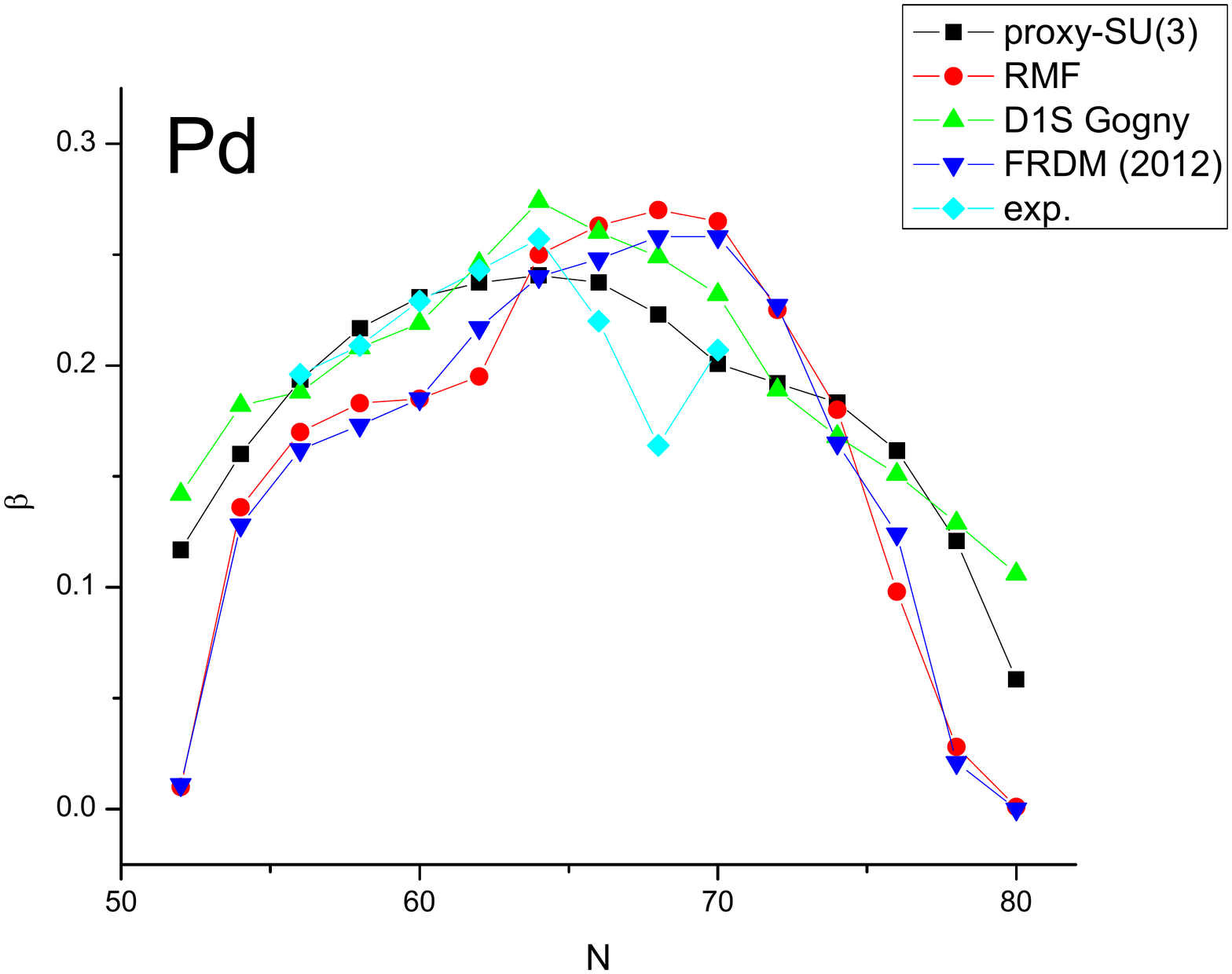,width=50mm}

\caption{Proxy SU(3) predictions for $Z=32$-46 for $\beta$, compared with
results by relativistic mean field theory (RMF) \cite{Lalazissis}, the D1S-Gogny interaction (D1S-Gogny) \cite{Gogny}, and the mass table FRDM(2012) \cite{Moller}, as well as to the data (exp.) \cite{Raman}. See Section \ref{Z2850} for further discussion.} 

\end{figure}



\begin{figure}[b]
\epsfig{file=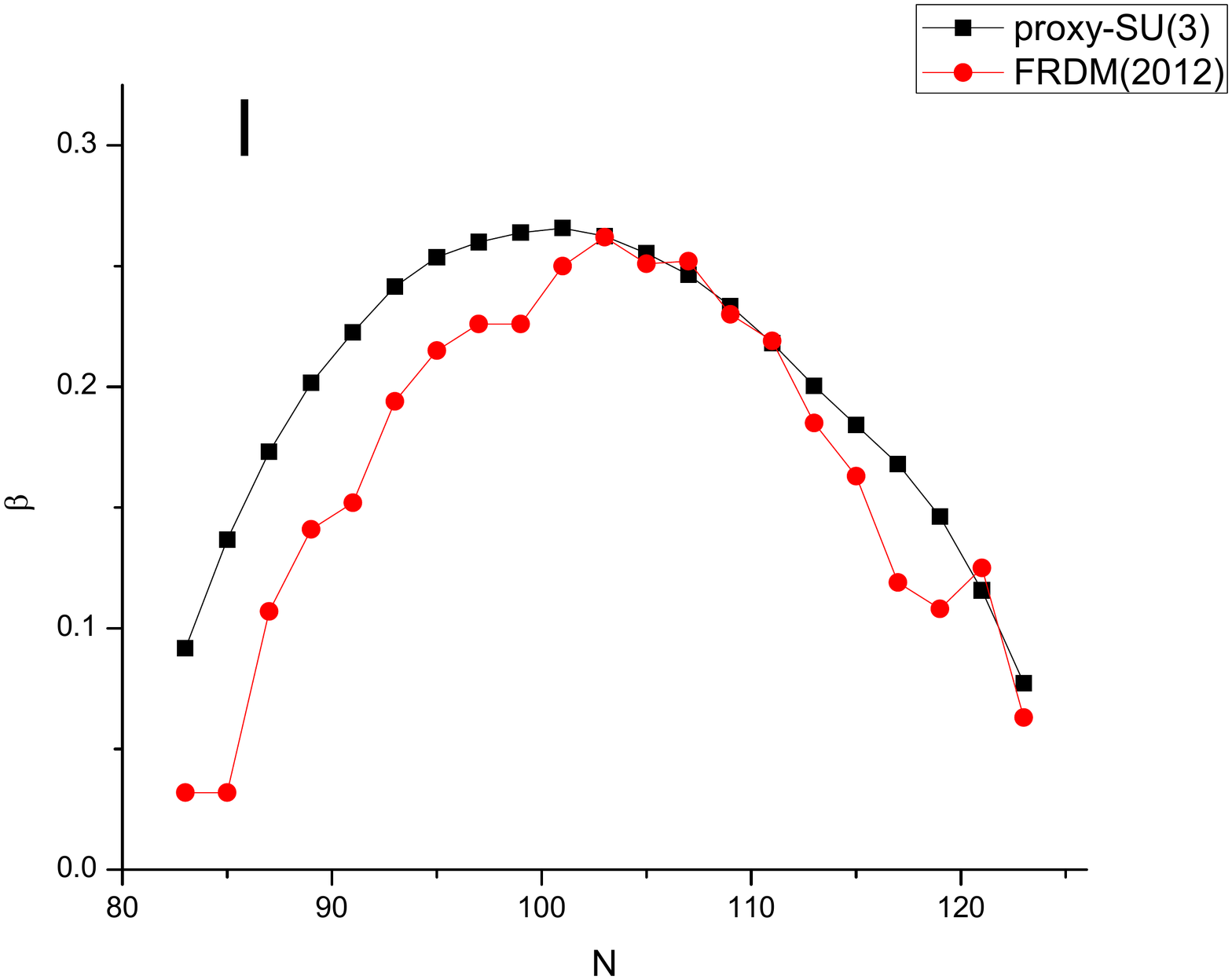,width=55mm}
\epsfig{file=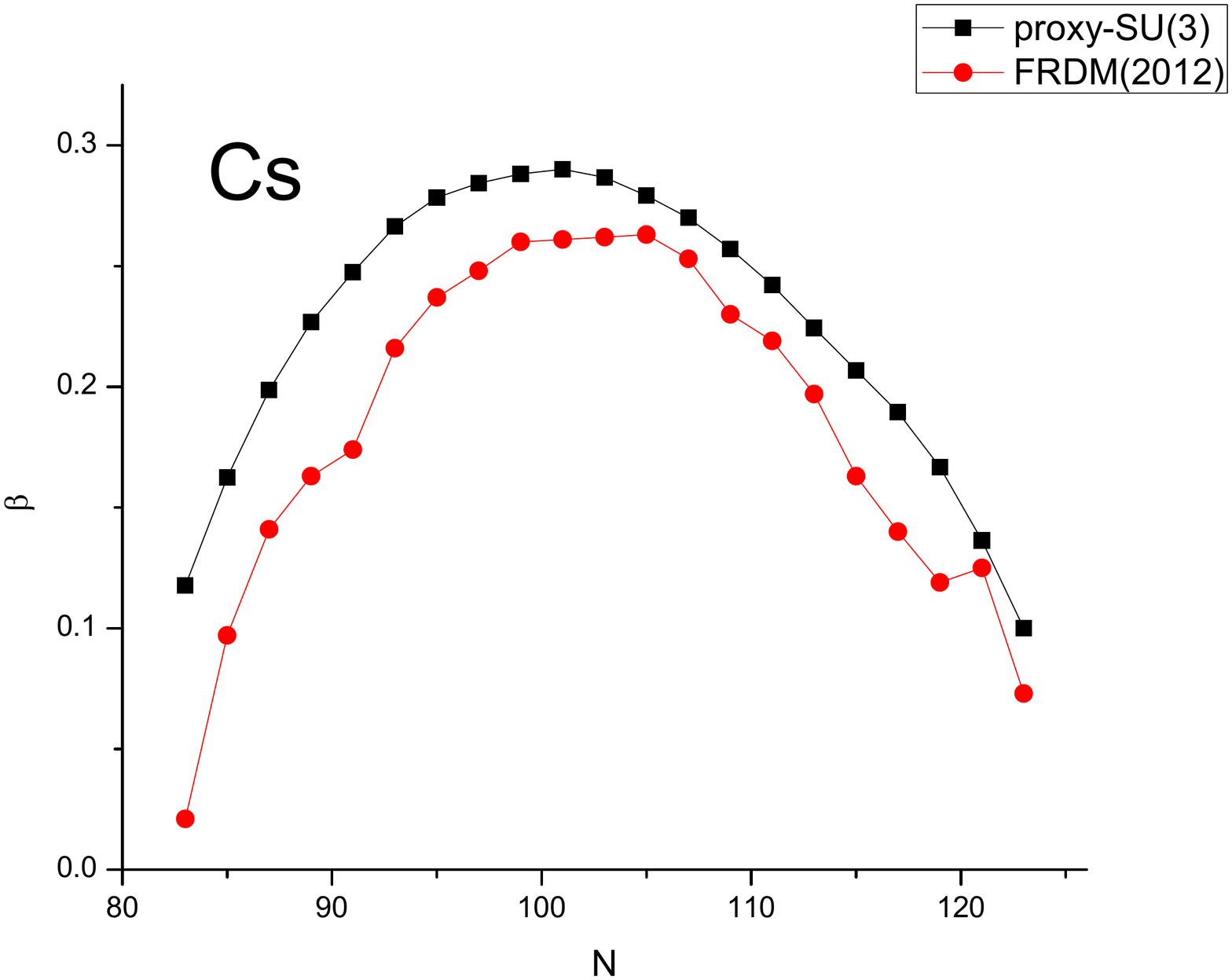,width=55mm}

\epsfig{file=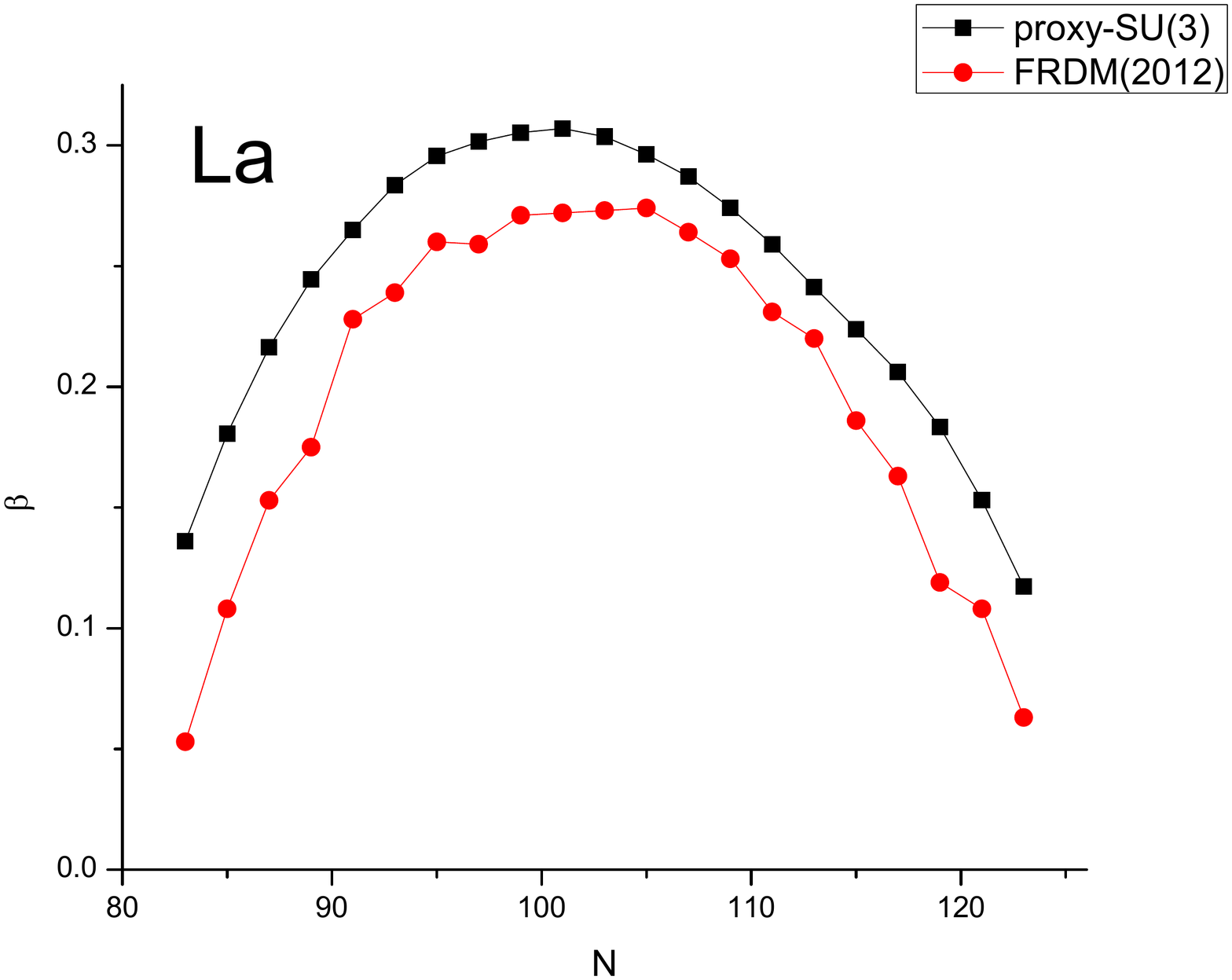,width=55mm}
\epsfig{file=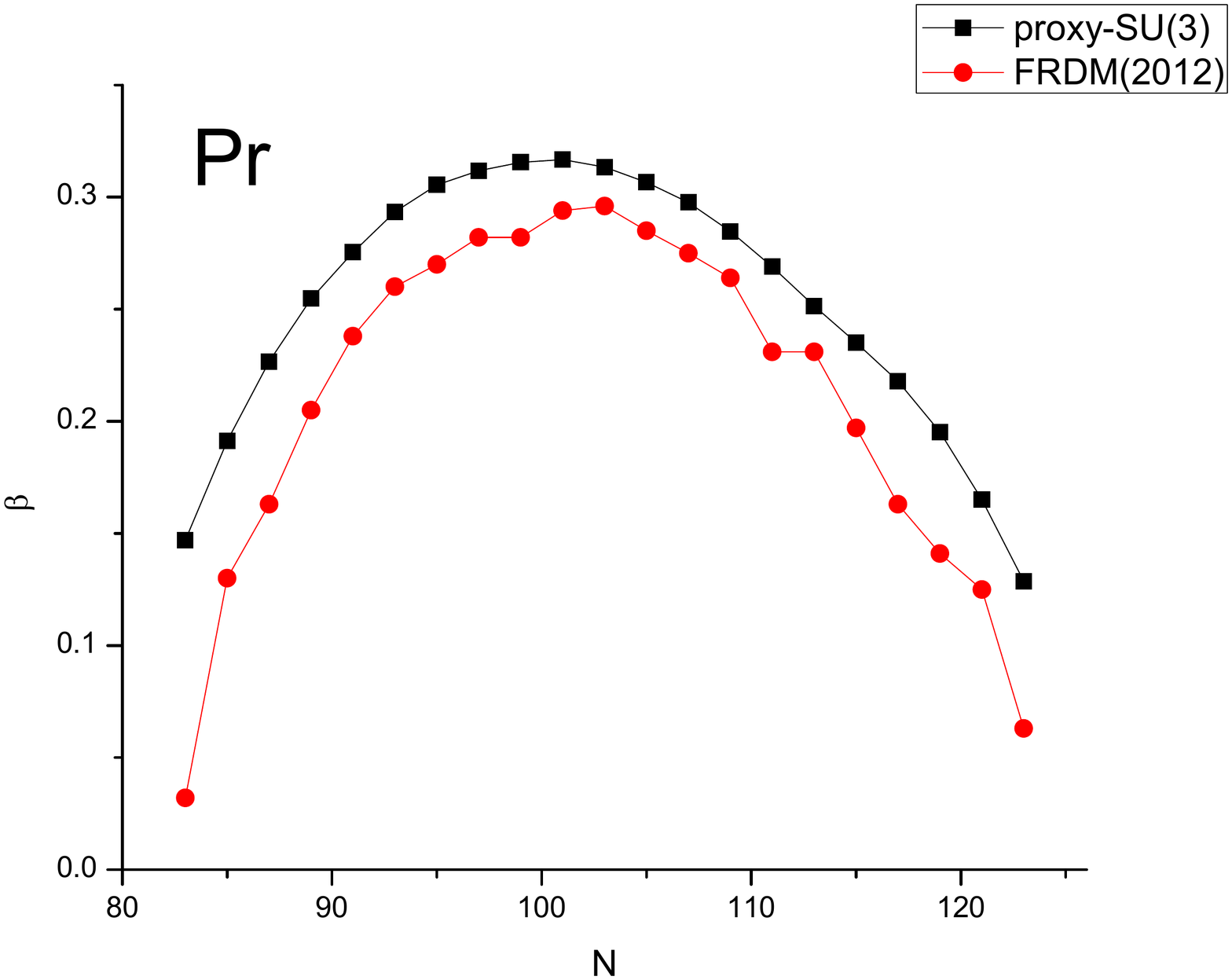,width=55mm}

\epsfig{file=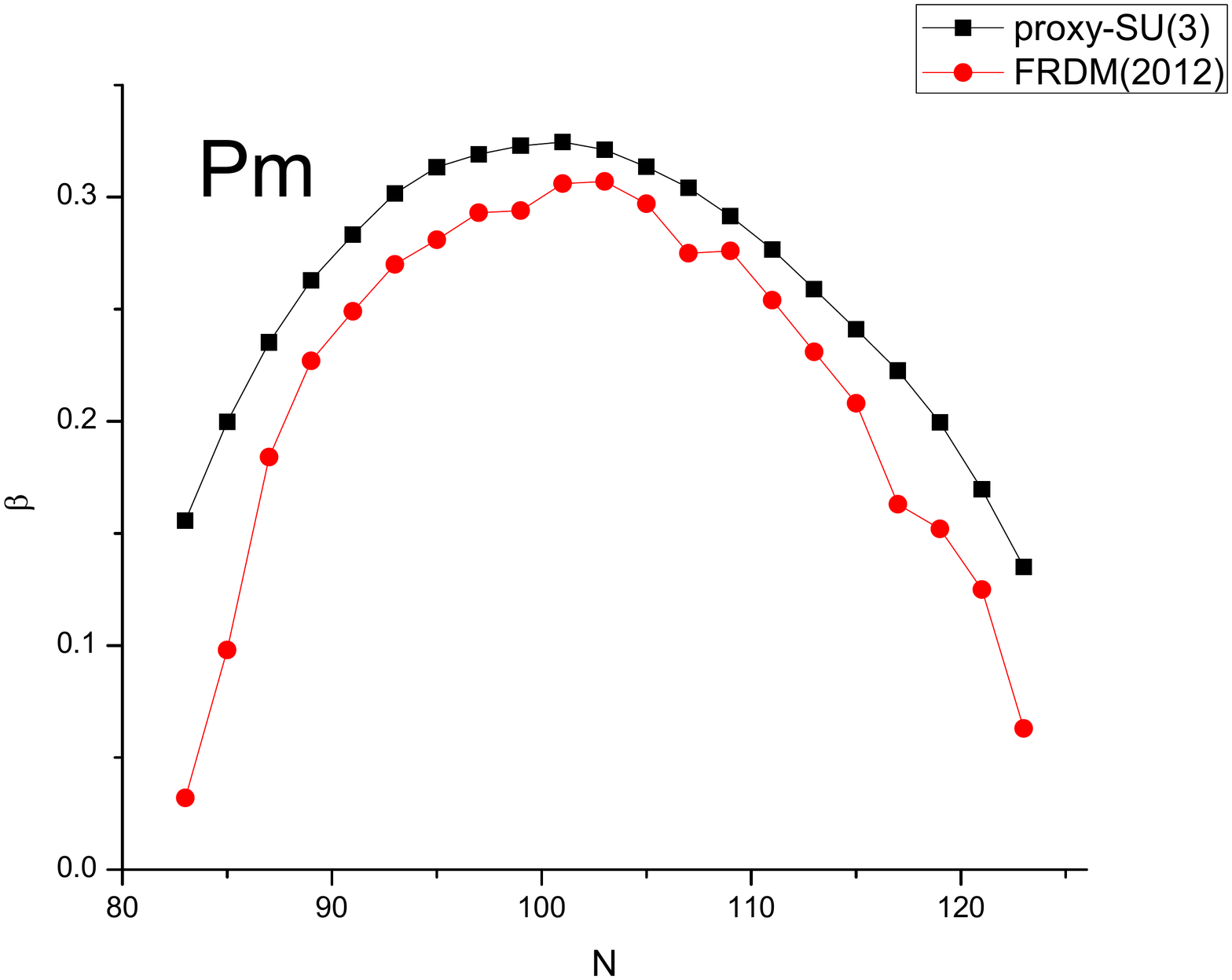,width=55mm}
\epsfig{file=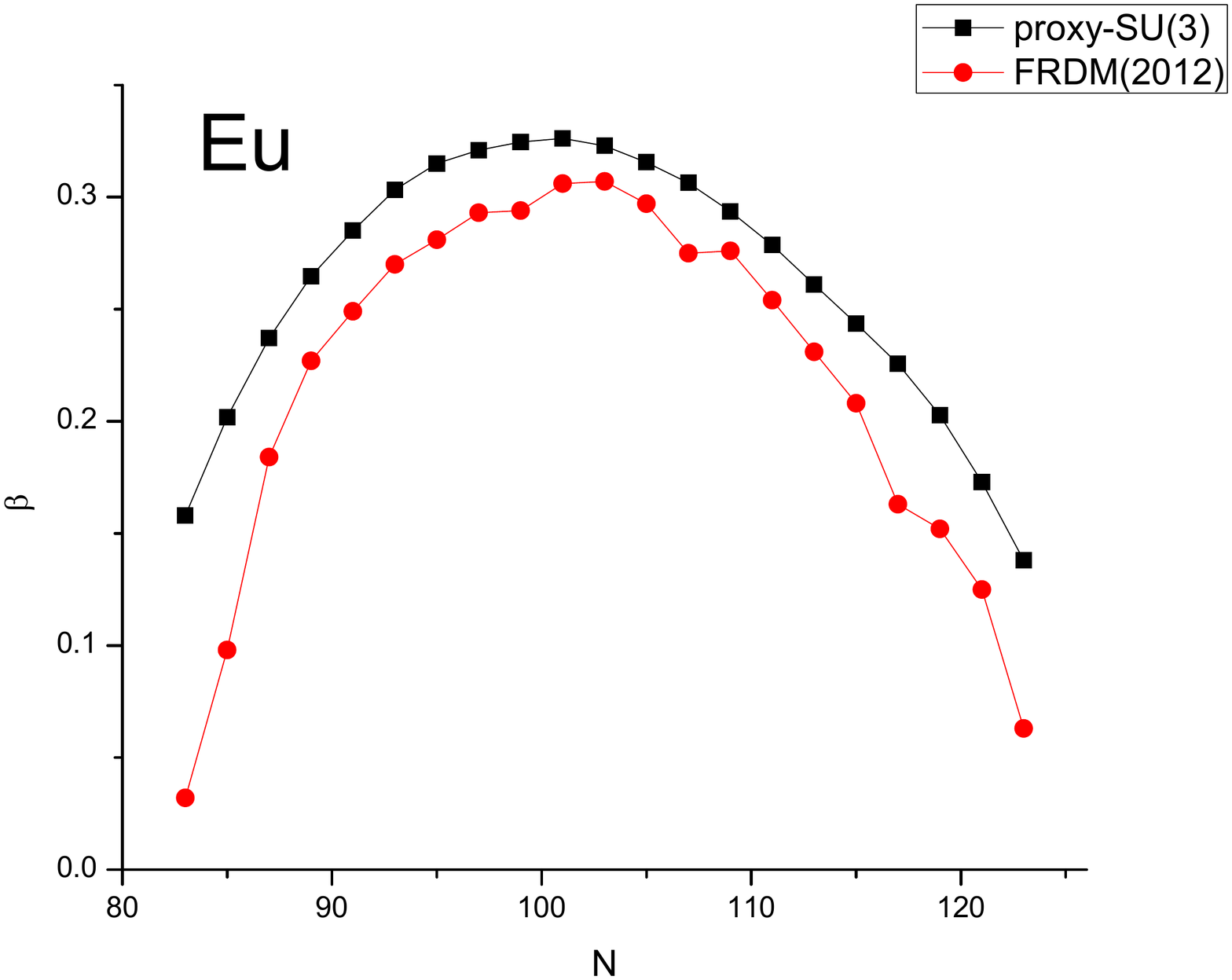,width=55mm}

\epsfig{file=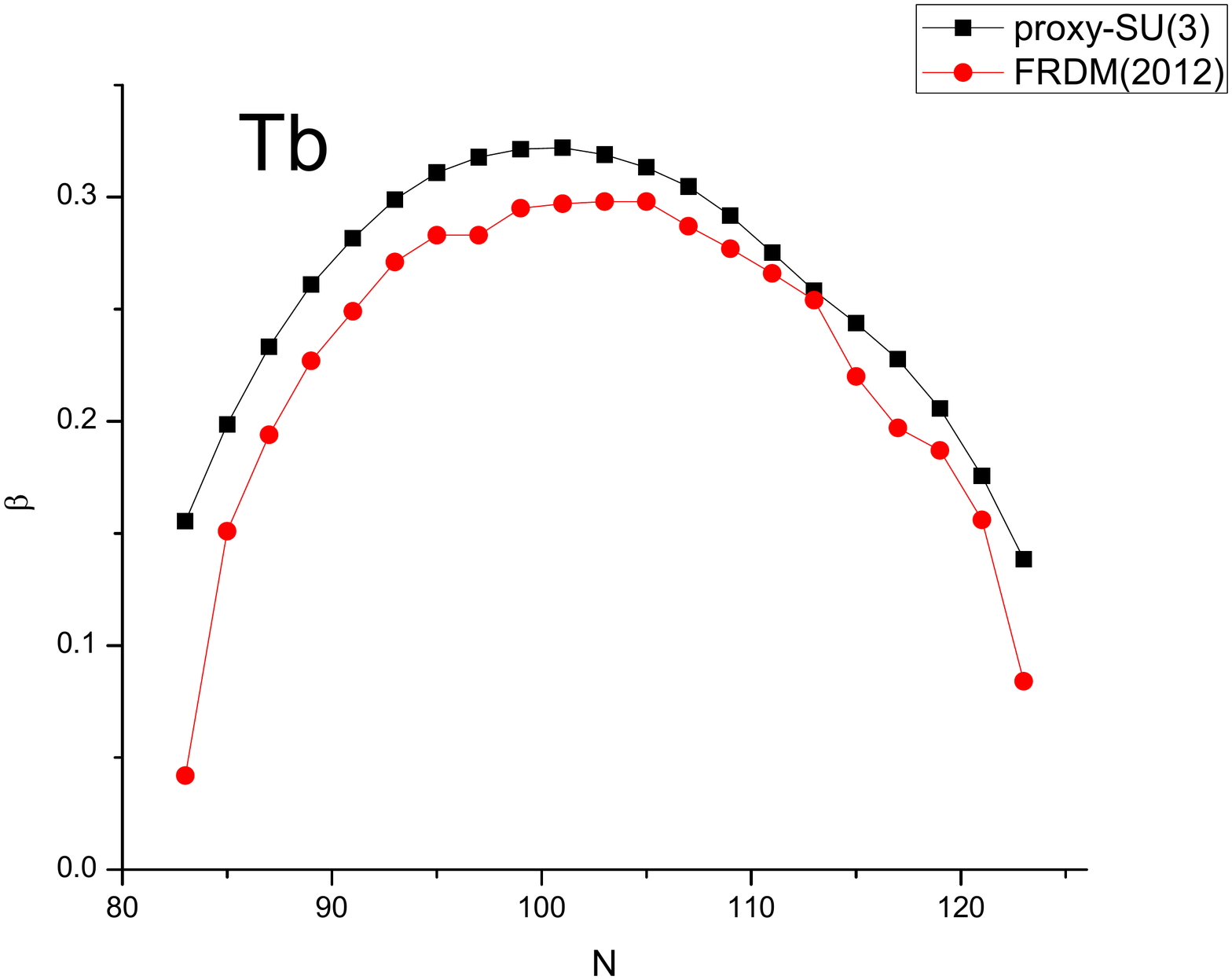,width=55mm}
\epsfig{file=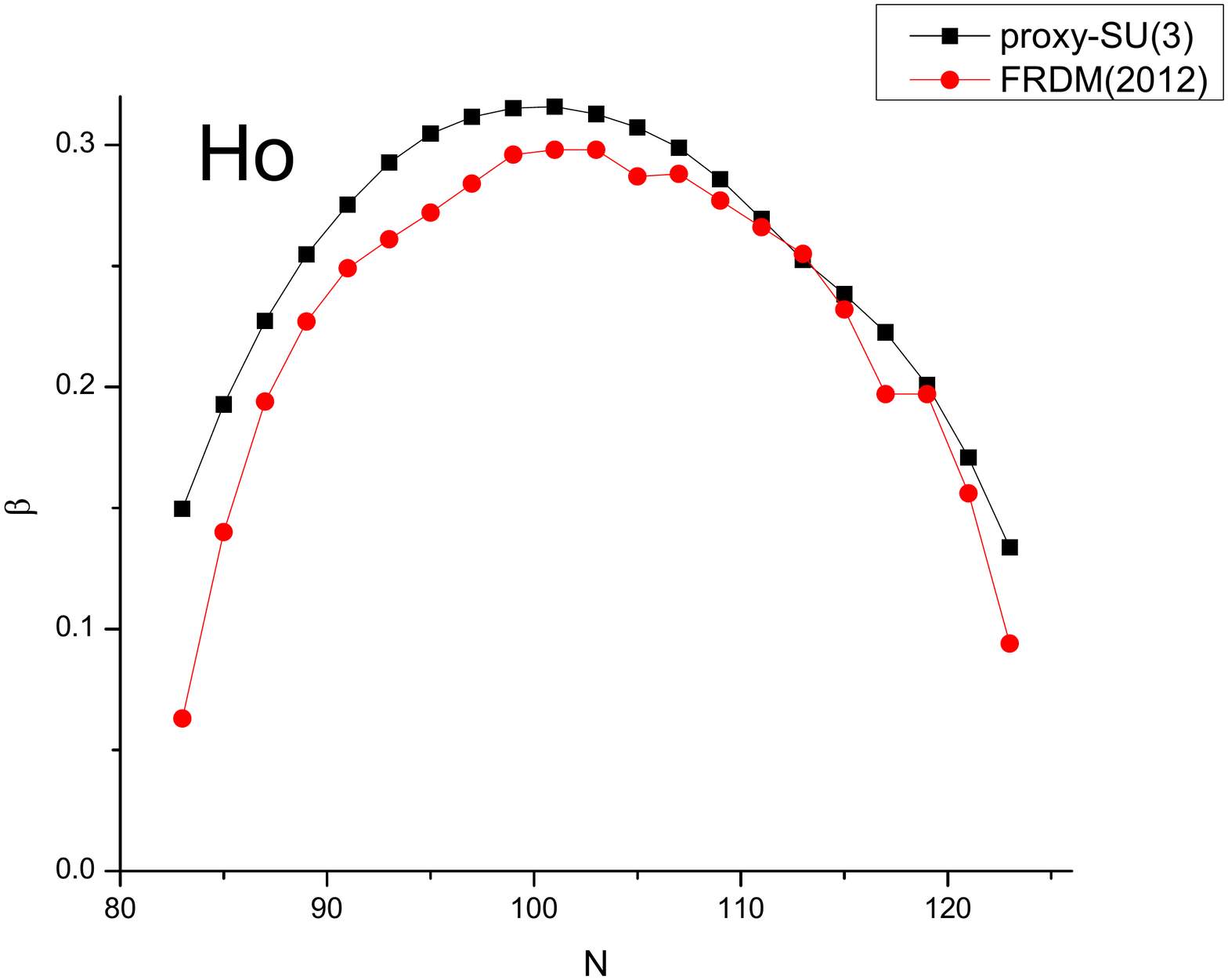,width=55mm}

\caption{Proxy SU(3) predictions for $\beta$ for odd-odd rare earths with $Z=53$-67, compared with results reported in the mass table FRDM(2012)\cite{Moller}.  See Section \ref{oddodd} for further discussion.} 

\end{figure}


\begin{figure}[tb]

\epsfig{file=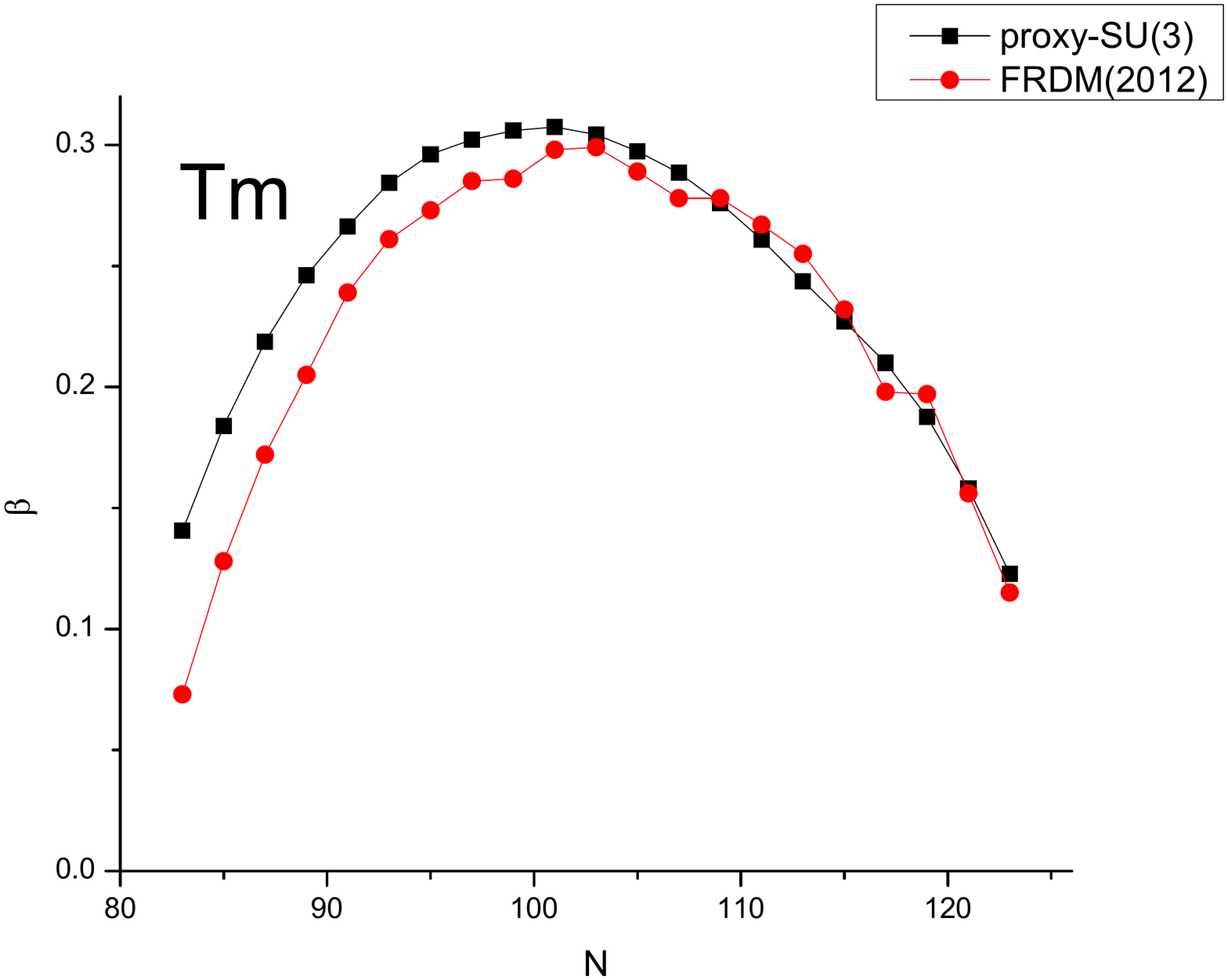,width=55mm}
\epsfig{file=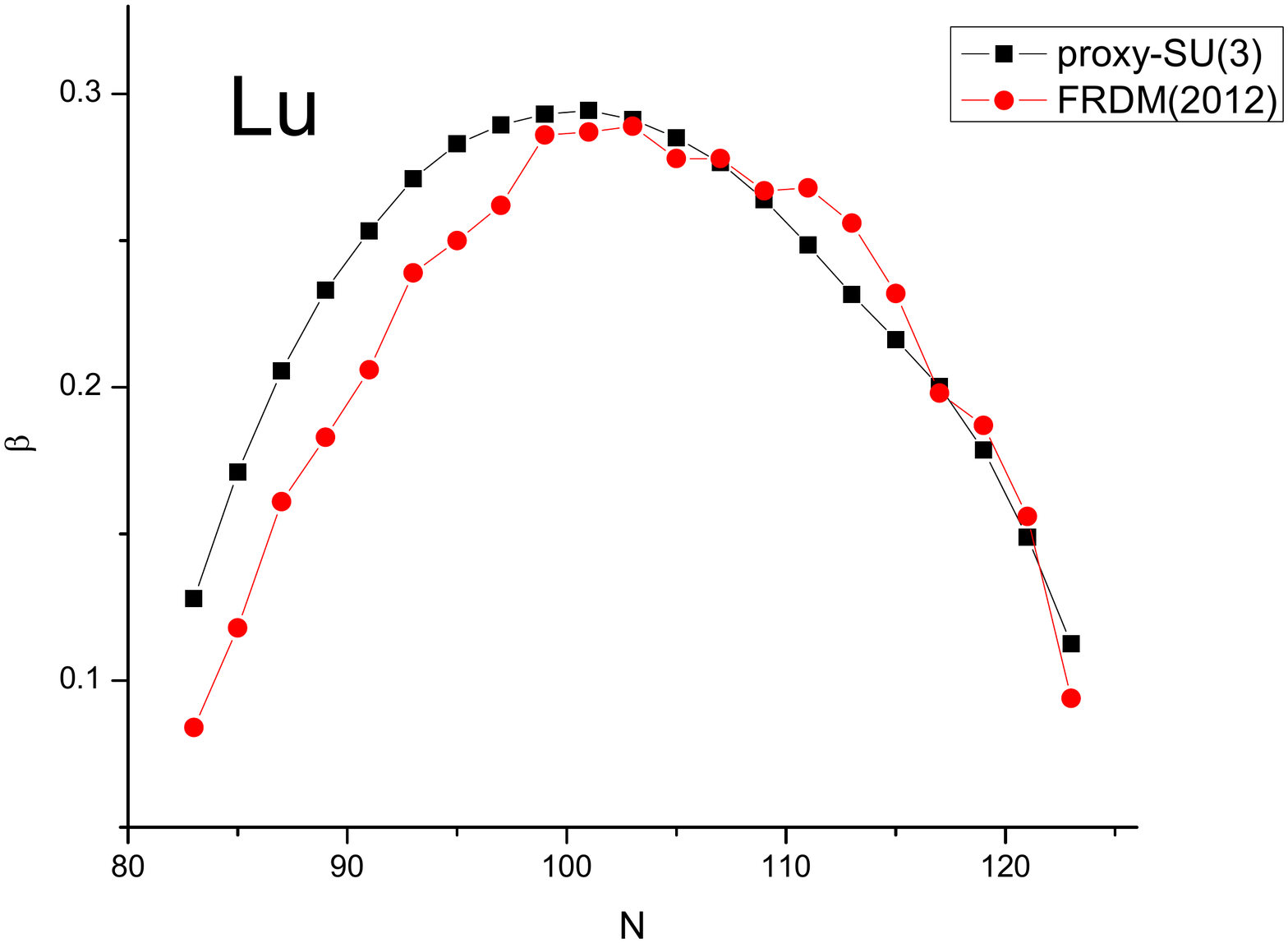,width=55mm}

\epsfig{file=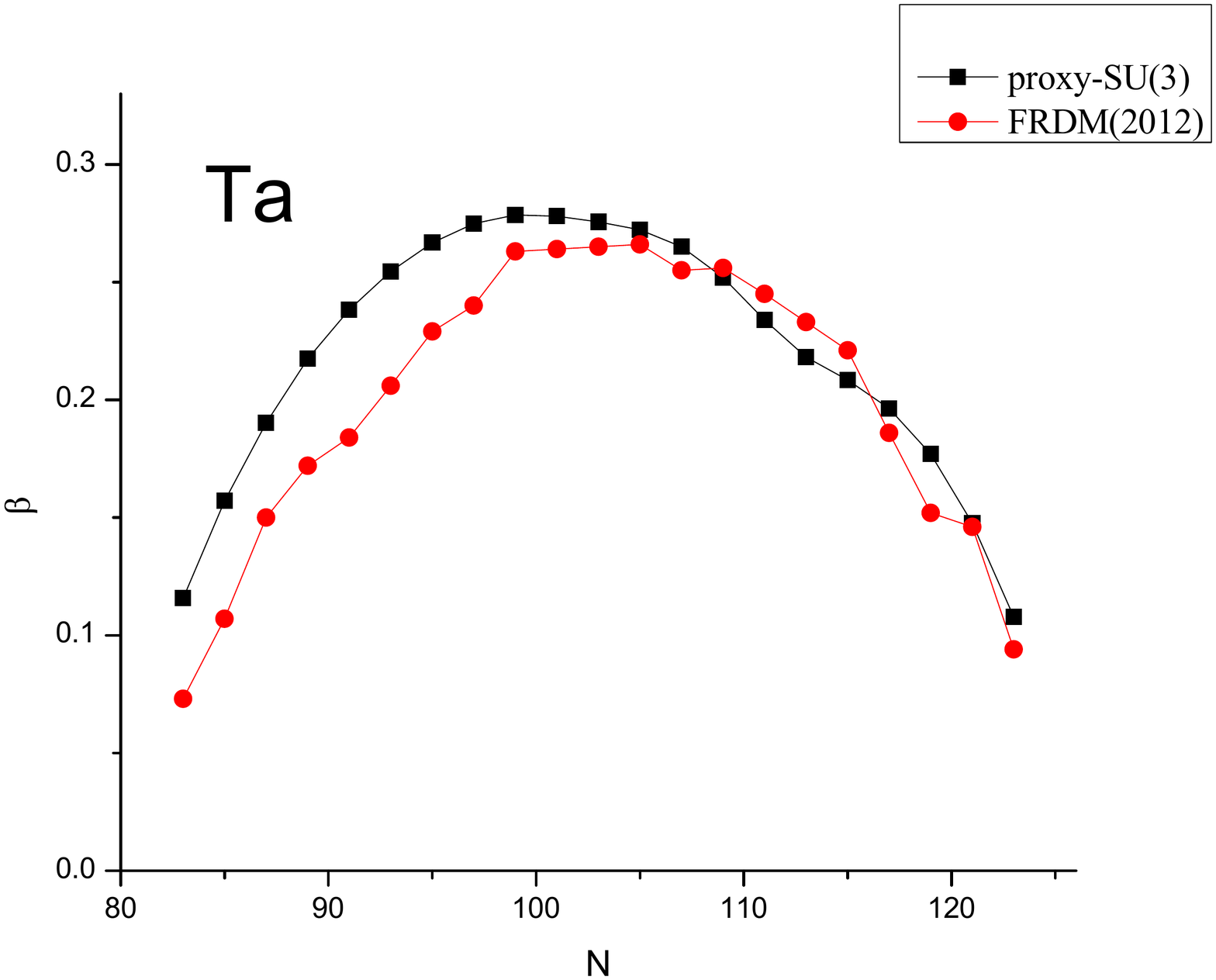,width=55mm}
\epsfig{file=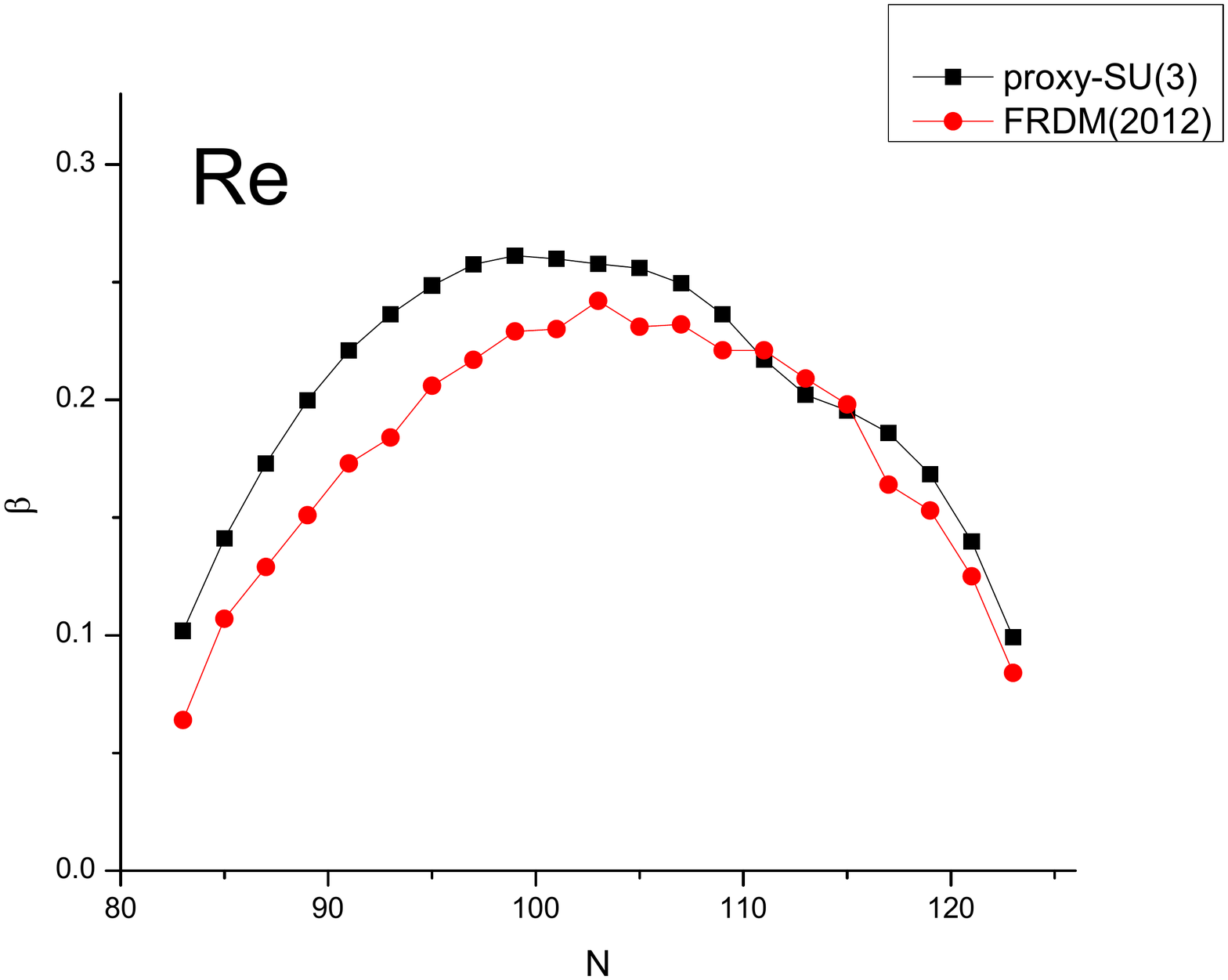,width=55mm}

\epsfig{file=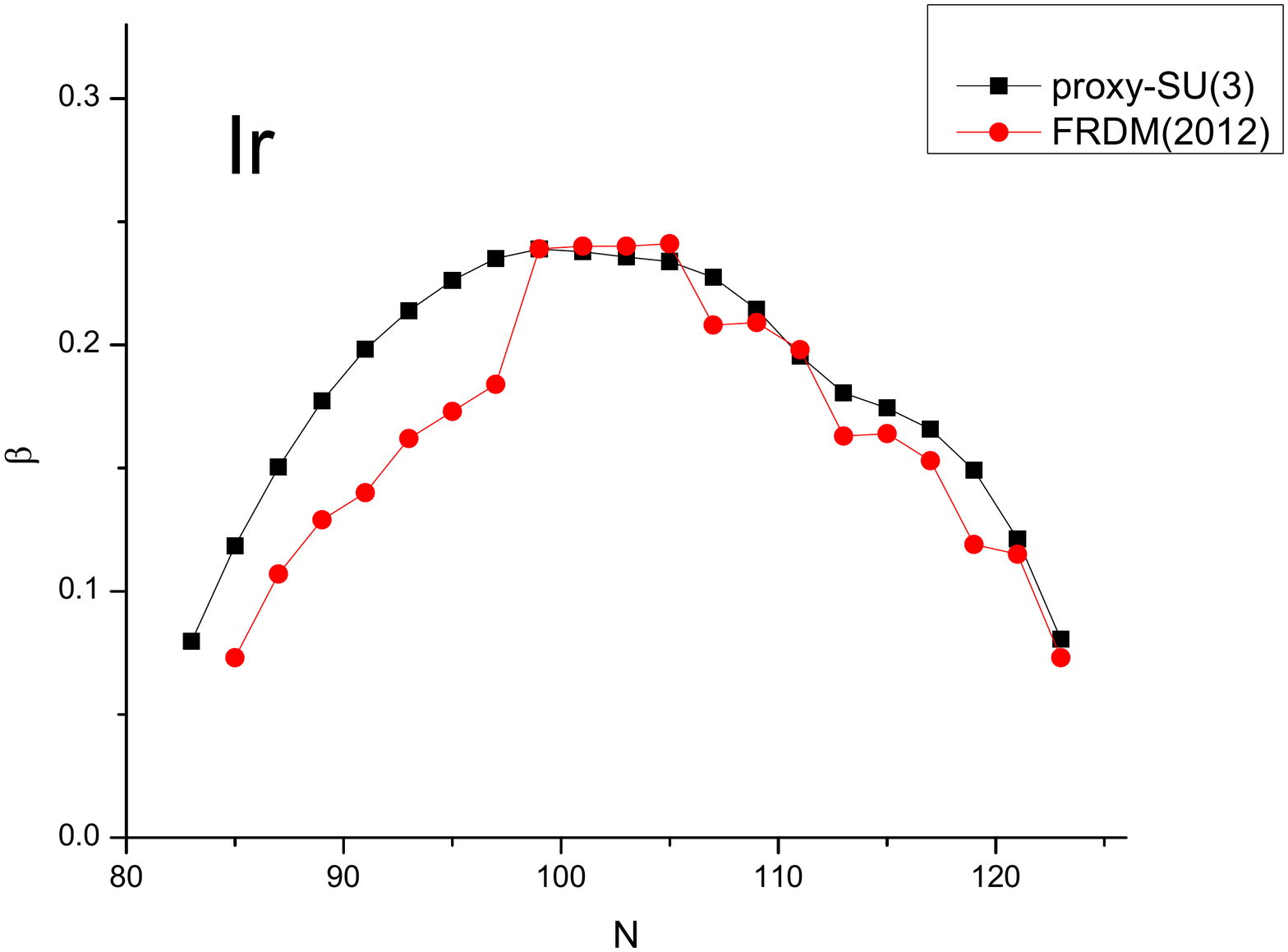,width=55mm}
\epsfig{file=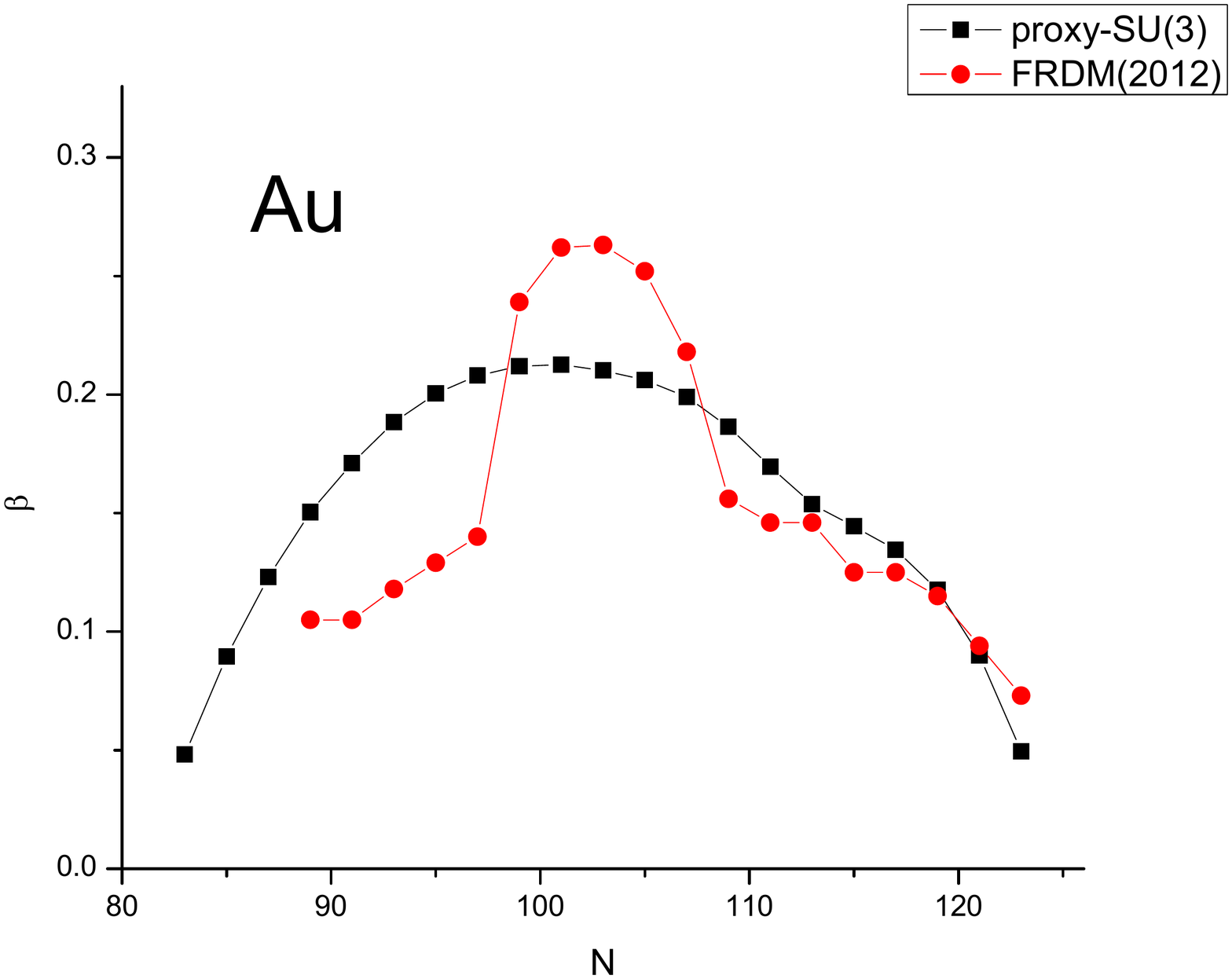,width=55mm}

\epsfig{file=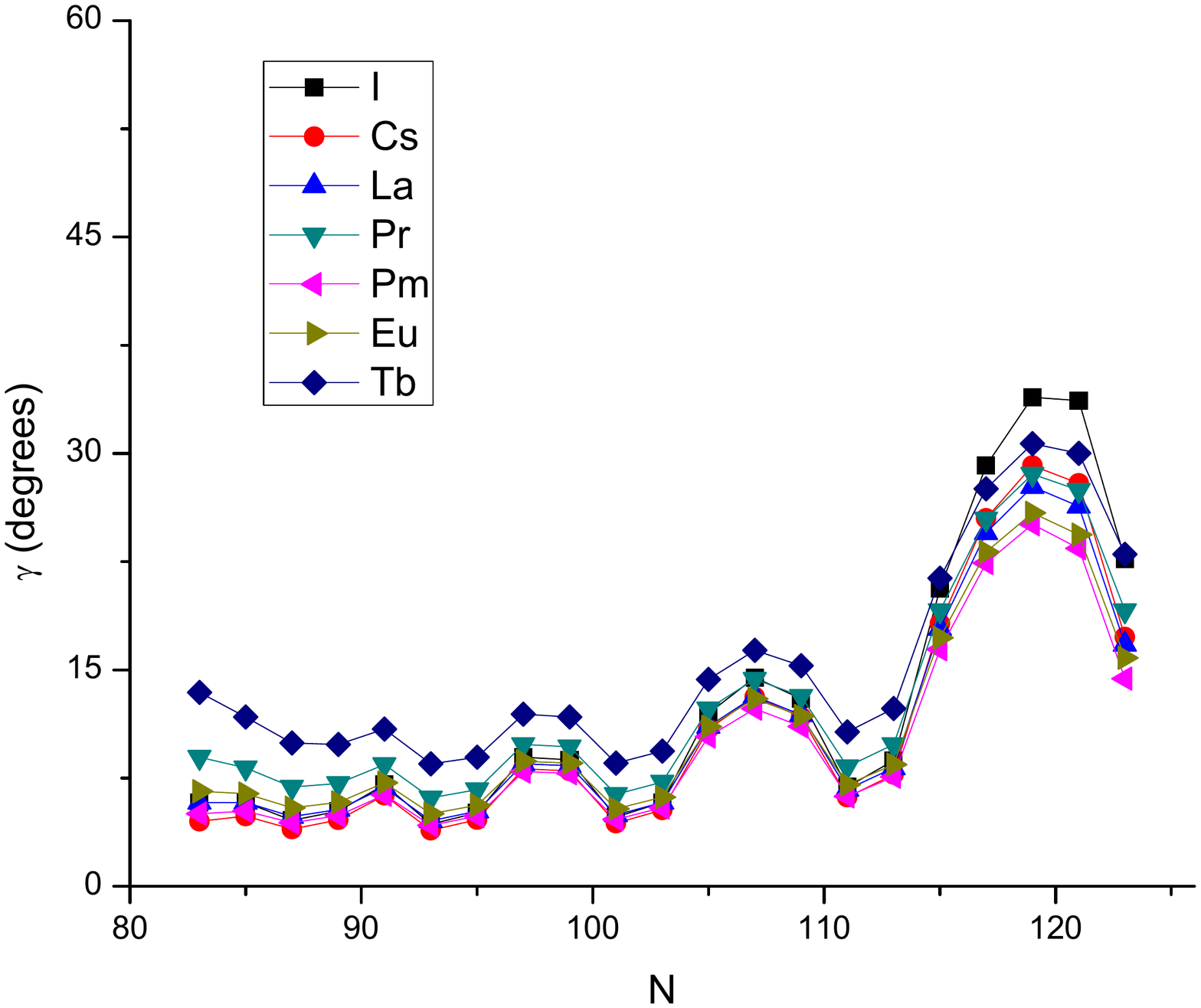,width=55mm}
\epsfig{file=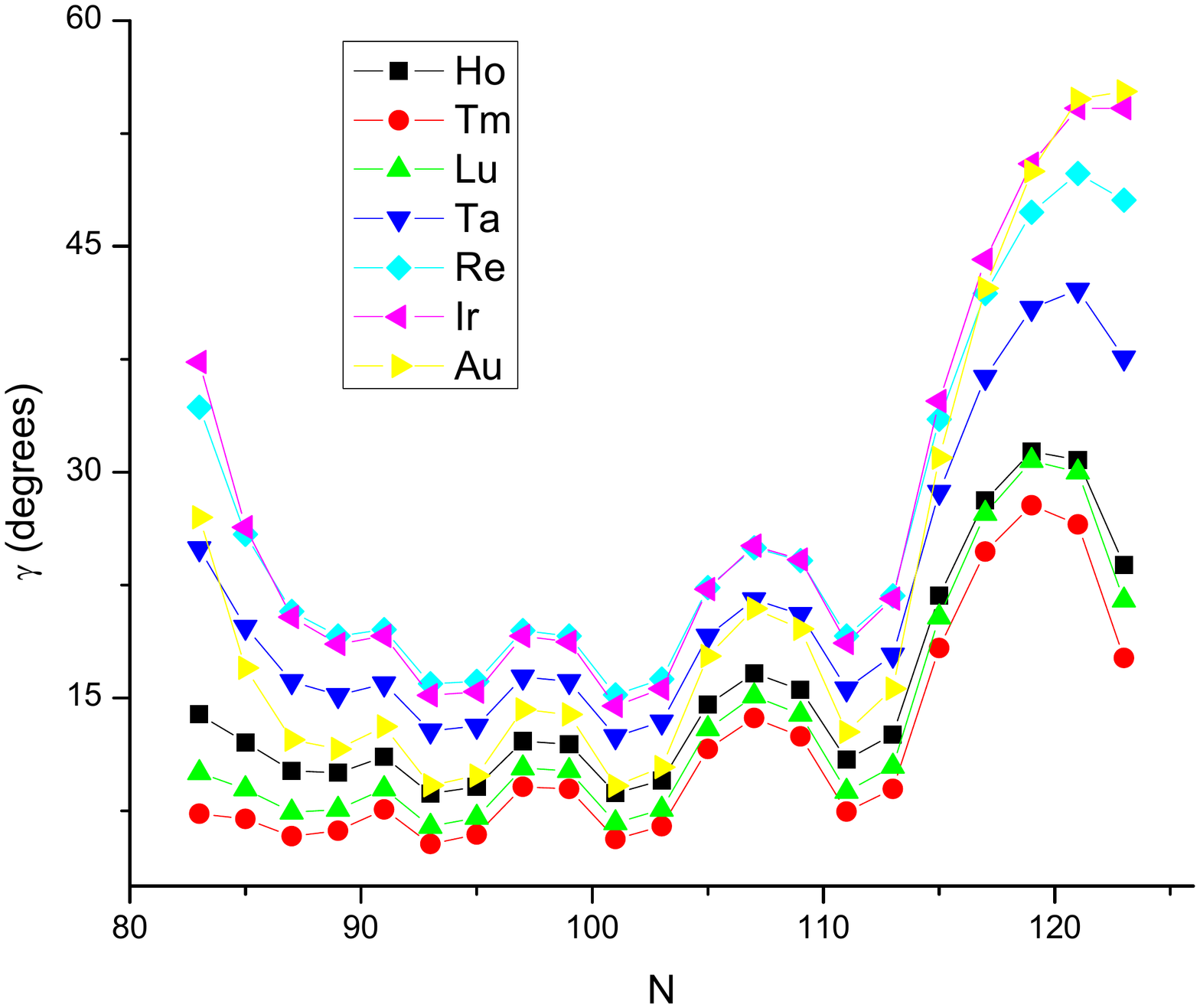,width=55mm}
 
\caption{Proxy SU(3) predictions for $\beta$ for odd-odd rare earths with $Z=69$-79, compared with results reported in the mass table FRDM(2012)\cite{Moller}. In the two bottom panels,
the proxy-SU(3) predictions for $\gamma$ are reported for $Z=53$-79. See Section \ref{oddodd} for further discussion.}

\end{figure}



\begin{figure}[b]
\epsfig{file=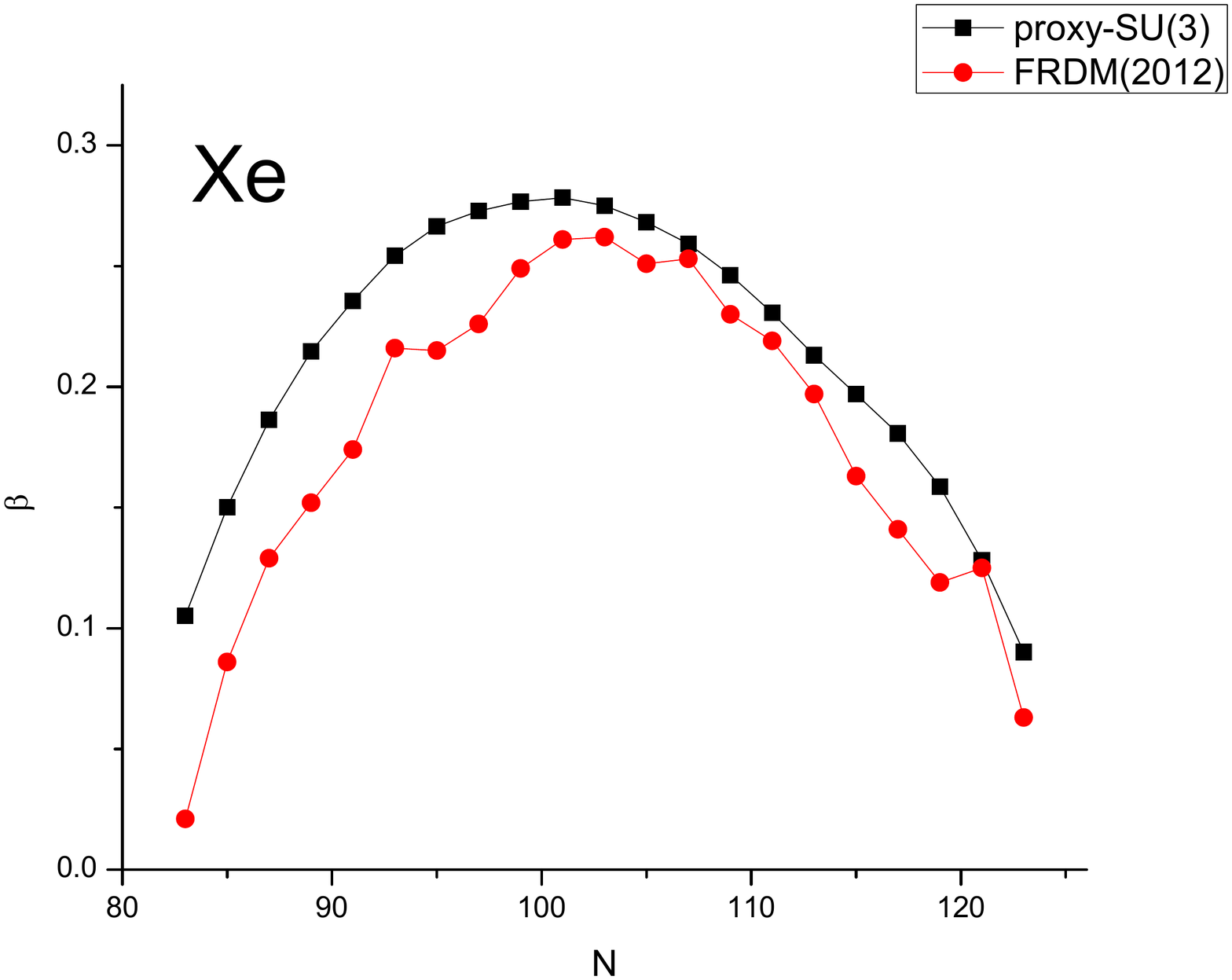,width=55mm}
\epsfig{file=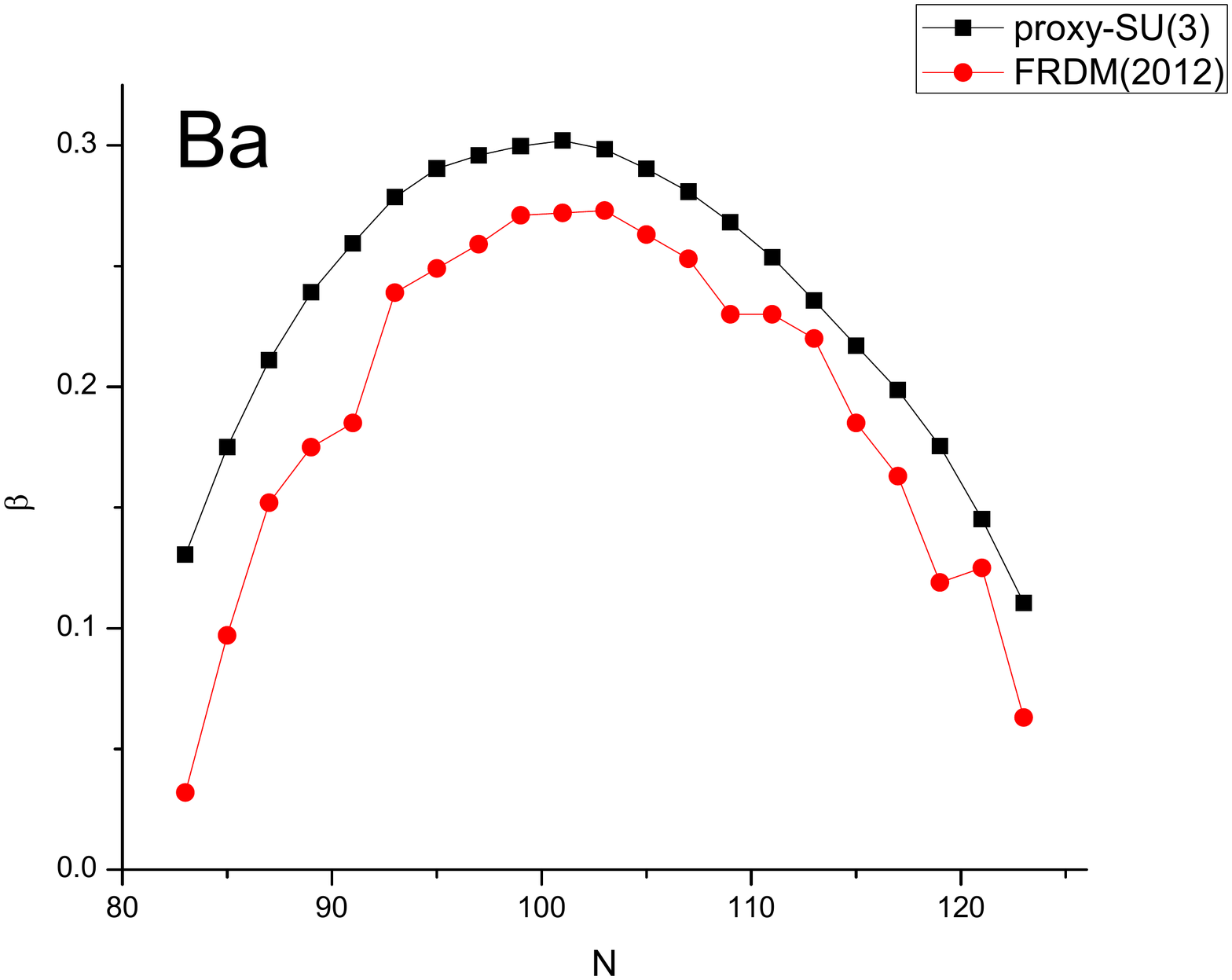,width=55mm}

\epsfig{file=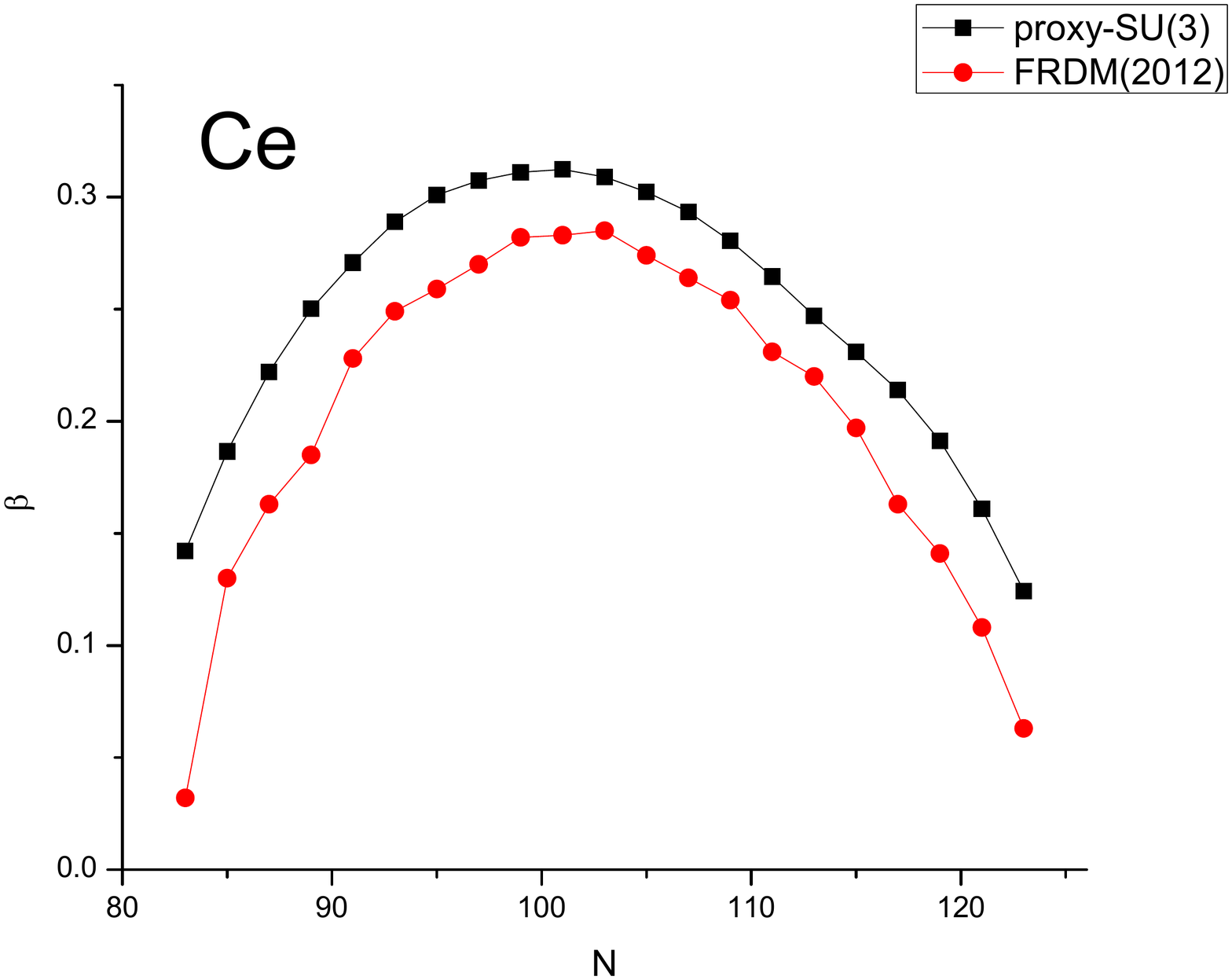,width=55mm}
\epsfig{file=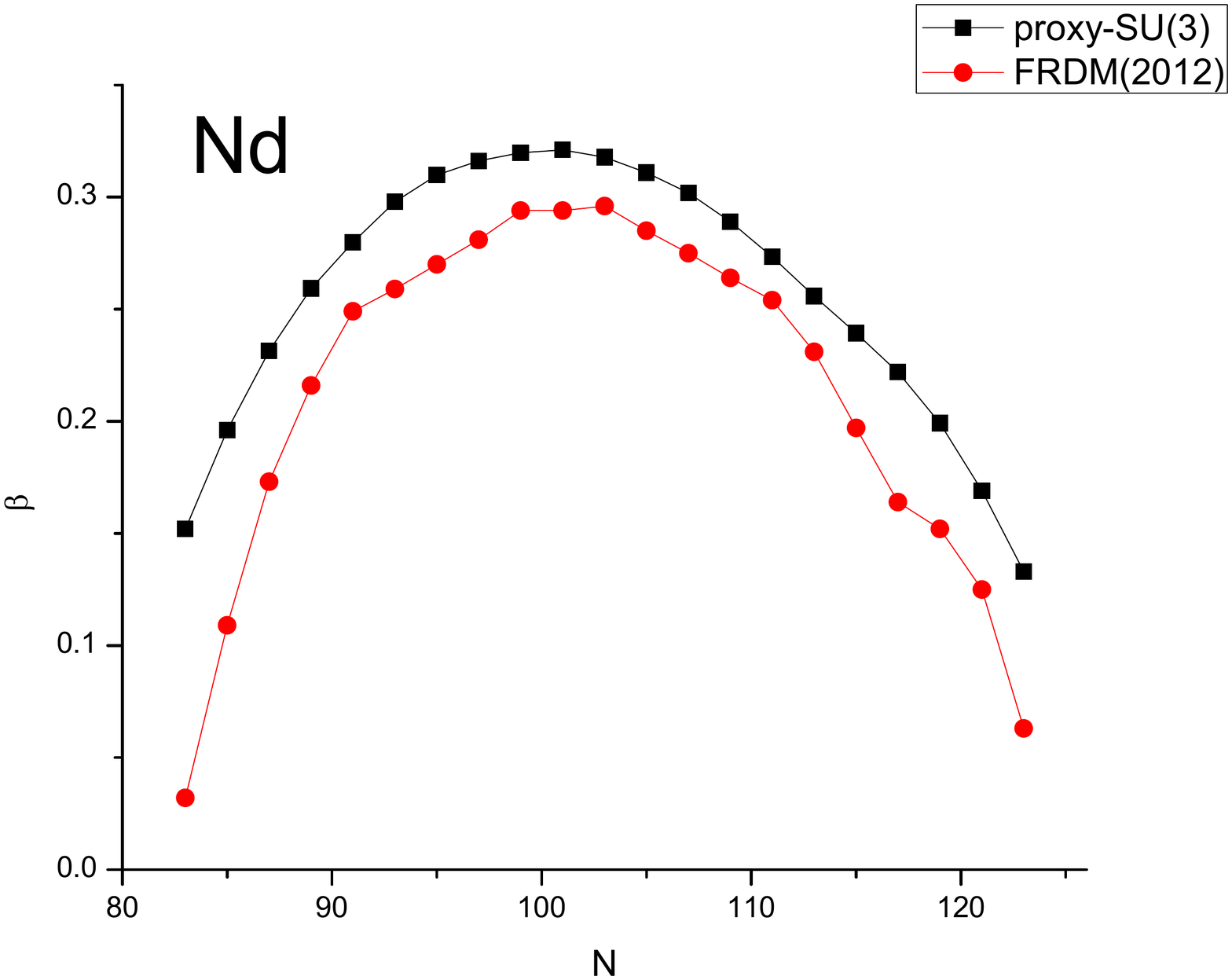,width=55mm}

\epsfig{file=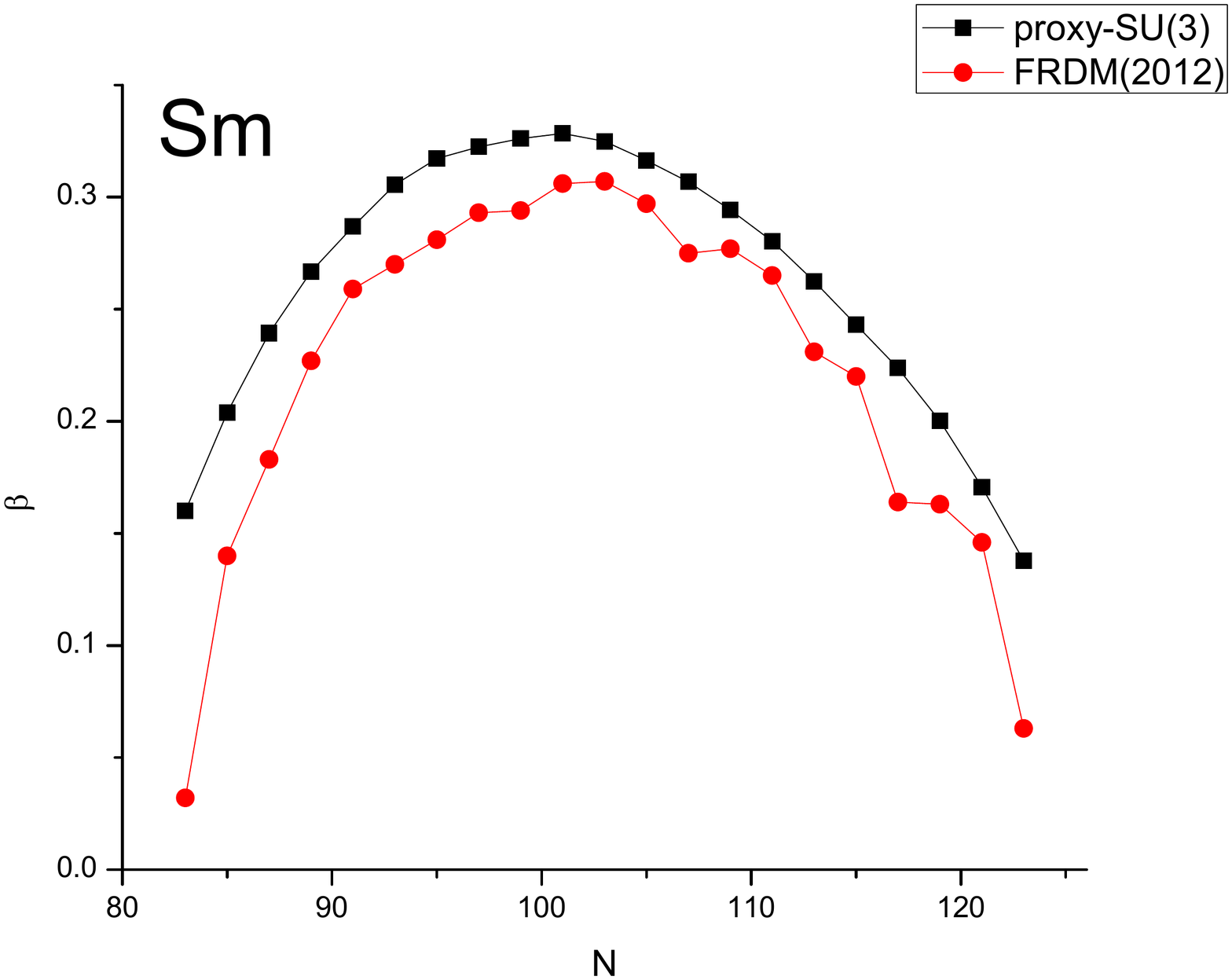,width=55mm}
\epsfig{file=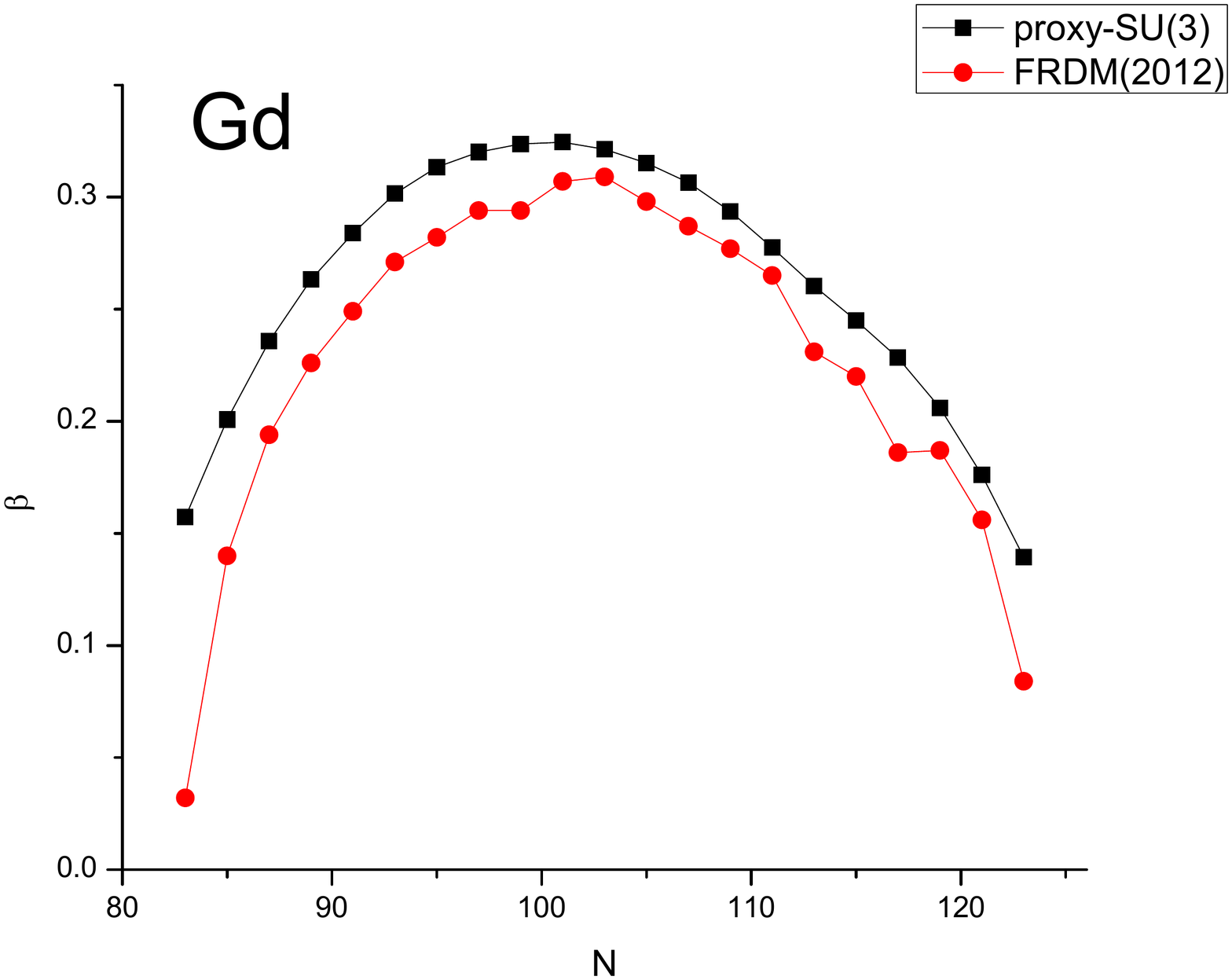,width=55mm}

\epsfig{file=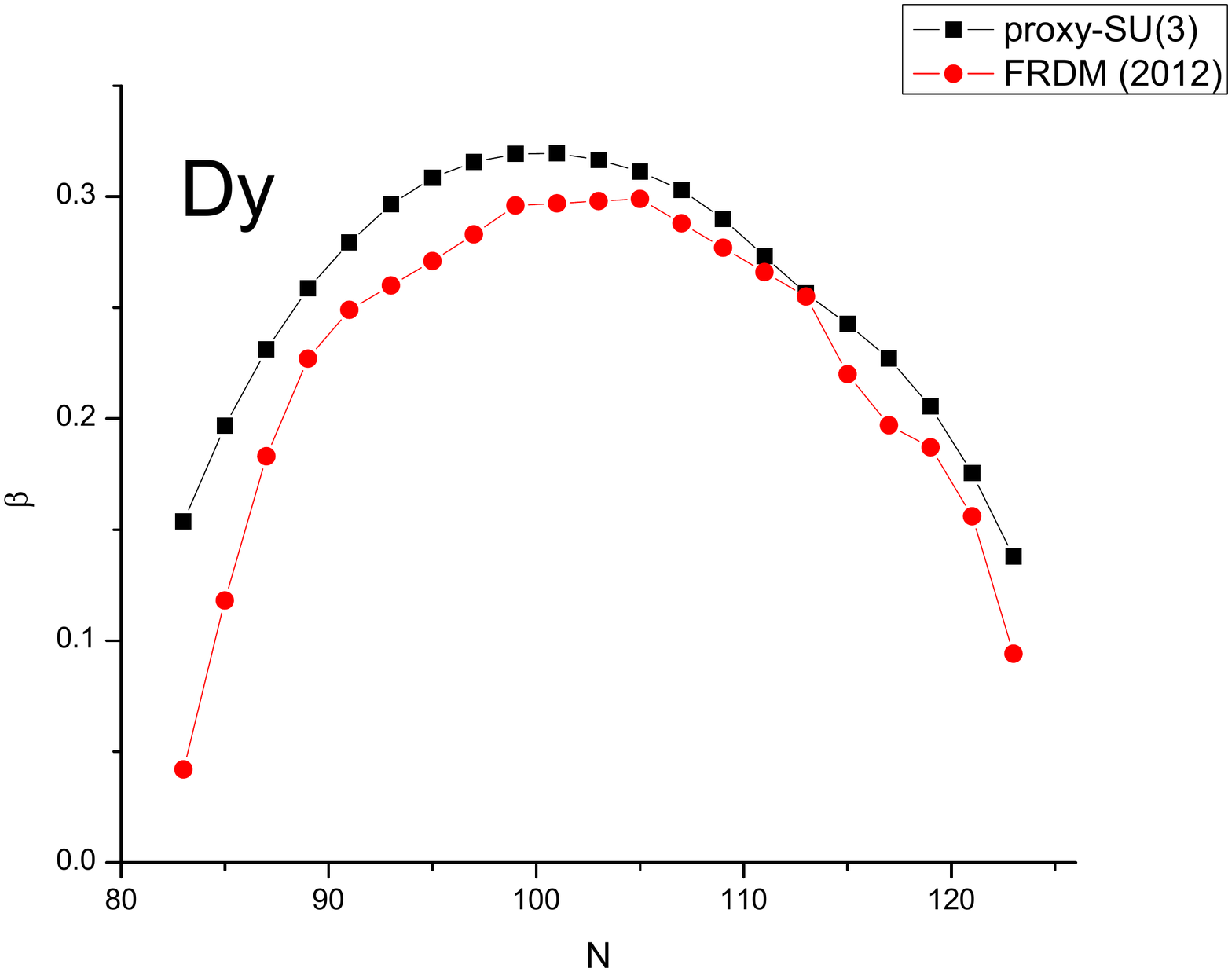,width=55mm}
\epsfig{file=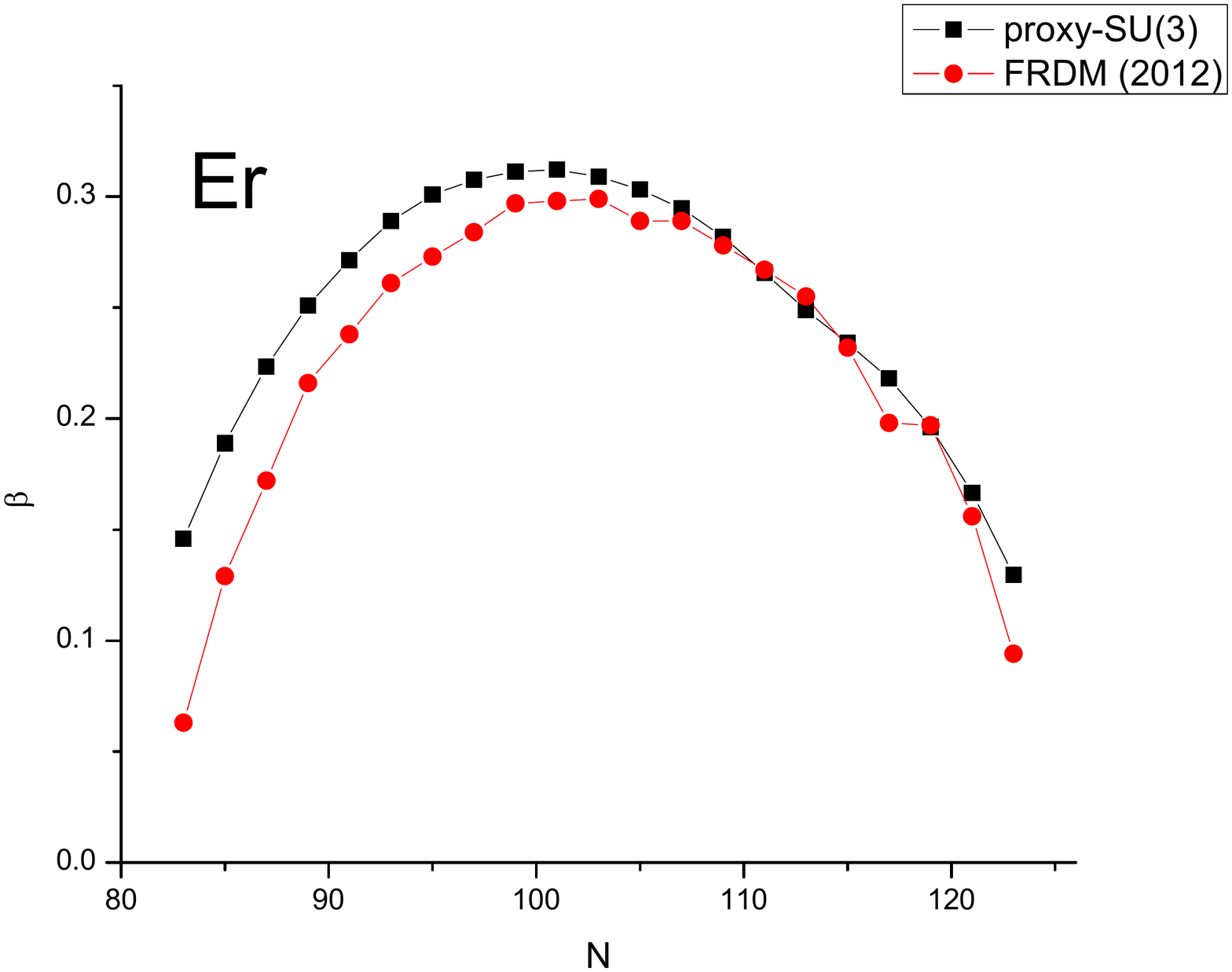,width=55mm}

\caption{Proxy SU(3) predictions for $\beta$ for even-odd rare earths with $Z=54$-68, compared with results reported in the mass table FRDM(2012) \cite{Moller}. See Section \ref{oddodd} for further discussion.}

\end{figure}


\begin{figure}[b]

\epsfig{file=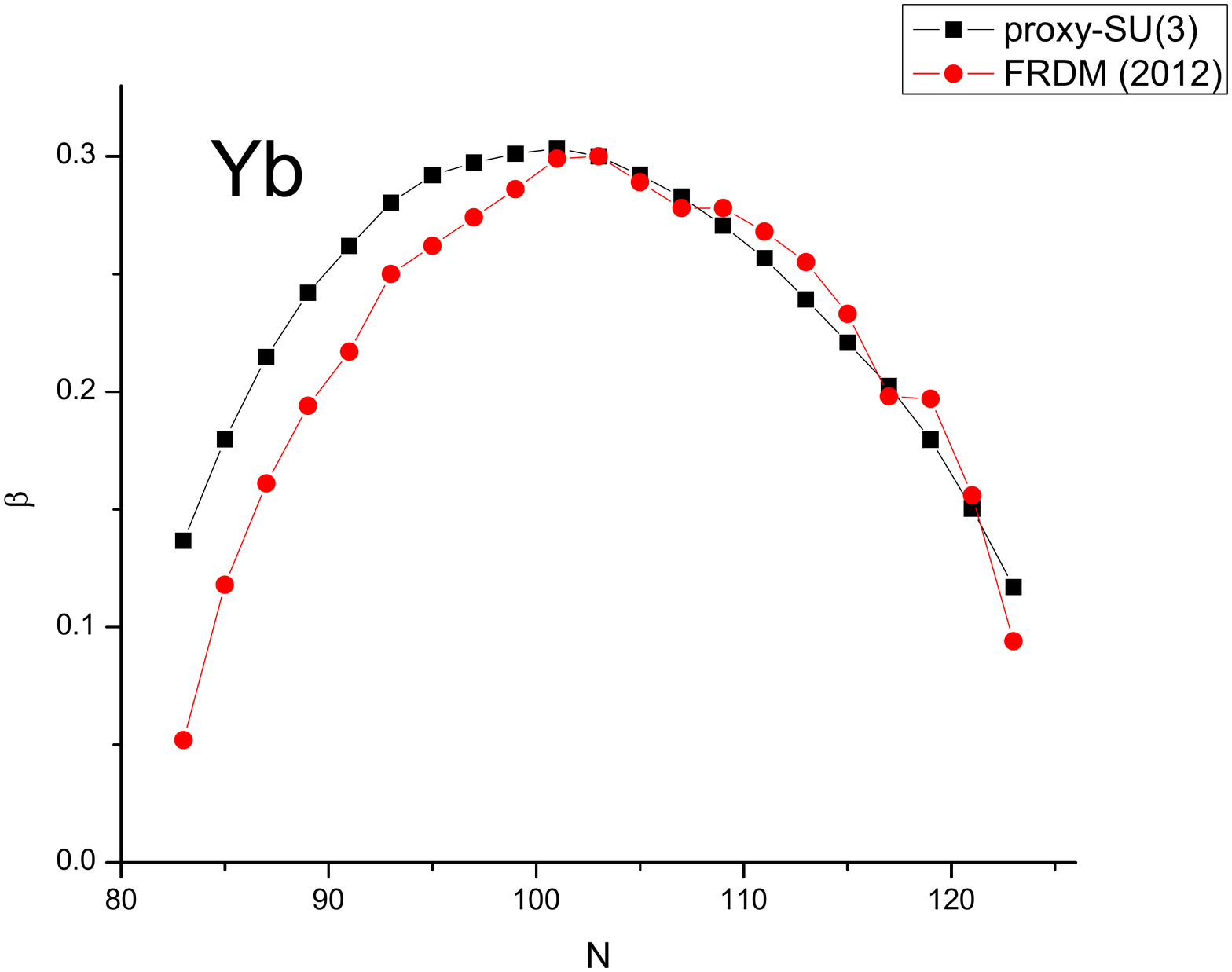,width=55mm}
\epsfig{file=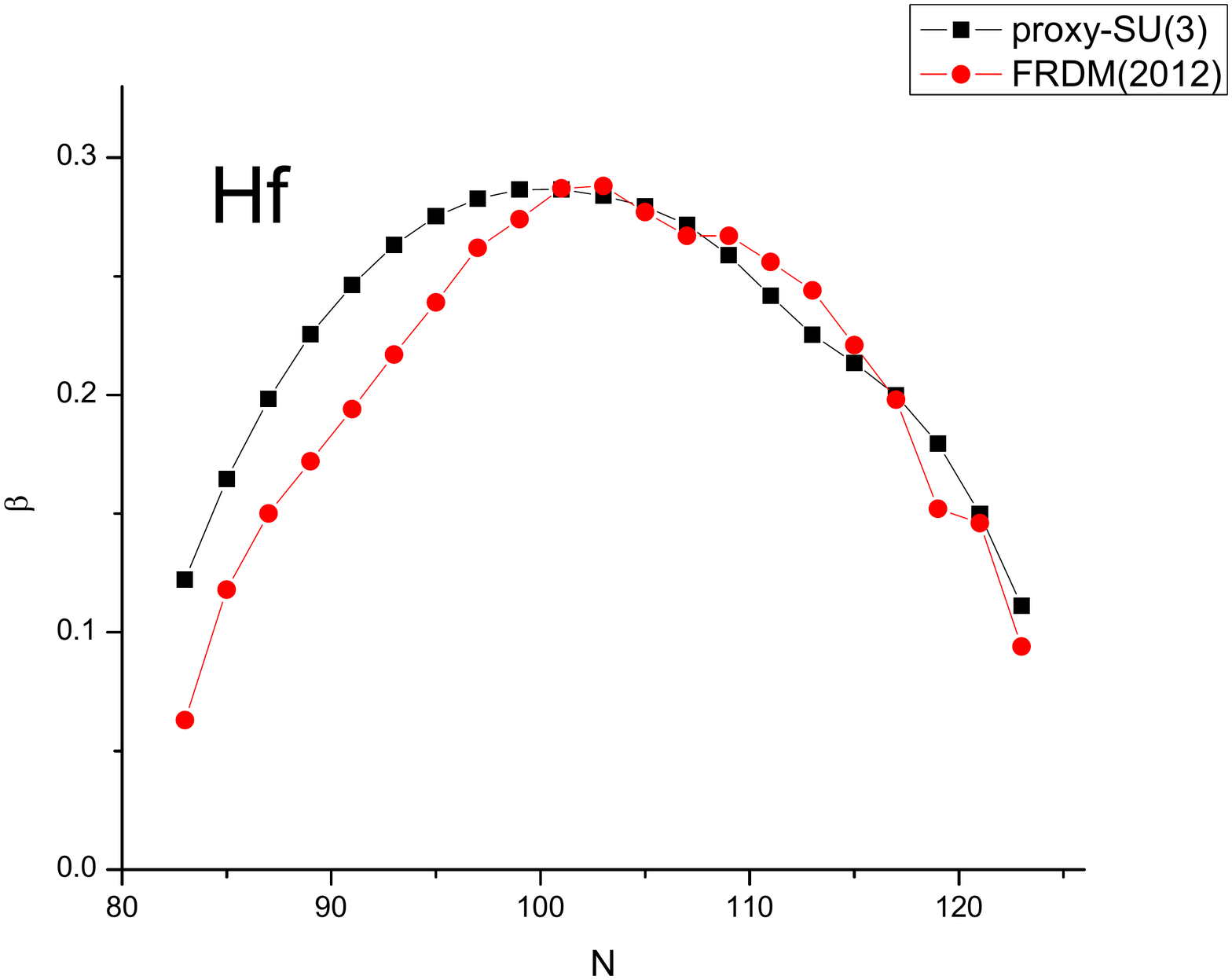,width=55mm}

\epsfig{file=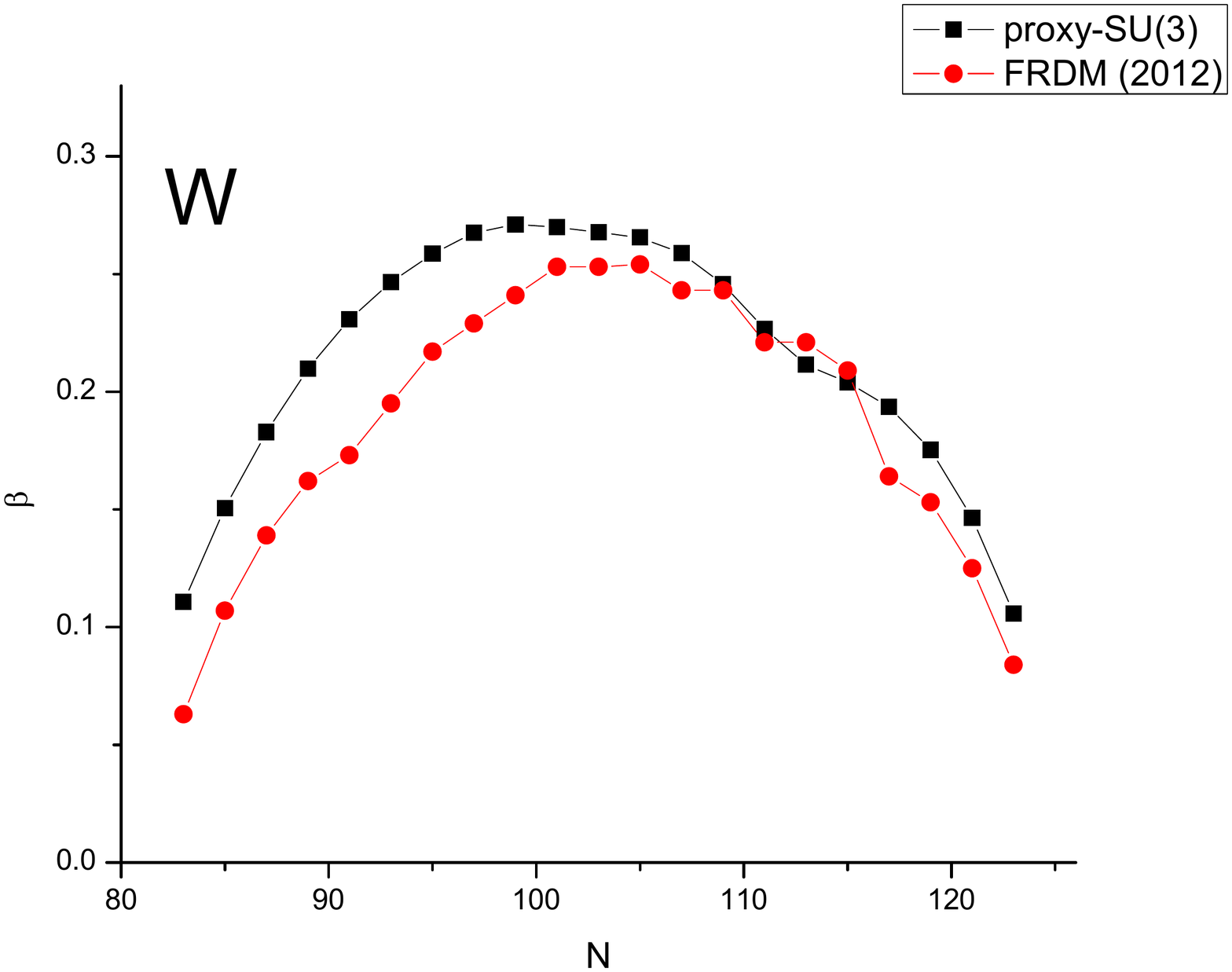,width=55mm}
\epsfig{file=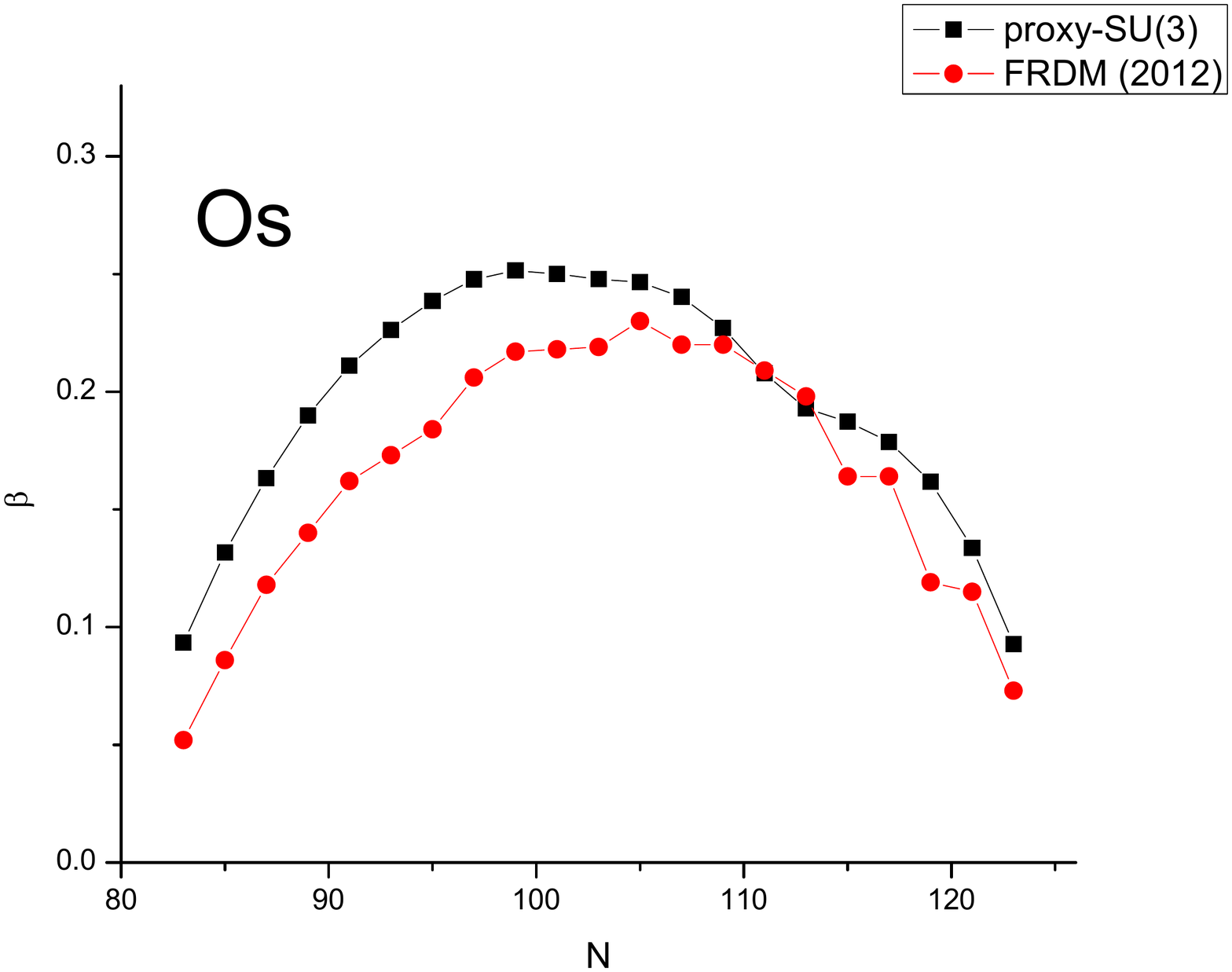,width=55mm}

\epsfig{file=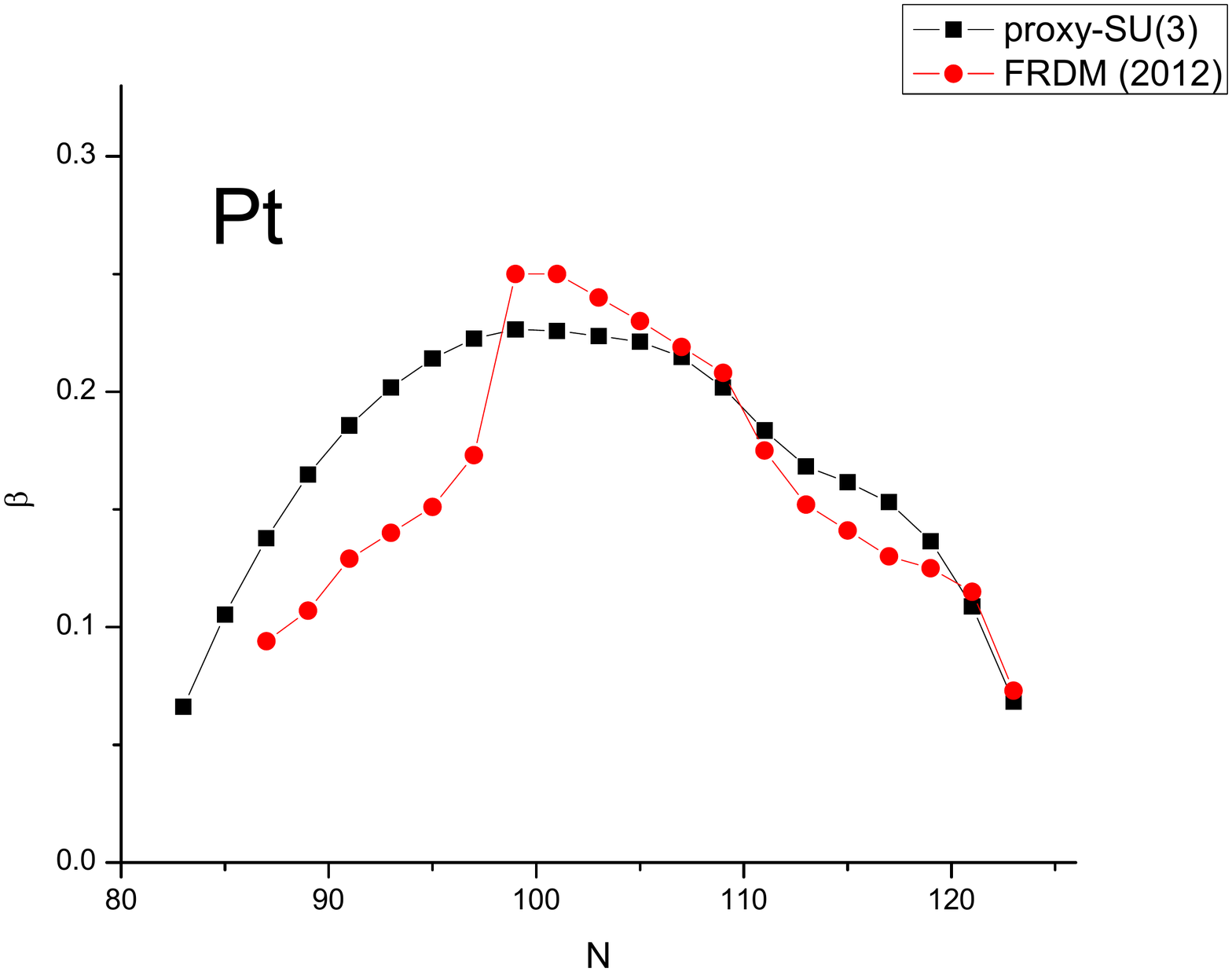,width=55mm}

\epsfig{file=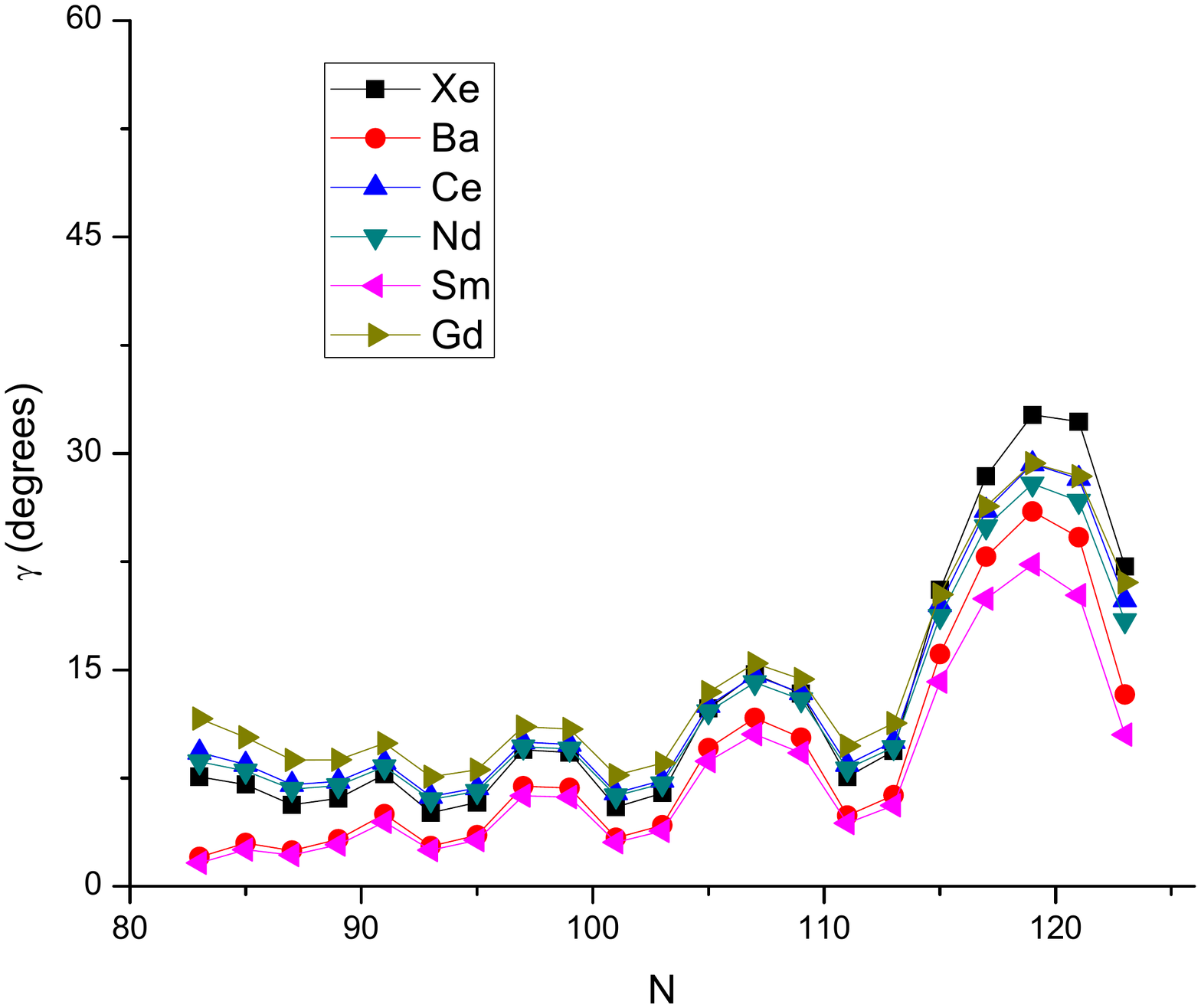,width=55mm}
\epsfig{file=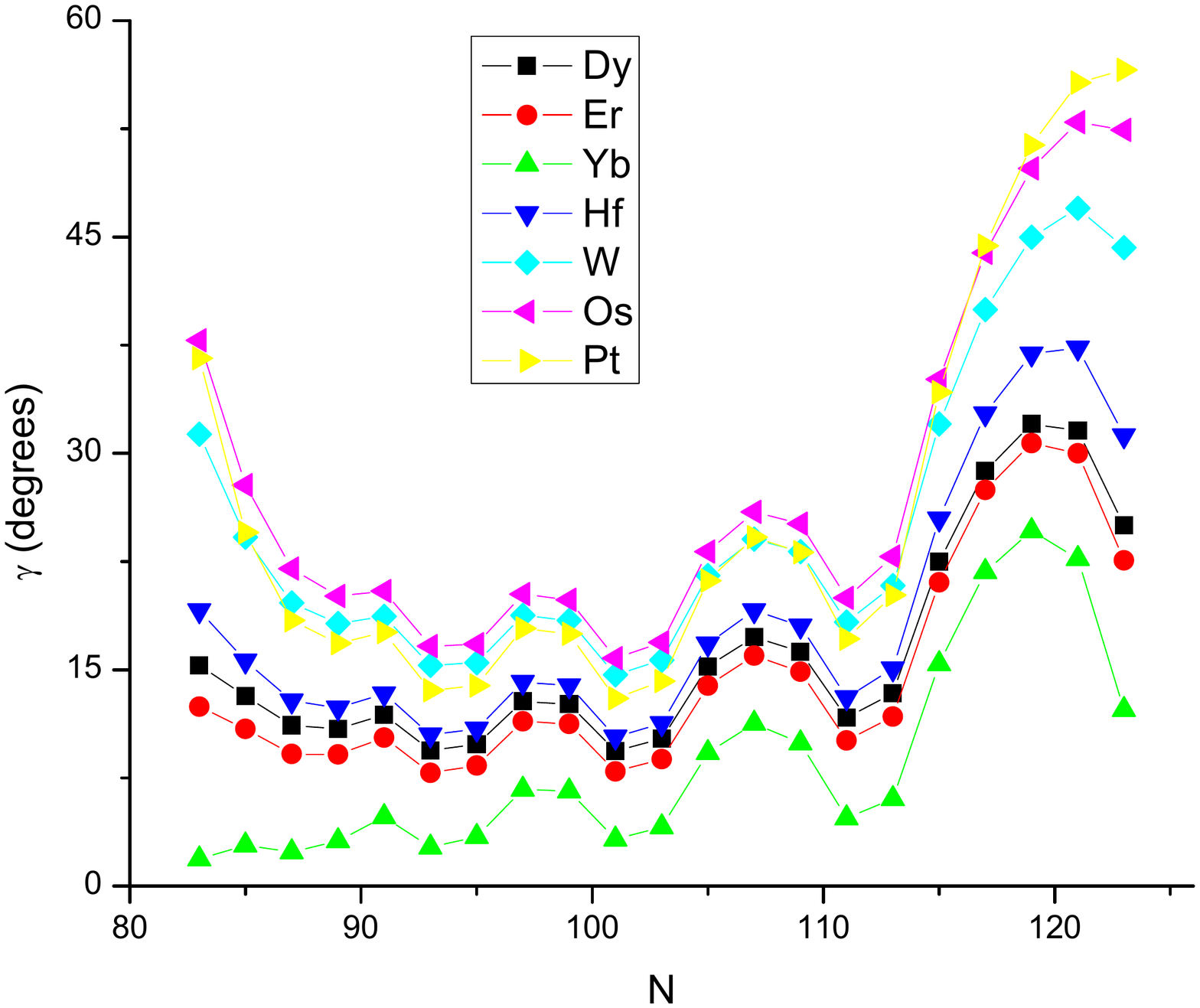,width=55mm}

\caption{Proxy SU(3) predictions for $\beta$ for even-odd rare earths with $Z=70$-78, compared with results reported in the mass table FRDM(2012)\cite{Moller}. In the two bottom panels,
the proxy-SU(3) predictions for $\gamma$ are reported for $Z=54$-78. See Section \ref{oddodd} for further discussion.}

\end{figure}



\begin{figure}[b]
\epsfig{file=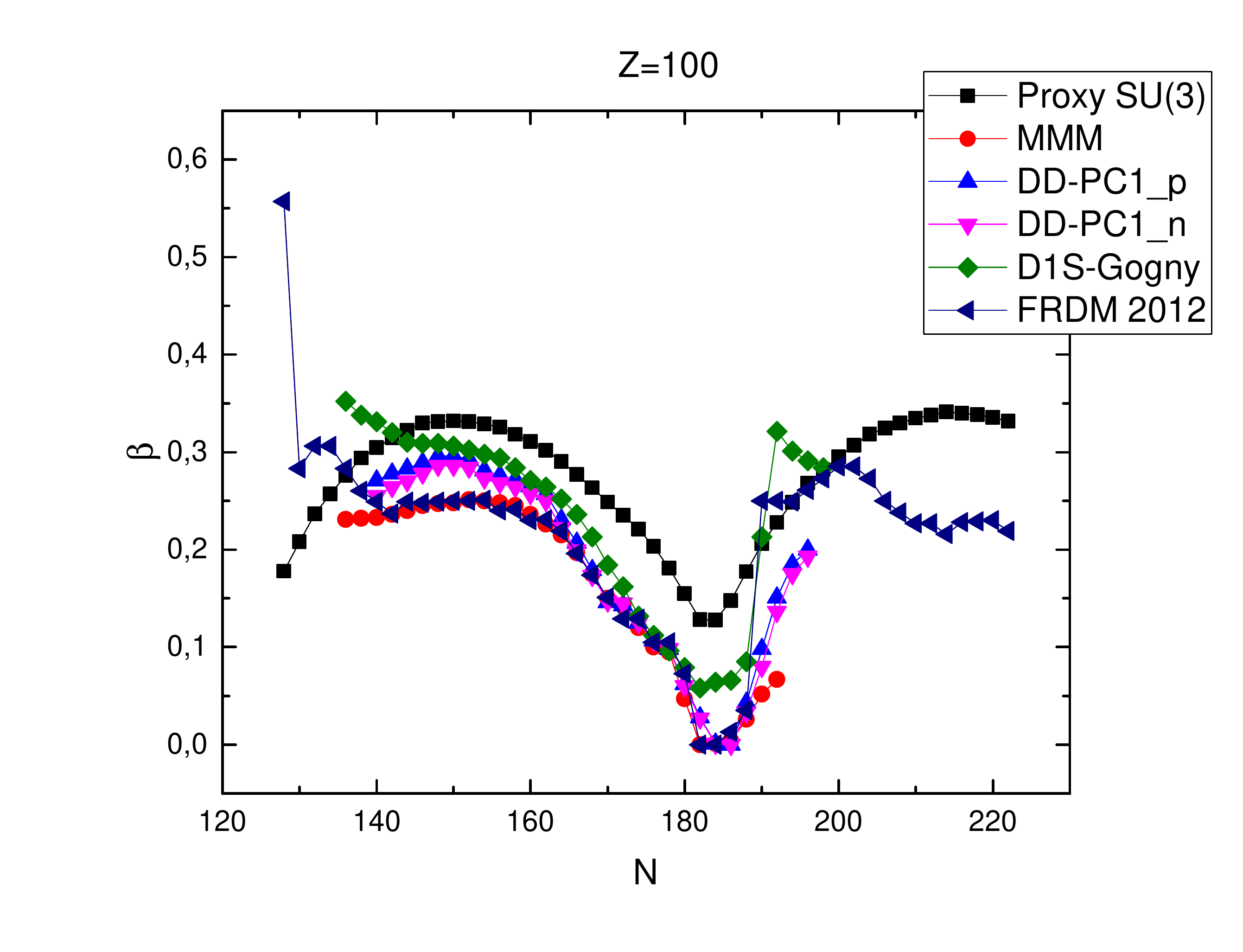,width=55mm}
\epsfig{file=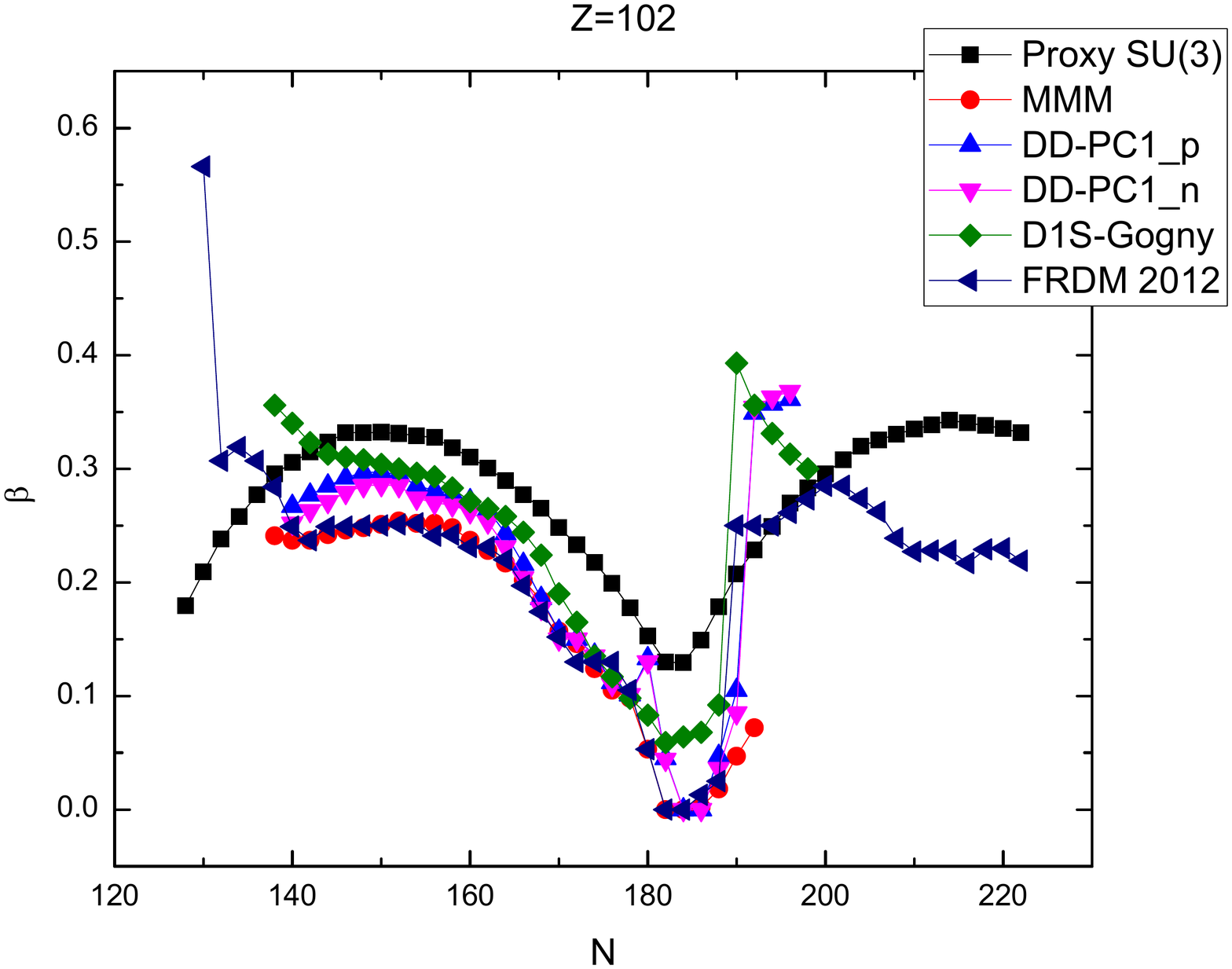,width=55mm}

\epsfig{file=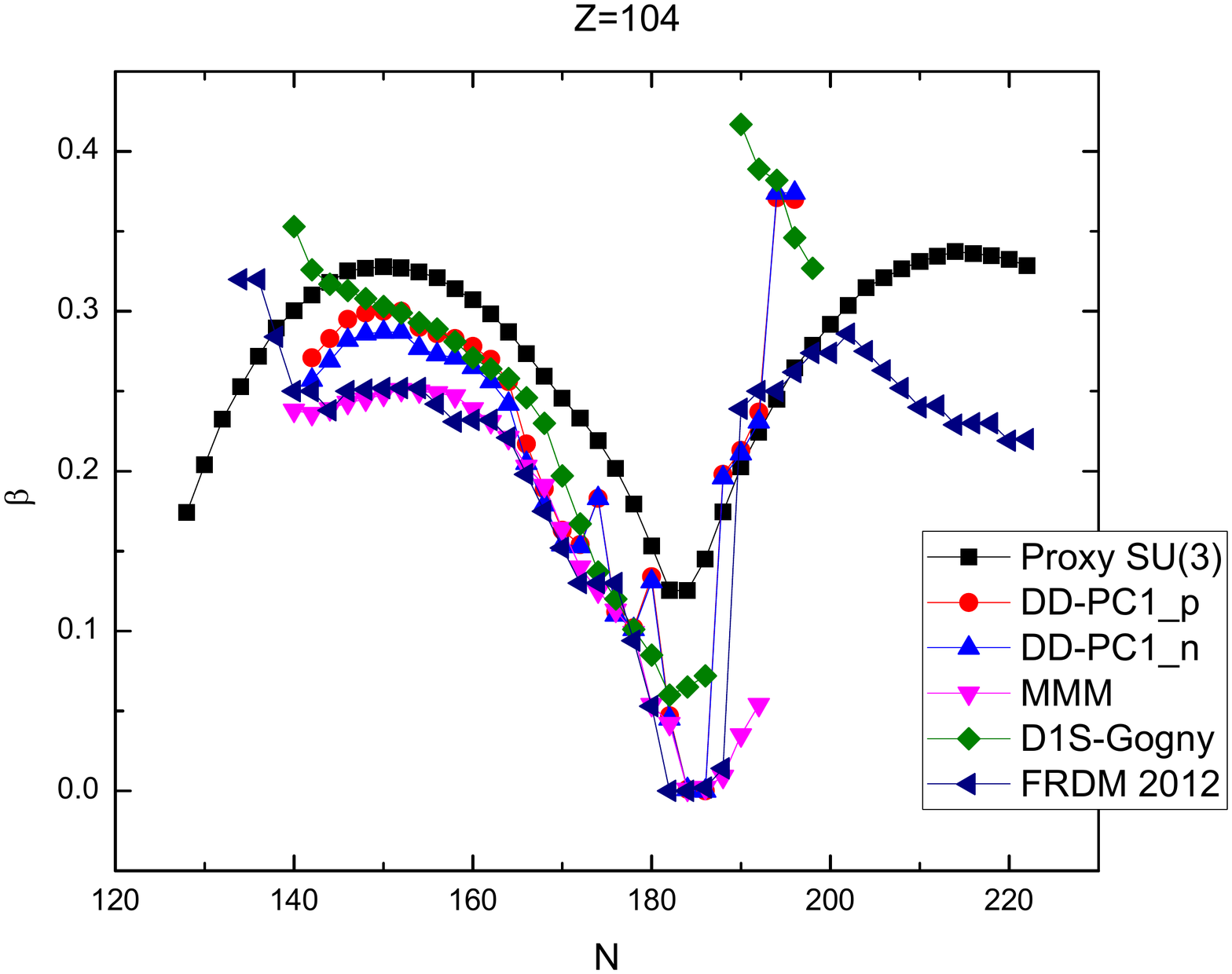,width=55mm}
\epsfig{file=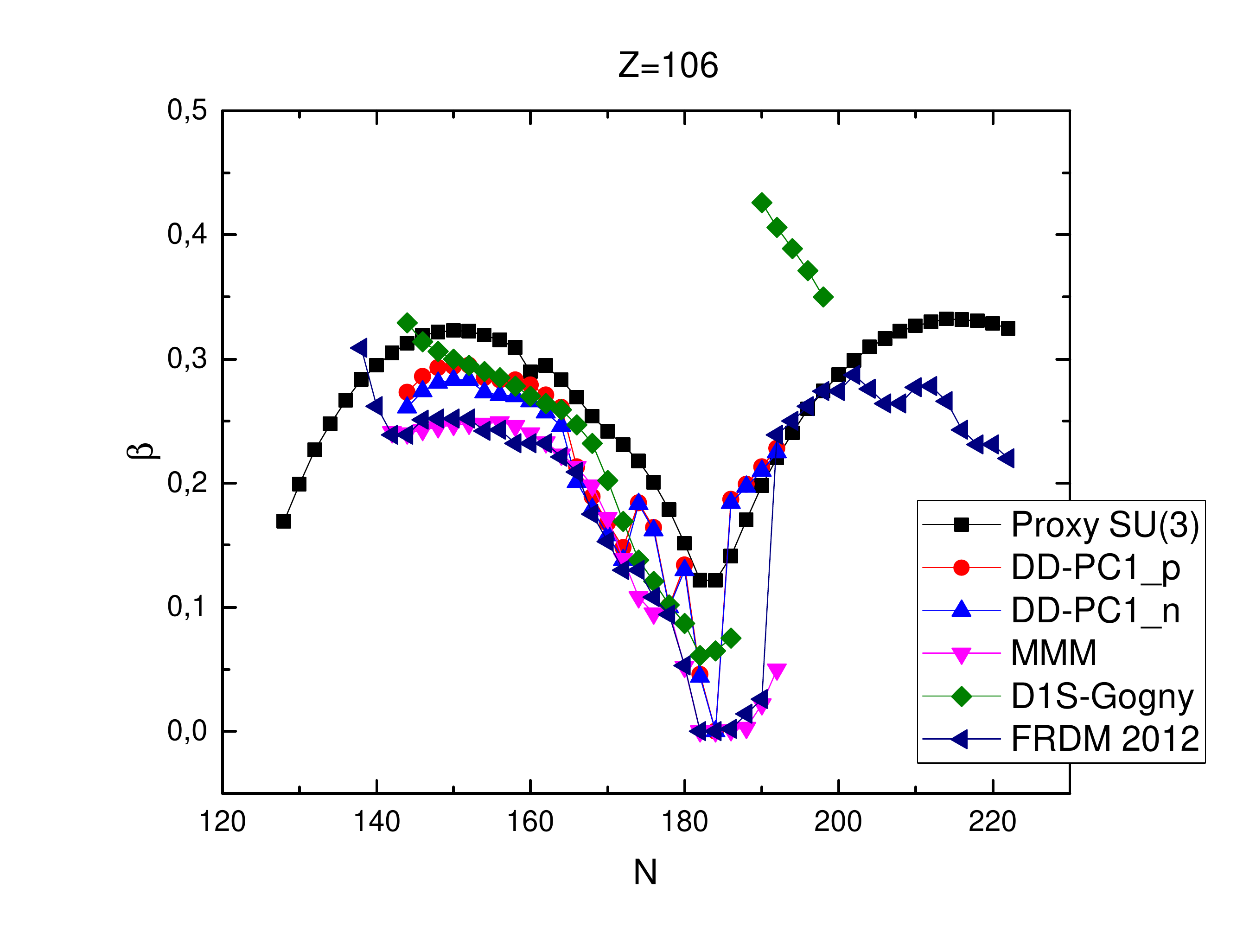,width=55mm}

\epsfig{file=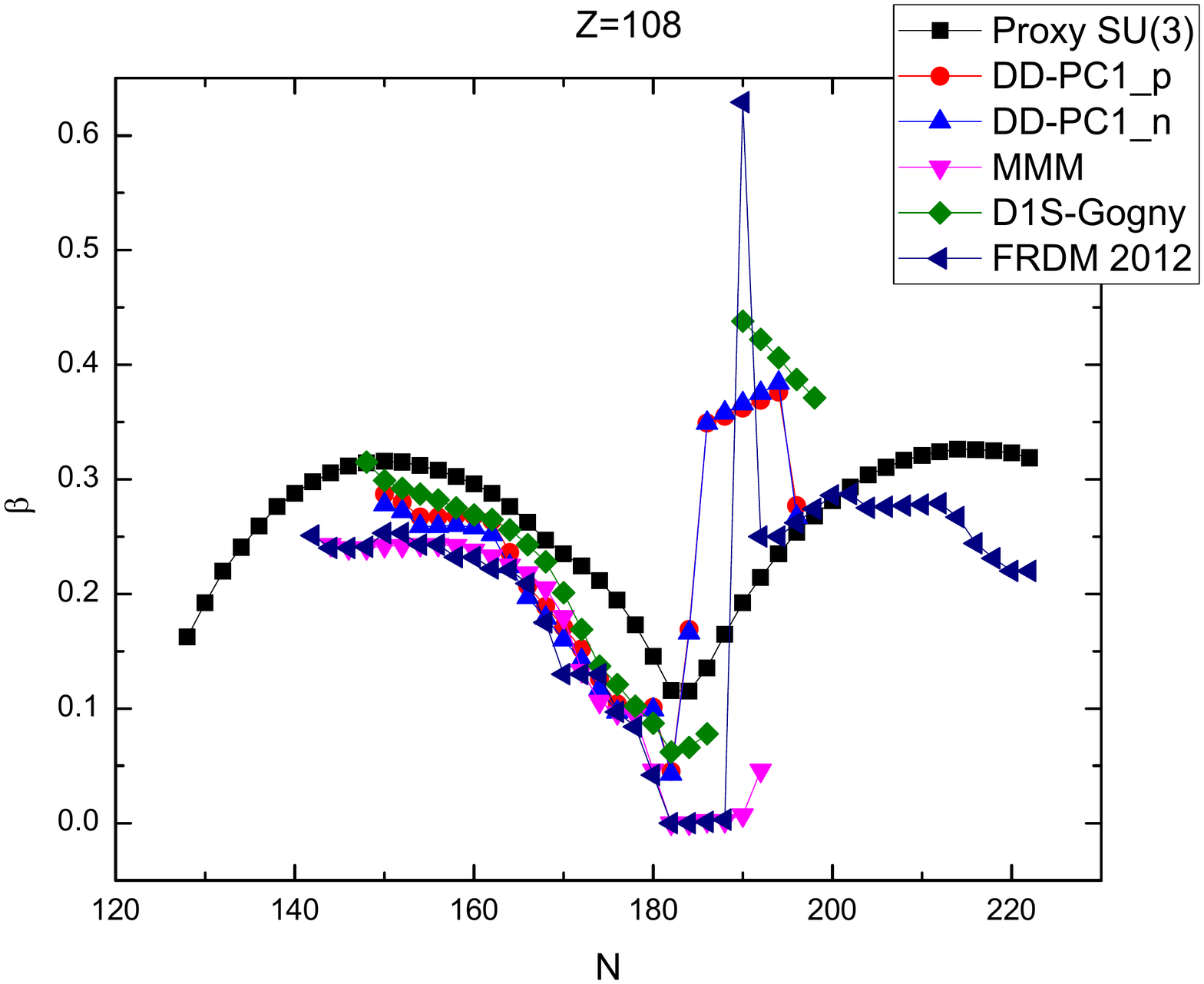,width=55mm}
\epsfig{file=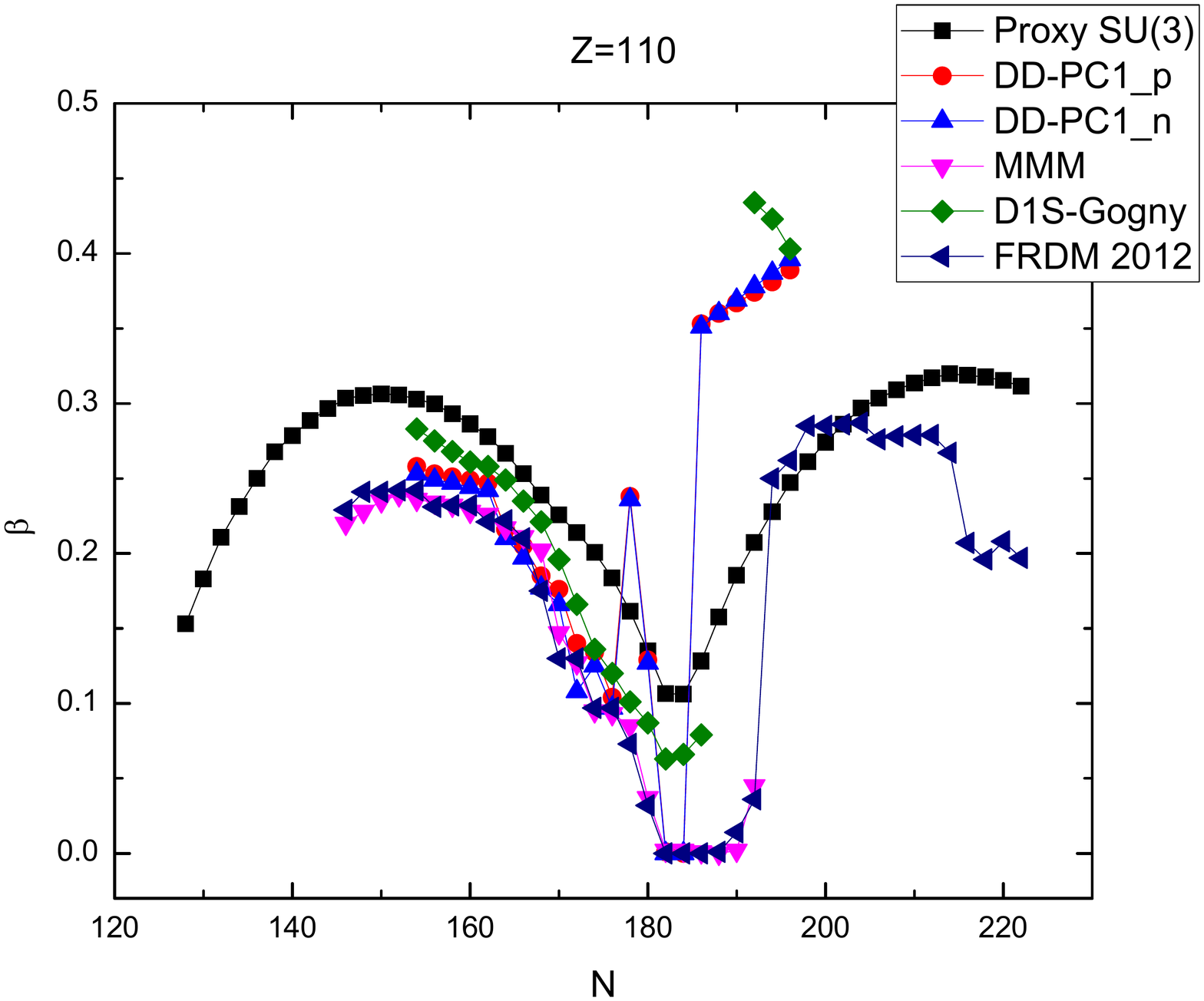,width=55mm}

\epsfig{file=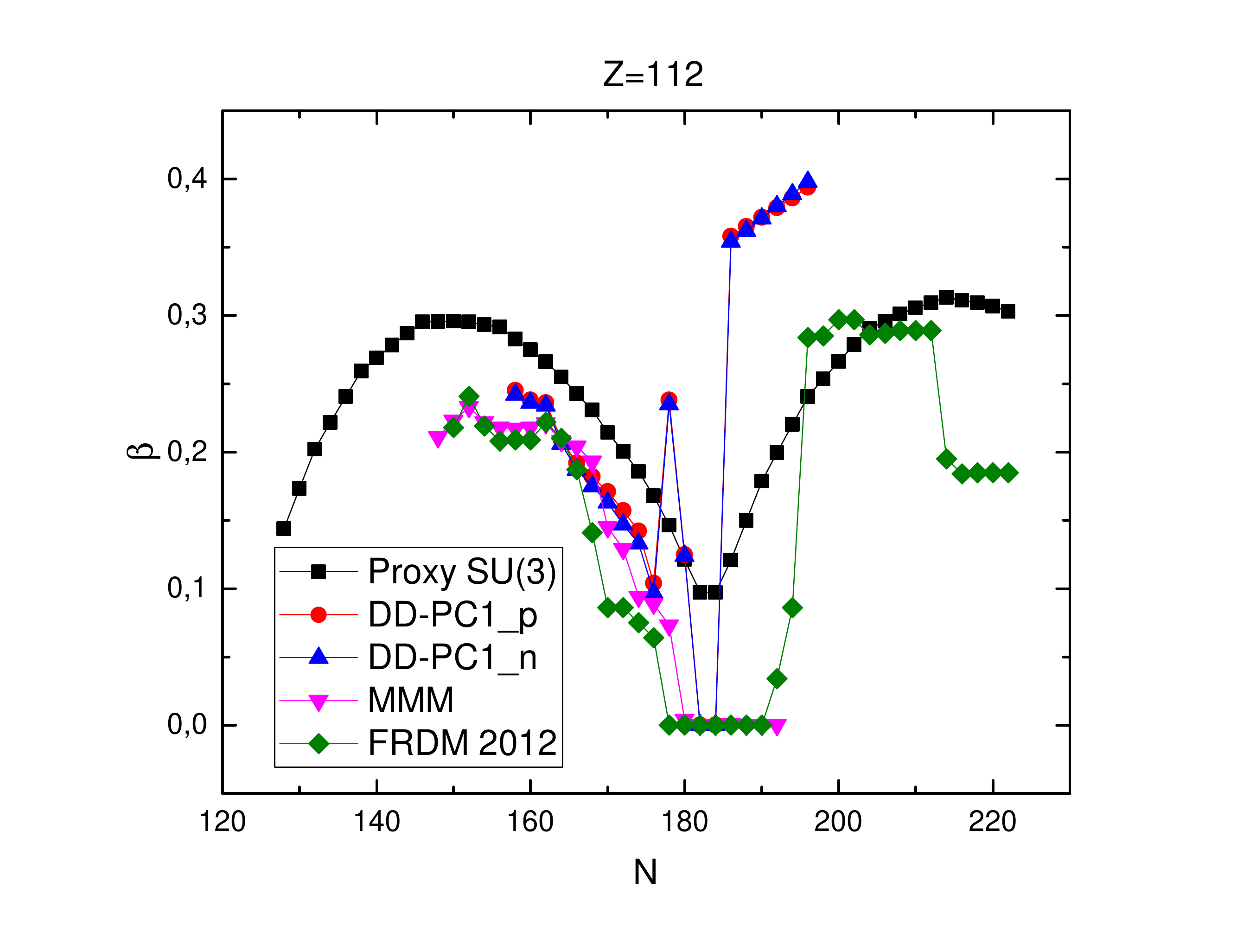,width=55mm}
\epsfig{file=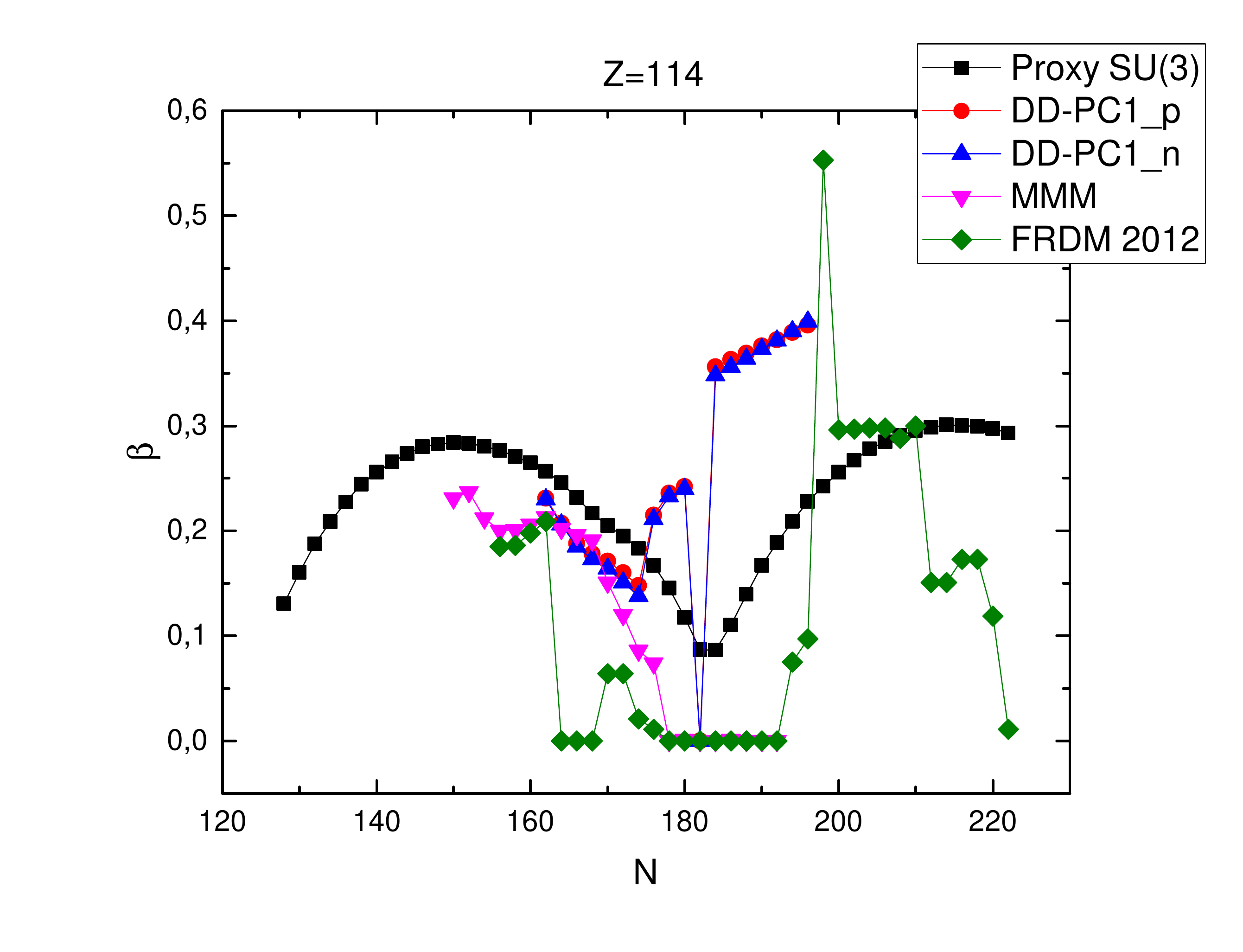,width=55mm}

\caption{Proxy SU(3) predictions for $Z$=100-114 for $\beta$, compared with
 covariant density functional theory with the DD-PC1 functional (DD-PC1) \cite{Ring} (in which case different values for protons (DD-PC1\_p) and neutrons (DD-PC1\_n) are reported), the microscopic-macroscopic method (MMM) \cite{Skalski}, the D1S-Gogny interaction (D1S-Gogny) 
\cite{Gogny},  and the mass table FRDM(2012)\cite{Moller}. See section \ref{eSHE} for further discussion.}

\end{figure}


\begin{figure}[b]
\epsfig{file=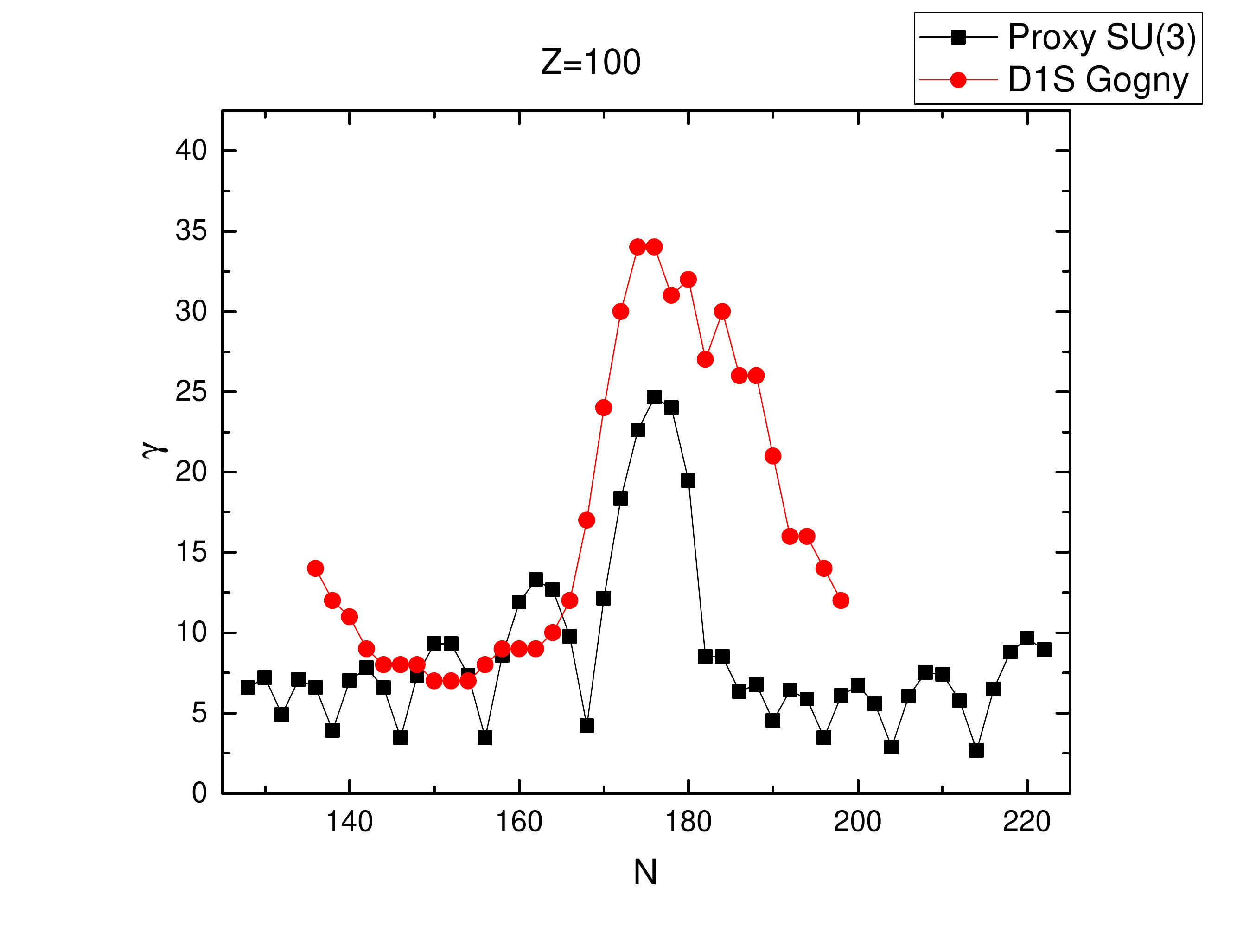,width=55mm}
\epsfig{file=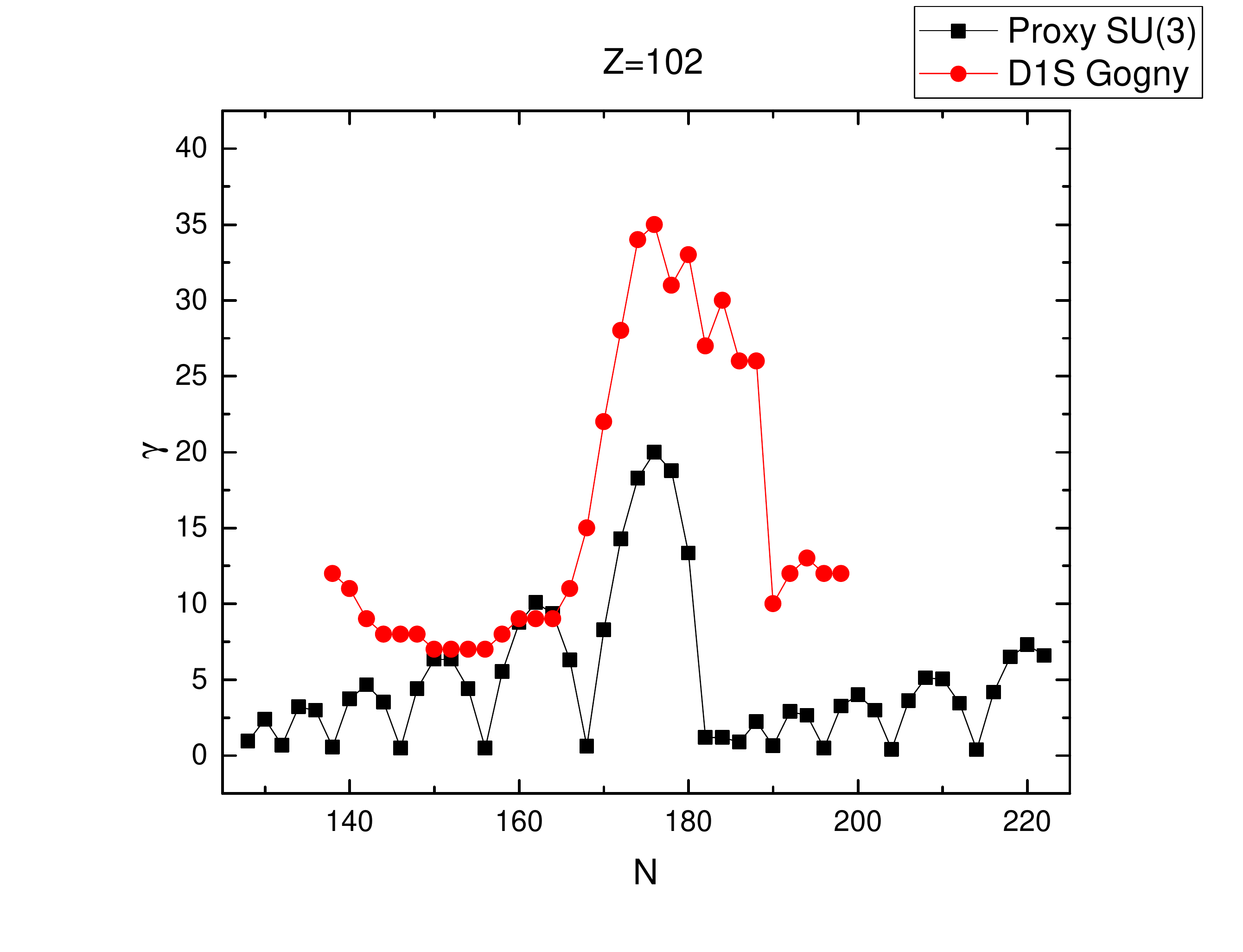,width=55mm}

\epsfig{file=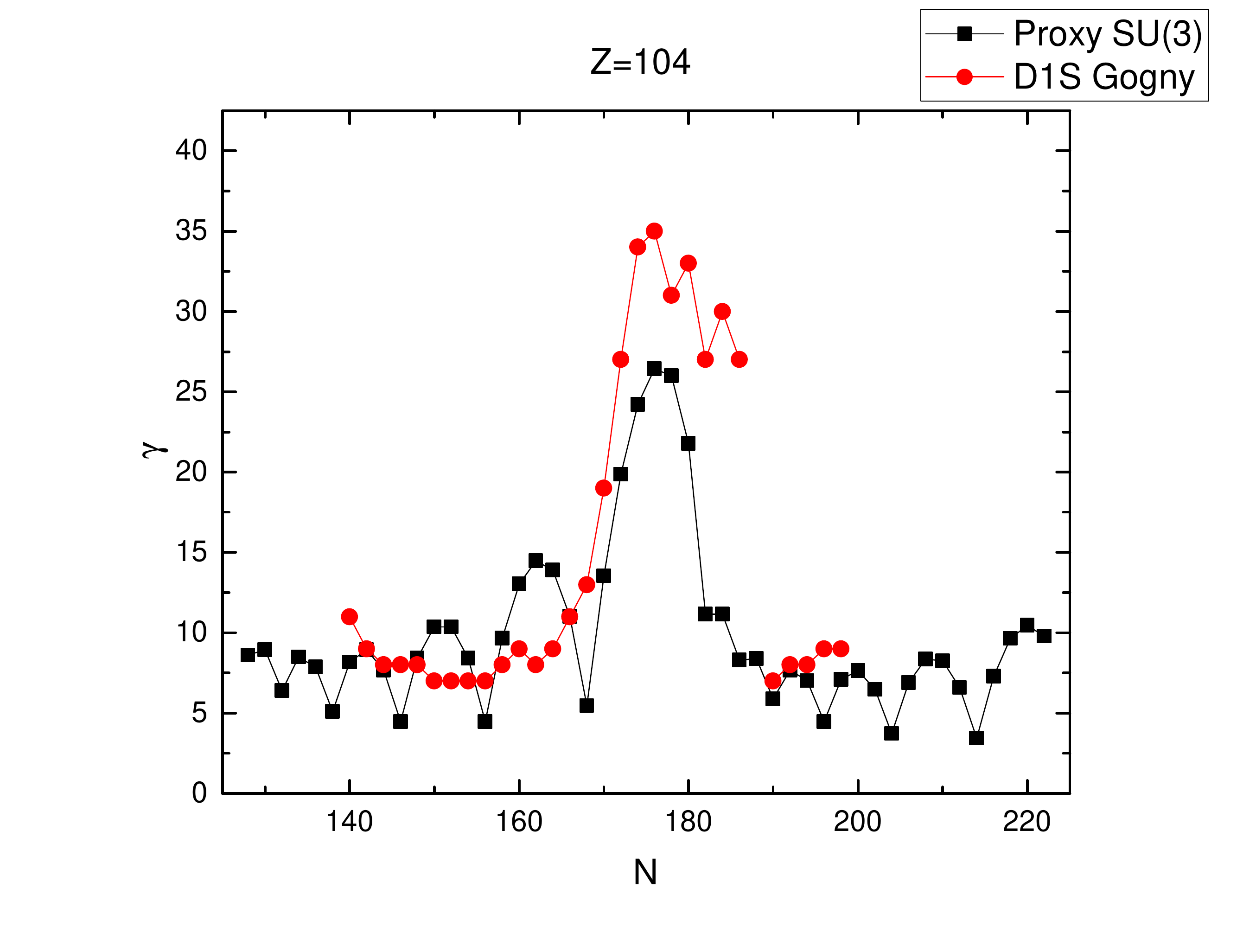,width=55mm}
\epsfig{file=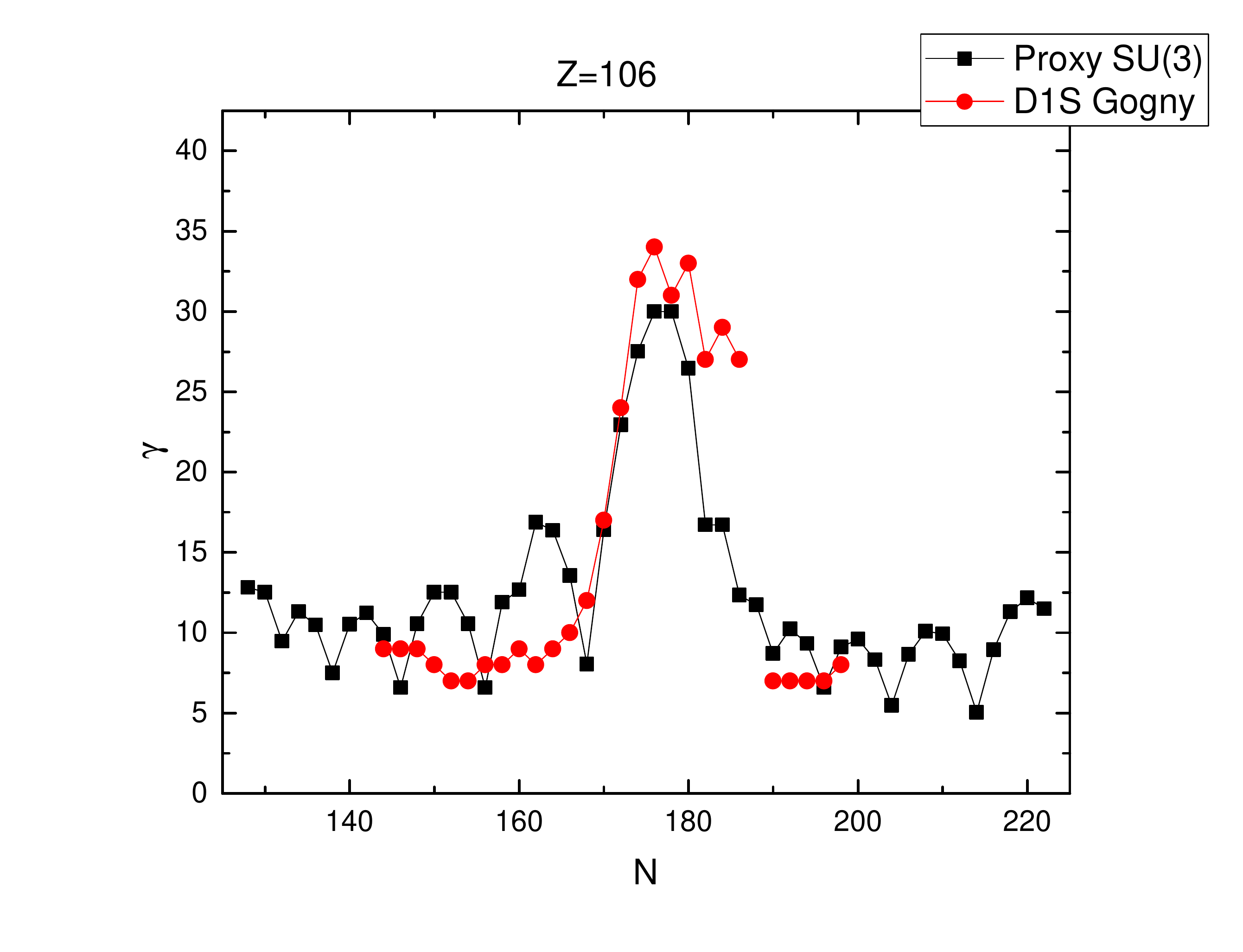,width=55mm}

\epsfig{file=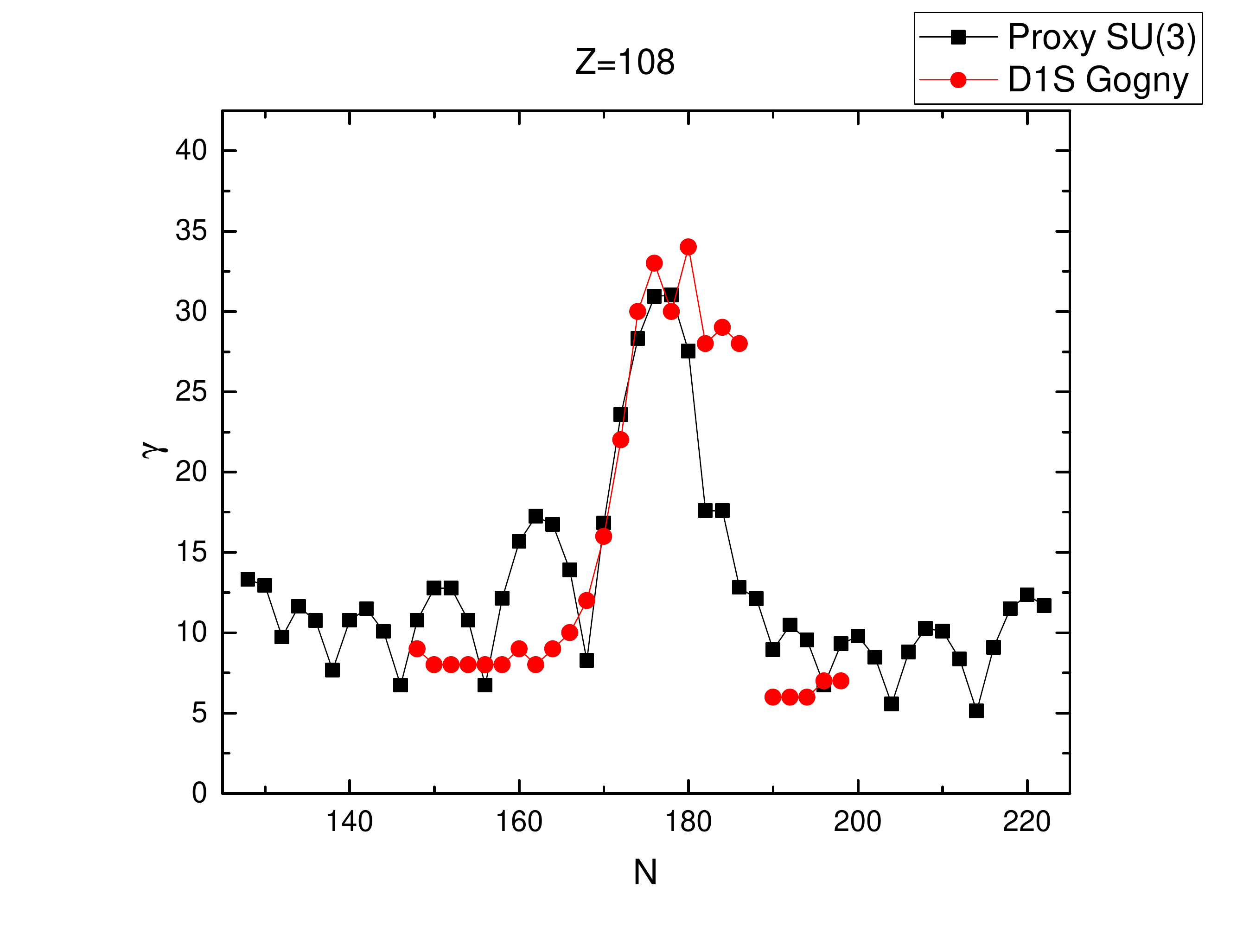,width=55mm}
\epsfig{file=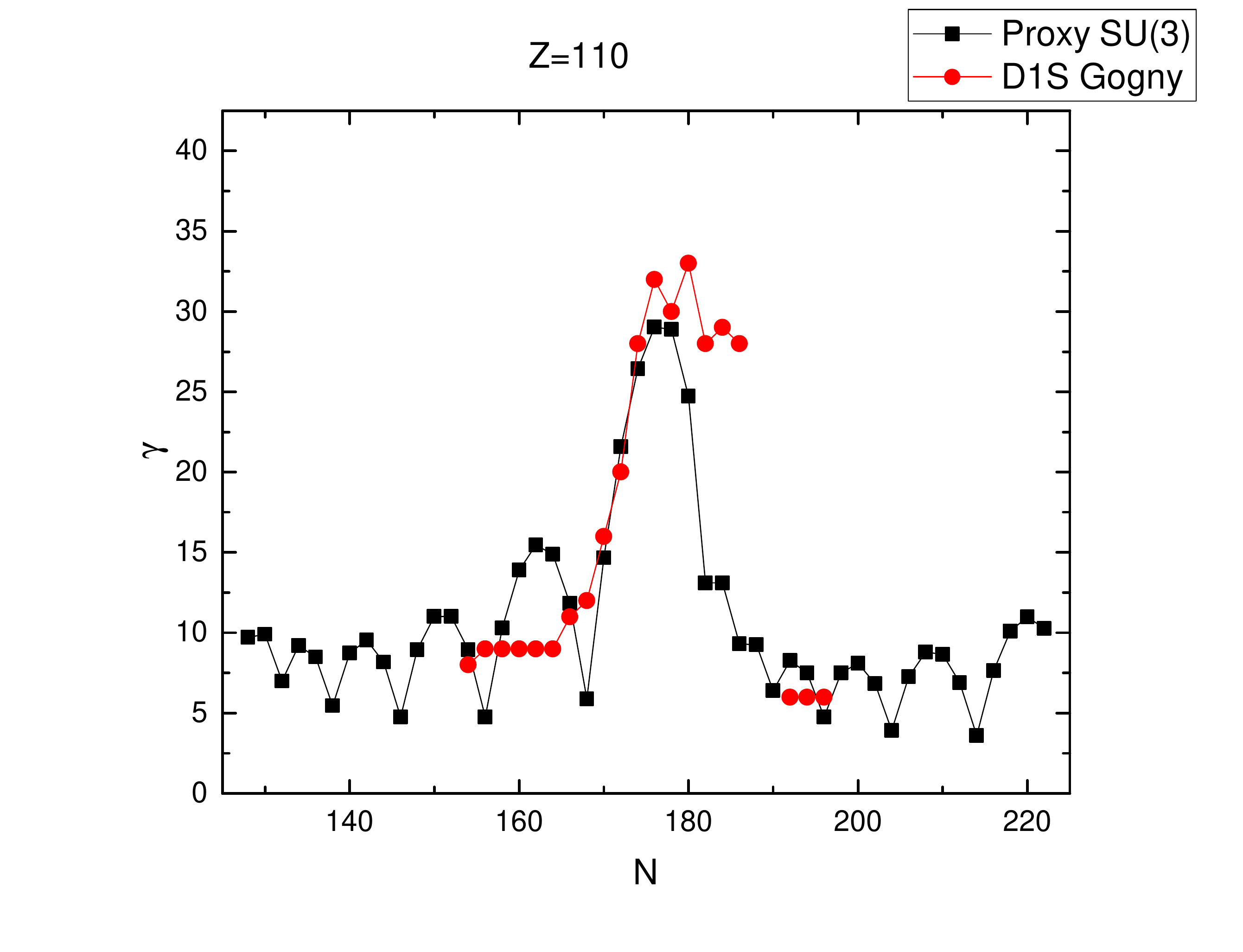,width=55mm}

\epsfig{file=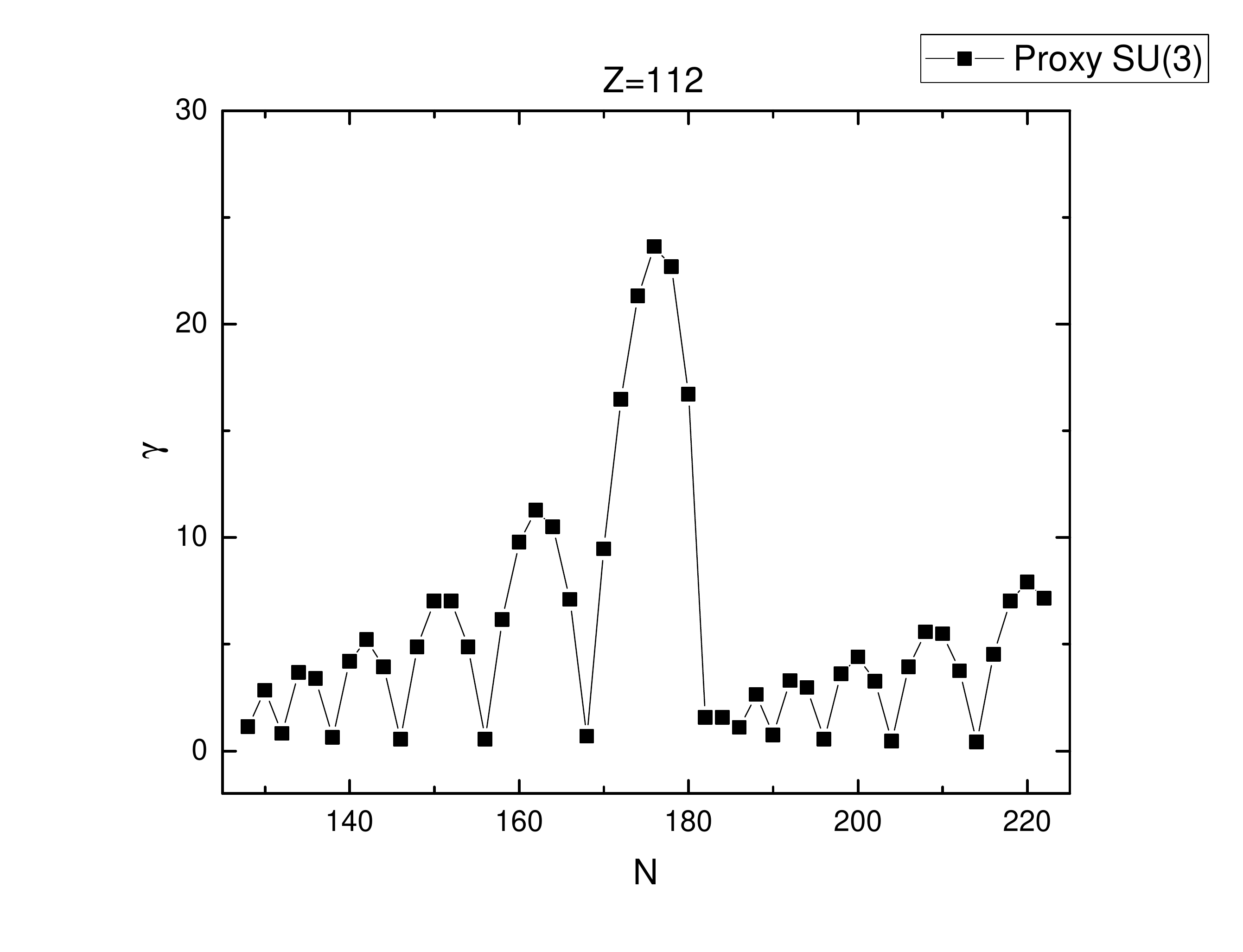,width=55mm}
\epsfig{file=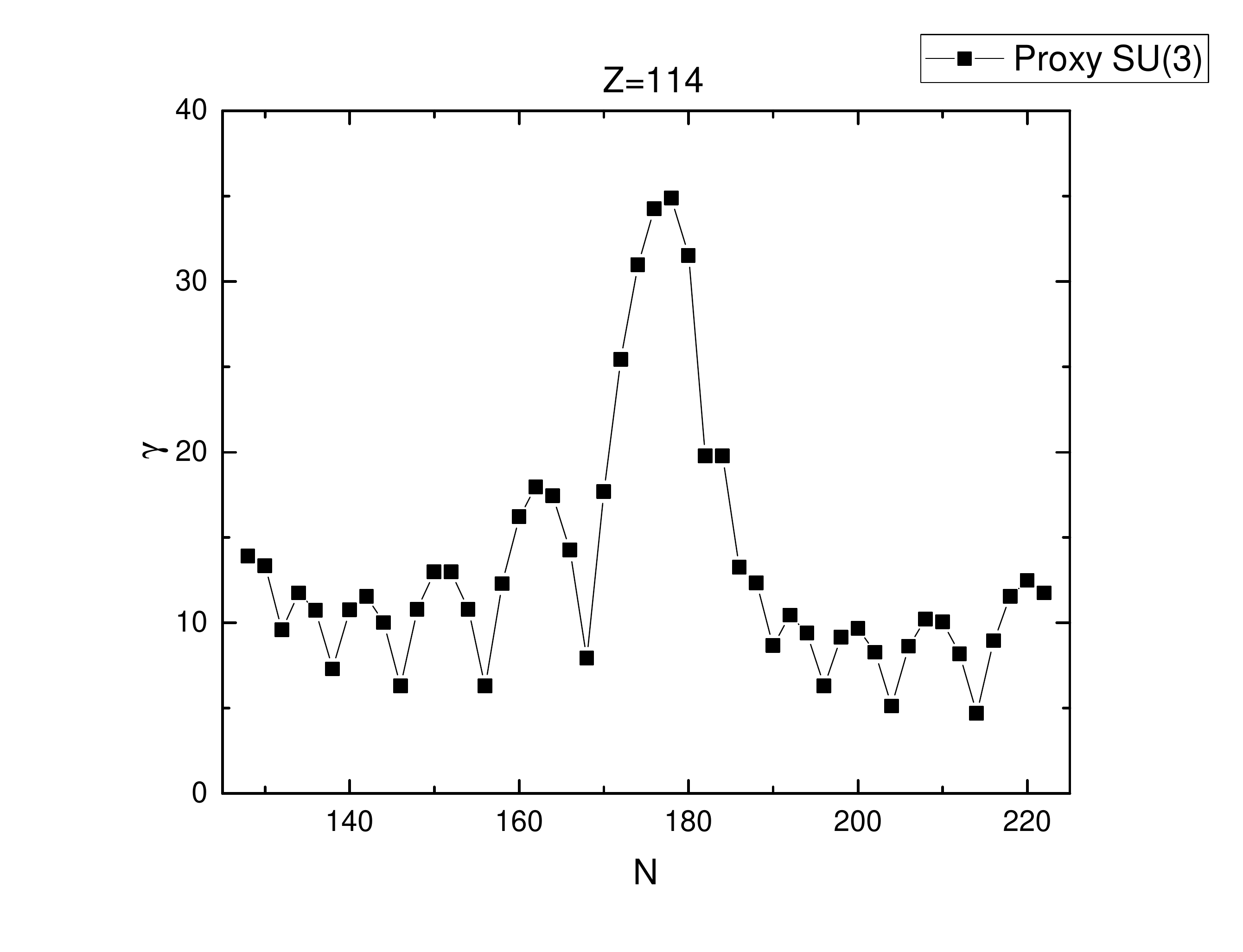,width=55mm}

\caption{ Proxy SU(3) predictions for $Z=100$-114 for $\gamma$, compared with results by the D1S-Gogny interaction (D1S-Gogny) \cite{Gogny}. See section \ref{eSHE} for further discussion.} 

\end{figure}


\begin{thebibliography}{99}

\bibitem{proxy1}
D. Bonatsos, I. E. Assimakis, N. Minkov, A. Martinou, R. B. Cakirli, R. F. Casten, and K. Blaum, Proxy SU(3) symmetry in heavy deformed nuclei, Phys. Rev. C \textbf{95}, 064325 (2017).


\bibitem{proxy2}
D. Bonatsos, I. E. Assimakis, N. Minkov, A. Martinou, S. Sarantopoulou, R. B. Cakirli, R. F. , and K. Blaum, Analytic predictions for nuclear shapes, prolate dominance and the prolate-oblate shape transition in the proxy-SU(3) model, Phys. Rev. C \textbf{95}, 064326 (2017).


\bibitem{Assimak}
I. E. Assimakis, D. Bonatsos, N. Minkov, A. Martinou, R.B. Cakirli, R.F. Casten, and K. Blaum, Foundations of the proxy-SU(3) symmetry in heavy nuclei, Proceedings of the International Workshop on Shapes and Dynamics of Atomic Nuclei: Contemporary Aspects (SDANCA17), Bulg. J. Phys. (2017) in press. 

\bibitem{EPJA} 
D. Bonatsos, Prolate over oblate dominance in deformed nuclei as a consequence of the SU(3) symmetry and the Pauli principle, Eur. Phys. J. A \textbf{53}, 148 (2017). 

\bibitem{Saranto}
S. Sarantopoulou, D. Bonatsos, I. E. Assimakis, N. Minkov, A. Martinou, R.B. Cakirli, R.F. Casten, and K. Blaum, Proxy-SU(3) symmetry in heavy nuclei: Prolate dominance and prolate-oblate shape transition, Proceedings of the International Workshop on Shapes and Dynamics of Atomic Nuclei: Contemporary Aspects (SDANCA17), Bulg. J. Phys. (2017) in press. 

\bibitem{Lalazissis}
G. A. Lalazissis, S. Raman, and P. Ring, Ground-state properties of even-even nuclei in the relativistic mean-field theory, At. Data Nucl. Data Tables \textbf{71}, 1 (1999). 

\bibitem{Raman}
S. Raman, C. W. Nestor, Jr., and P. Tikkanen, Transition probability from the ground to the 
first-excited $2^+$ state of even-even nuclides, At. Data Nucl. Data Tables \textbf{78}, 1 (2001). 

\bibitem{Wood}
J. L. Wood, K. Heyde, W. Nazarewicz, M. Huyse and P. van Duppen, Coexistence in even-mass nuclei, Phys. Rep. \textbf{215}, 101 (1992). 

\bibitem{Heyde}
K. Heyde and J. L. Wood, Shape coexistence in atomic nuclei,  Rev. Mod. Phys. \textbf{83}, 1467 (2011). 

\bibitem{Gogny}
J. -P. Delaroche, M. Girod, J. Libert, H. Goutte, S. Hilaire, S. P\'eru, N. Pillet, and G. F. Bertsch, Structure of even-even nuclei using a mapped collective Hamiltonian and the D1S Gogny interaction,  Phys. Rev. C {\bf 81}, 014303 (2010).

\bibitem{Moller} 
P. M\"oller, A. J. Sierk, T. Ichikawa, and H. Sagawa, 
Nuclear ground-state masses and deformations: FRDM(2012), Ad. Data Nucl. Data Tables 
{\bf 109-110}, 1 (2016). 

\bibitem{ENSDF}
Brookhaven National Laboratory ENSDF database http://www.nndc.bnl.gov/ensdf/


\bibitem{Ring}
S. E. Agbemava, A. V. Afanasjev, T. Nakatsukasa, and P. Ring, Covariant density functional theory: Reexamining the structure of superheavy nuclei, Phys. Rev. C {\bf 92}, 054310 (2015). 

\bibitem{Skalski} 
M. Kowal, P. Jachimowicz, and J. Skalski, Ground state and saddle point: masses and deformations for even-even superheavy nuclei with $98 \leq Z \leq 126$ and 
$134\leq  N \leq 192$,  arXiv:1203.5013 [nucl-th]. 

\end{thebibliography}
\end{document}